\documentclass[reprint,amsfonts, amssymb, amsmath,  showkeys,pra, superscriptaddress, twocolumn,longbibliography]{revtex4-2}
\usepackage{float}
\makeatletter
\let\newfloat\newfloat@ltx
\makeatother
\usepackage[english]{babel}
\usepackage[utf8]{inputenc}
\usepackage{graphics}
\usepackage{selinput}
\usepackage[normalem]{ulem}
\usepackage[shortlabels]{enumitem}

\usepackage{braket}
\usepackage{amsthm}
\usepackage{mathtools}
\usepackage{physics}
\usepackage{xcolor}
\usepackage{graphicx}
\usepackage[left=16mm,right=16mm,top=35mm,columnsep=15pt]{geometry} 
\usepackage{adjustbox}
\usepackage{placeins}
\usepackage[T1]{fontenc}
\usepackage{lipsum}
\usepackage{csquotes}
\usepackage{bm}

\usepackage[linesnumbered,ruled,vlined]{algorithm2e}
\SetKwInput{kwInit}{Init}

\def\HC{\mathcal{H}}

\def\LC{\mathcal{L}}

\def\ad{^{\dagger}}

\newcommand{\fsnull}[1]{}
\newcommand{\old}[1]{}

\usepackage[makeroom]{cancel}
\usepackage[toc,page]{appendix}
\usepackage[colorlinks=true,citecolor=blue,linkcolor=magenta]{hyperref}

\usepackage{tikz}
\tikzset{every picture/.style=remember picture}

\usepackage[utf8]{inputenc}
\usepackage{graphicx}
\usepackage{xcolor}
\usepackage{amsmath}
\usepackage{amsthm}
\usepackage{bm}
\usepackage{bbm}
\usepackage{comment}
\usepackage{appendix}
\usepackage{mathdots}
\usepackage{lipsum}
\usepackage{verbatim}
\usepackage{natbib}
\usepackage{nccmath}
\usepackage{amsfonts} 

\newcommand{\dya}[1]{\ket{#1}\!\bra{#1}}

\newcommand{\GC}{\mathcal{G}}

\newcommand{\OC}{\mathcal{O}}

\renewcommand{\geq}{\geqslant}
\renewcommand{\leq}{\leqslant}

\renewcommand{\Re}{\text{Re}}
\renewcommand{\Im}{\text{Im}}

\newcommand{\spn}{{\rm span}}

\DeclareMathOperator*{\argmin}{arg\,min}
\renewcommand{\vec}[1]{\boldsymbol{#1}}  

\newcommand*{\id}{\openone}

\newcommand{\bs}{\textsf{BS}}

\newcommand{\losalamos}{Theoretical Division, Los Alamos National Laboratory, Los Alamos, New Mexico 87545, USA}

\def\be{\begin{equation}}
\def\ee{\end{equation}}
\def\bs{\begin{split}}
\def\e{\end{split}}
\def\ba{\begin{eqnarray}}
\def\bea{\begin{eqnarray}}

\def\tea{\end{eqnarray}}
\def\ea{\end{eqnarray}}
\def\eea{\end{eqnarray}}

\newcommand\mbb[1]{\mathbb{#1}}
\newcommand\mf[1]{\mathfrak{#1}}

\def\Lie{\text{Lie}}

\newtheorem{theorem}{Theorem}

\newtheorem{corollary}{Corollary}

\newtheorem{proposition}{Proposition}

\usepackage{amssymb}
\usepackage{dsfont}

\def\be{\begin{equation}}
\def\te{\end{equation}}
\def\ee{\end{equation}}
\def\ba{\begin{eqnarray}}
\def\bea{\begin{eqnarray}}

\def\tea{\end{eqnarray}}
\def\ea{\end{eqnarray}}
\def\eea{\end{eqnarray}}

\begin{document}

\title{Parallel-in-time quantum simulation via Page and Wootters quantum time}  

\author{N. L. Diaz}
\affiliation{Information Sciences, Los Alamos National Laboratory, Los Alamos, New Mexico 87545, USA}
\affiliation{Departamento de F\'isica-IFLP/CONICET,
		Universidad Nacional de La Plata, C.C. 67, La Plata 1900, Argentina}
        \affiliation{Center for Non-Linear Studies, Los Alamos National Laboratory, Los Alamos, New Mexico 87545, USA}
  
\author{Paolo Braccia}
\affiliation{\losalamos}
\affiliation{Dipartimento di Fisica, Universit\`a di Firenze and INFN Sezione di Firenze,
Via G. Sansone 1, 50019 Sesto Fiorentino, Italy}

\author{Martin Larocca}
\affiliation{\losalamos}
\affiliation{Center for Nonlinear Studies, Los Alamos National Laboratory, Los Alamos, New Mexico 87545, USA}

\author{J.M. Matera}
\affiliation{Departamento de F\'isica-IFLP/CONICET,
		Universidad Nacional de La Plata, C.C. 67, La Plata 1900, Argentina}
  
\author{R. Rossignoli}
\affiliation{Departamento de F\'isica-IFLP/CONICET,
		Universidad Nacional de La Plata, C.C. 67, La Plata 1900, Argentina}
\affiliation{Comisi\'on de Investigaciones Cient\'{\i}ficas (CIC), La Plata 1900, Argentina}
 
\author{M. Cerezo}
\affiliation{Information Sciences, Los Alamos National Laboratory, Los Alamos, New Mexico 87545, USA}

\begin{abstract}
In the past few decades, researchers have created a veritable zoo of quantum algorithms by drawing inspiration from classical computing, information theory, and even from physical phenomena. Here we present quantum algorithms for parallel-in-time simulations that are inspired by the Page and Wootters formalism. In
this framework, and thus in our algorithms, the classical time-variable of quantum mechanics is promoted to the quantum realm by introducing a Hilbert space of ``clock'' qubits which are then entangled with the ``system'' qubits. 
We show that our algorithms can compute  temporal properties over $N$ different times of many-body systems by only using $\log(N)$ clock qubits. As such, we  achieve  an exponential trade-off between time and spatial complexities. In addition, we rigorously prove that the entanglement created between the system qubits and the clock qubits has operational meaning, as it encodes valuable information about the system's dynamics.
We also provide a circuit depth estimation of all the protocols, showing a running time advantage in computation times over traditional sequential-in-time algorithms. In particular, for the case when the dynamics are determined by the Aubry–Andre model, we present a hybrid method for which our algorithms have a depth that only scales as $\OC(\log(N)n)$. 
As a by-product, we can relate the previous schemes to the problem of equilibration of an isolated quantum system, thus indicating that our framework enables a new dimension for studying dynamical properties of many-body systems.

\end{abstract}

\maketitle

\section{Introduction}

The field of quantum foundations  studies the fundamental principles of quantum theory, such as the nature of quantum states, the interpretation of measurements, the equilibration and thermalization of isolated systems, and the emergence of classicality~\cite{auletta2001foundations,zurek2003decoherence,schlosshauer2005decoherence,reimann2008foundation,wheeler2014quantum}. 
Another important question that has recently attracted wide attention within this field is that of the role of time in quantum mechanics (QM) \cite{mcclean2013feynman, giovannetti2015quantum,boette2016system,boette2018history,horsman2017can,diaz2019history, diaz2019historystate, diaz2021spacetime, diaz2021path,pabon2019parallel,mendes2019time,favalli2020time,valdes2020emergent,castro2020quantum,mendes2021non,foti2021time,hohn2021trinity,hohn2021equivalence,lomoc2022history,loc2022time,paiva2022non,paiva2022flow,favalli2022peaceful,rijavec2022heisenberg,baumann2022noncausal,apadula2022quantum,barison2022variational,giovannetti2023geometric,chester2023covariant}: It is clear that ever since its inception, time in QM has been treated as an external classical parameter, in asymmetry with other quantum observables. 
For instance, in the canonical quantization procedure, one promotes the position and momentum variables to operators and the Poisson bracket to a commutator~\cite{sakurai1995modern}. 
This quantization is implemented at a fixed time value so  
the variable $t$ appearing in Schr\"{o}dinger equation is the same as that appearing in the classical 
equations of motion. While seemingly innocuous, it is believed that the imbalance between time and space could be a critical issue in developing a quantum theory of gravity~\cite{isham1993canonical,gambini2009conditional,kuchavr2011time, hohn2021trinity}. 
At the same time, such asymmetry inevitably limits the range of applicability of quantum information and computation tools, as asking questions like ``\textit{what is the entanglement between the space and time coordinates?}'' is an entirely moot point within the conventional quantum mechanical framework.

It is tempting to fix the space-time asymmetry by promoting $i\hbar\frac{d}{dt}$ to a quantum operator conjugate to some quantum time observable $T$, such that $[T,H]= i\hbar$. However, the previous approach has the critical issue that it forces $T$ and $H$ to have exactly the same eigenspectrum, which is generally incompatible (such argument is often attributed to Pauli \cite{pauli1933allgemeinen}).
Despite this apparent difficulty and other subtleties, there are  
several proposals 
to treat time on equal footing with other physical quantities 
\cite{page1983evolution,connes1994neumann,isham1994quantum,fitzsimons2015quantum,cotler2018superdensity,diaz2021spacetime,diaz2021path}. Here, we will focus on the so-called Page and Wootters (PaW) mechanism~\cite{page1983evolution}. 
In this framework the universe is composed by a quantum system of interest plus an ancillary \textit{clock} quantum system, such that the joint state of the universe, the \textit{history state}, is in a stationary state. The previous ``Pauli's objection'' is circumvented since the operator $T$ acts on the clock system implying $[T,H]=0$. Remarkably, as long as 
the system and the clock are correlated in a specific way, the unitary evolution of the system can be restored by  conditioning over the clock states. 
In this way, measures of \textit{system-time entanglement} become rigorous 
quantifiers of the amount of distinguishable evolution undergone by the system during its history~\cite{boette2016system}.

The foundational discussion surrounding the role of time has many intriguing ramifications, even when focusing solely on the PaW mechanism.
The interested reader can refer to the Appendix \ref{sec:literat} for pertinent discussions. However, the primary objective of this work  is computational: In this manuscript, we provide a translation of the PaW mechanism into a useful quantum computational scheme where the quantum aspects of time are captured by clock qubits.  This allows us to
develop quantum algorithms for studying temporal averages of several dynamical properties of a quantum system. Specifically, given an $n$-qubit quantum system that is evolving under the action of a time-independent Hamiltonian $H$, we consider the problem of approximating the infinite-time average of some time-dependent dynamical quantity by a discrete sum over $N$ different times. In a standard setting, we can estimate said discrete average  by sequentially running $N$ different quantum circuits (one for each time in the  average).  However, by  leveraging the history state of the Page and Wootters formalism we propose a quantum algorithm for parallel-in-time simulations that uses $\log(N)$ (any logarithm in this manuscript is taken in base $2$) ancillary clock qubits and that allows us to evaluate the temporal average with a single quantum circuit. The previous shows that using the history state leads to an exponential trade-off between  temporal complexity (running multiple circuits) and spatial complexity (using more qubits). 

In addition, we also show that the entanglement between the system and the clock qubits carries operational meaning since it serves as a bound for the infinity time average of the  Loschmidt echo and for the temporal variance of expectation values. These results imply that the history state encodes valuable information in its correlations that can be used to study and understand the system's dynamics and equilibration. Given this operational meaning of the entanglement, we present two different schemes to compute the linear entropy of the history state, one based on the state-overlap circuit~\cite{cincio2018learning}, and another one leveraging classical shadows and randomized measurements~\cite{huang2020predicting,brydges2019probing}

We also present a depth study of the circuits showing a clear advantage of using parallel-in-time protocols over the conventional sequential-in-time approaches where time is not mapped to clock qubits. Moreover, we   propose a scheme to further reduce the circuit depth needed to prepare the history state via  Hamiltonian diagonalization~\cite{commeau2020variational,kokcu2021fixed}. Here, we show that by leveraging tools from variational quantum algorithms~\cite{cerezo2020variationalreview,peng2020simulating,bharti2021noisy, ollitrault2022quantum} and quantum machine learning~\cite{biamonte2017quantum,schuld2015introduction,cerezo2022challenges} to variationally diagonalize $H$, one can significantly  further reduce the required circuit depth. For the special case when $H$ is given by the Aubry–Andre model, we show that all of our algorithms can be implemented with a depth that only scales as $\OC(\log(N)n)$, i.e., as the product of the number of clock and system qubits. Finally, we perform simulations which showcase the performance of our algorithms for studying temporal averages of systems evolving under an interacting and non-interacting Aubry–Andre model \cite{vstrkalj2021many}, illustrating the possibility of studying many-body localization.

\begin{figure*}[t!]
\centering
    \includegraphics[width=1\linewidth]{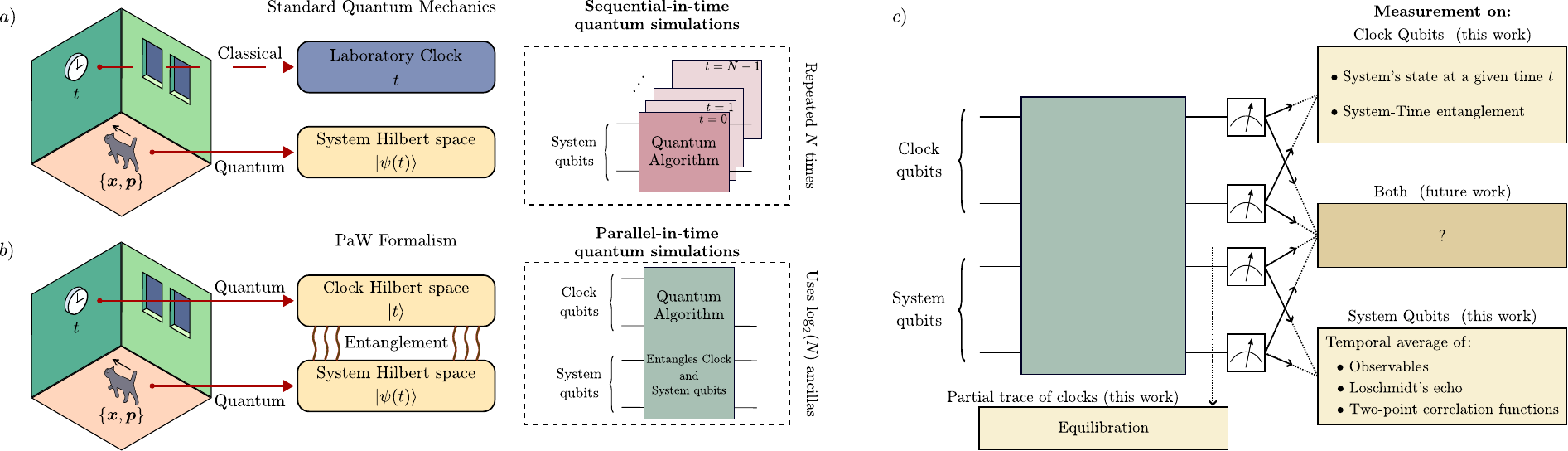}
\caption{\textbf{Quantum algorithms based on the standard quantum mechanics or the PaW formalism.} 
a) In Hamiltonian classical mechanics, dynamical variables are functions of the phase space coordinates position $\vec{x}$ and momentum $\vec{p}$. In standard quantum mechanics, one promotes $\vec{x}$ and  $\vec{p}$ to quantum operators, but the time variable $t$ is treated as a classical parameter that is external to the quantum system being studied. Quantum algorithms for studying dynamical properties based on this framework are  implemented for some fixed time $t$. If we want to compute an average of $N$ times, we need to repeat the run $N$ different sequential-in-time experiments.  b) In the PaW formalism, time is treated as a quantum variable, with its own associated Hilbert space. In this work we present quantum algorithms for parallel-in-time simulations which  trade circuit repetitions for ancilla clock qubits. c) After a proper entangling protocol one has access not only to properties of the system at a given time, but also to their complete history. This information can be retrieved by performing measurements at the end of the circuit, which now can involve either the clock qubits, the system qubits or both.  Measurements on the system which are conditioned to a certain time value give properties of the system at a given time. More interestingly, if one only measures on the system and complete ignore the clock's values, temporal averages  are obtained. This is a consequence of the entanglement between the system and the clock which induces a useful quantum channel when the clock is treated as an environment.  Because of the quantum nature of the simulated clock and system, many other measurements can be proposed, meaning that the different protocols we discuss in this manuscript do not exhaust all the possibilities opened by this computational framework. }
\label{fig:1}
\end{figure*}

\section{Quantum time formalism and its discretization} 

Let us consider an $n$-qubit quantum system with associated Hilbert space $\HC_S$. Then, let $H$ be a time-independent Hamiltonian under which the system evolves. The dynamical evolution of the system  is determined by the Schr\"{o}dinger equation
\begin{equation}\label{eq:schrodinger}
    i \frac{d}{dt}\ket{\psi(t)} =H\ket{\psi(t)}\,,
\end{equation}
where we have set $\hbar=1$. It is well known that  the solution of Eq.~\eqref{eq:schrodinger} is given by  
\begin{equation}\label{eq:unitary-t-si}
    \ket{\psi(t)}=U(t)\ket{\psi_0}\,, \quad \text{with}\quad  U(t)=e^{-iHt}\,,
\end{equation}
and where $\ket{\psi_0}$ is some initial state of the system.

As previously discussed, and as shown in Fig.~\ref{fig:1}(a), there exists an inherent asymmetry between the space and time variables in quantum mechanics. Namely, the $t$ variable over which we take a derivative is a fully classical parameter that is external to the quantum system. An alternative to fully incorporate time in a quantum framework is 
to introduce a new Hilbert space  $\mathcal{H}_T$ 
spanned by some states $|t\rangle$ (see Fig.~\ref{fig:1}(b)) such that  $T|t\rangle=t |t\rangle$ and $    [T,P_T]=i\hbar$, which in the time basis leads to $P_T\equiv -i\hbar\frac{d}{dt}$. Note that $P_T$ is not the Hamiltonian of the system and in fact $[T,H]=0$ (as they act on distinct Hilbert spaces). 
Evolution is then recovered from an extended Schrodinger equation, involving both the system and the clock Hilbert spaces, which is given by  $\mathcal{J}|\Psi\rangle=0$, 
for $\mathcal{J}=   P_T\otimes \id_S + \id_T\otimes H$ and $|\Psi\rangle \in \mathcal{H}_T\otimes \mathcal{H}_S$. Here $\id_T$ and $\id_S$ respectively denote the identities on $\mathcal{H}_T$ and $ \mathcal{H}_S$. In general, the extended Schrodinger equation, together with an initial condition, leads to entanglement between the system and the time Hilbert space.

The previous scheme can also be regarded as the mathematical basis of the Page and Wootters (PaW) mechanism. Under this framework the \textit{universe state} $|\Psi\rangle$ is stationary (as $\mathcal{J}|\Psi\rangle=0$) while the unitary evolution of the subsystem $S$ emerges by conditioning on the rest. In our previous notation this means that given a universe state 
\begin{equation}
    |\Psi\rangle=\int dt\, \ket{t}|\psi(t)\rangle \,,
\end{equation}
 we can recover the state of the system as  $|\psi(t)\rangle=\langle t|\Psi\rangle$ (assuming $\langle t|t'\rangle=\delta(t-t')$). More notably, one can readily see that $\mathcal{J}|\Psi\rangle=(i\partial_t-H)|\psi(t)\rangle=0$ precisely recovers the standard Sch\"{o}dinger equation with the index $t$ being  demoted from a quantum state label to a time parameter.

In order to make the states $|\Psi\rangle$ accessible to conventional (discrete) qubit-based quantum computers, one needs a proper discrete time framework. Fortunately, it is easy to guess the form of a discrete time \textit{history state}. Namely,  we start by introducing a finite dimensional Hilbert space $\mathcal{H}_T$, which we denote as the  \textit{time} or \textit{clock}-Hilbert space with basis $|t\rangle$ satisfying $\langle t'|t\rangle=\delta_{tt'}$ for $t=0,\dots ,N-1$.
A discrete history state is then defined as the state 
\begin{equation}\label{eq:phystate}
    |\Psi\rangle=\frac{1}{\sqrt{N}}\sum_{t=0}^{N-1}|t\rangle|\psi(\varepsilon t)\rangle \,,
\end{equation}
with $|\psi( \varepsilon t)\rangle=U( \varepsilon t) |\psi_0\rangle \in \mathcal{H}_S$. Here, we have $\varepsilon= T/N$ the time-spacing for a given time-window $T$, while $t$ denotes a discrete dimensionless index (so that $\varepsilon t$ is a physical time interval).

In analogy with the continuum case one can recover the state of the system at a given time by conditioning as $    |\psi(\varepsilon t)\rangle\langle \psi(\varepsilon t)|={\rm Tr}_T[|\Psi\rangle \langle \Psi| \Pi_t]/\langle \Psi|\Pi_t|\Psi\rangle$ for $\Pi_t=   |t\rangle \langle t|\otimes \id_S$. In this way the unitarily evolved state is recovered for the time values allowed by $\mathcal{H}_T$.  Notice that this operation is different from a direct partial trace over the clock states which generally yields a mixed state. It turns out that the partial trace induces a quantum channel which also encodes useful information about the system's dynamics and its (eventual) equilibration. In fact, one can think about the history states as a purification of that particular quantum channel. This is related to the system-time entanglement  as we discuss in Section \ref{sec:stent} and Appendix \ref{sec:equilibr}.

Notice also that if the system itself is composite, namely $\mathcal{H}_S=\mathcal{H}_A\otimes \mathcal{H}_B$, a partial trace over system $B$ induces a mixed history state of the form $\Tr_B{|\Psi\rangle \langle \Psi|}=\frac{1}{N}\sum_{t,t'}|t\rangle\langle t'|\otimes \Tr_B\{|\psi(\epsilon t)\rangle \langle \psi(\epsilon t')|\}$ (see also \cite{boette2016system}). Conversely, the history of a mixed state $\rho$ can always be obtained from \eqref{eq:phystate} by a purification of the initial state leading to $\Tr_B{|\Psi\rangle \langle \Psi|}=\frac{1}{N}\sum_{t,t'}|t\rangle \langle t'|\otimes U(\epsilon t)\rho  U^\dag(\epsilon t')$ which is just the convex combination of histories. The standard evolved reduced density matrix is again obtained by conditioning, namely $\rho_A(t)={\rm Tr}_T[|\Psi\rangle \langle \Psi| \Pi_t]/\langle \Psi|\Pi_t|\Psi\rangle$. For these reasons we mainly focus on pure initial states, as the history states of mixed states are just obtained by standard convex combinations.

\begin{figure}[t!]
\centering
\includegraphics[width=.8\columnwidth]{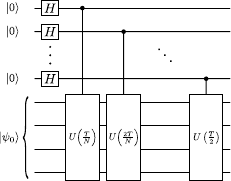}
\caption{\textbf{Circuit for preparing history states.} As shown above, the initial state to the clock qubits  is $\ket{0}^{\otimes \log(N)}$ while that of the system is $ \ket{\psi_{0}}$. The action of the Hadamard gates is to map the initial state to $\ket{+}^{\otimes \log(N)}\otimes \ket{\psi_{0}}$. Here, we find it convenient to write $\ket{+}^{\otimes \log(N)}=\frac{1}{\sqrt{N}}\otimes_{j=1}^{\log{N}}(\ket{0_j}+\ket{1_j})=\frac{1}{\sqrt{N}}\sum_{t=0}^{N-1}\ket{t}$ where we have expressed $t$ in its binary form $t=\sum_{j=1}^{\log{N}}t_j2^{j-1}$. Next, the  $\log{N}$ controlled gates $U(2^{j-1}\frac{T}{N})=U(\frac{T}{N})^{2^{j-1}}$ for $j=1,\ldots,\log{N}$ perform the operations $U(\varepsilon t)\ket{\psi_0}=\ket{\psi(\varepsilon t)}$ for $U(\varepsilon t)=U(\frac{T}{N})^{\sum_{j=1}^{\log{N}}t_j2^{j-1}}$. }
\label{fig:hist-state}
\end{figure}

Here we note that for the case of $N$ being a power of two, the discrete history state can be prepared with the quantum phase estimation-like circuit of Fig.~\ref{fig:hist-state}. For $N$ being a power of two, one requires $\log(N)$ ancillary or clock qubits (we henceforth assume the logarithms to be base 2). As such, the clock Hilbert space $\HC_T$ is of dimension $\dim(\HC_T)=2^{\log(N)}=N$. This result has been reported recently in \cite{boette2016system} and \cite{boette2018history}, where the
discrete history state of Eq.~\eqref{eq:phystate} has also been extensively studied.

The advantages of encoding history states in a quantum computer become clear once one starts considering measurements on the end of the circuit which are different from simple conditioning: while conditioned measurements allow one to recover properties of the system at a given time, new genuinely quantum possibilities become accessible through the clock qubits. A small summary of such possibilities is provided in Fig.~\ref{fig:1}(c).

\section{From qubit-clocks to parallel-in-time simulations}\label{sec:qclockandsims}

Here we discuss how the mathematical formalism of qubit-clocks presented in the previous section can be leveraged to create novel quantum algorithms aimed at studying averages of dynamical-in-time properties of quantum systems. In particular, in this section we focus in developing parallel-in-time-type algorithms that estimate time averages of physical quantities.

\subsection{Setting}

Given a time-independent Hamiltonian $H$ acting on $n$-qubits, and its associated time evolution operator $
    U(t)=e^{-i H t}$, 
we consider the problem of estimating general quantities of the form
\begin{align}\label{eq:formula-F}
    \overline{F}(O_1,O_2,\omega)&=\lim_{T\rightarrow \infty}\int_{0}^T \frac{dt}{T} e^{-i\omega t} \langle O_1(t) O_2 \rangle_{\rho}\\
    &=\lim_{T\rightarrow \infty}\int_{0}^T \frac{dt}{T} e^{-i\omega t} \Tr[\rho O_1(t) O_2]\,,
\end{align}
where $O_1(t) =U\ad(t)O_1U(t)$. Here, $\rho$ is an $n$-qubit state acting on the $d$-dimensional Hilbert space $\HC_S$ (with $d=2^n$), $O_1$ and $O_2$ are two operators, and $\omega\in\mathbb{R}$. 

To illustrate the relevance of the quantity $\overline{F}(\rho,O_1,O_2,\omega)$ in Eq.~\eqref{eq:formula-F} let us consider several special cases. First, let  $\omega=0$ and $O_2=\id$, which leads to 
\begin{align}\label{eq:temp-averages}
    \overline{F}(O_1)&:=\overline{F}(O_1,\id,0)\nonumber\\
    &=\lim_{T\rightarrow \infty}\int_{0}^T \frac{dt}{T}  \langle O_1(t) \rangle_{\rho}\,.
\end{align}
We can see that $\overline{F}(\rho,O_1)$ simply corresponds to an infinite temporal average of the observable $O_1$. These quantities are crucial to understanding the dynamical properties of closed quantum systems and in particular their equilibration \cite{reimann2008foundation,linden2009quantum, malabarba2014quantum}. They are also relevant to the study of quantum quench processes in field theories~\cite{mussardo2013infinite} and signatures of non-equilibrium quantum phase transition 
 through infinite-time averages of Loschmidt  echoes~\cite{venuti2010universality,venuti2011exact,goussev2012loschmidt,yang2017many,zhou2019signature}.  Next, when $\omega=0$, we have 
 \begin{align}\label{eq:correlation-function-averages}
    \overline{F}(O_1,O_2)&:=\overline{F}(O_1,O_2,0)\nonumber\\
    &=\lim_{T\rightarrow \infty}\int_{0}^T \frac{dt}{T}  \langle O_1(t) O_2 \rangle_{\rho}\,.
\end{align}
Here we can recognize $\langle O_1(t) O_2 \rangle_{\rho}$ as a two-point correlation function (also known as a dynamical Green’s function). Two-point correlation functions are used to describe the behavior of a system under perturbations, and are a widely used tool in quantum many-body systems
and condensed matter physics~\cite{rickayzen2013green,khatami2013fluctuation,luitz2016anomalous, kokcu2023linear}. The infinite-time average of $\langle O_1(t) O_2 \rangle_{\rho}$ has been recently considered in~\cite{alhambra2020time} to study thermodynamic properties of closed quantum systems  such as the emergence of dissipation at late times. 

Finally, we note that the general function $\overline{F}(O_1,O_2,\omega)$ corresponds to a Fourier transform of the two-point correlation function, which is commonly referred to as the dynamical structure factor in the condensed matter community~\cite{pedernales2014efficient,baez2020dynamical}. Crucially, the dynamical structure factors are used to study dynamical properties of a given system and have the properties of being experimentally accessible~\cite{coldea2010quantum,jia2014persistent}, and usually being hard to compute via classical simulations~\cite{baez2020dynamical}.

\begin{figure}[t!]
\centering
\includegraphics[width=.8\columnwidth]{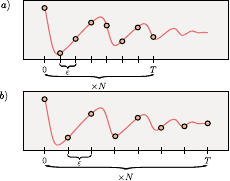}
\caption{\textbf{Trade-off between accuracy and resolution.} Consider the  approximation of the infinite-time average of Eq.~\eqref{eq:formula-F} given by the discrete sum in Eq.~\eqref{eq:discrete-time}. For some fix number of time-steps $N$, there exists a trade-off between the window size $\varepsilon$ and the final time $T$. Namely, larger $T$ implies larger window size $\varepsilon$, and hence less accuracy. On the other hand, smaller final time $T$ implies more resolution in the temporal average at the cost of less accuracy.  }
\label{fig:tradeoff}
\end{figure}

While the importance of Eq.~\eqref{eq:formula-F} is clear, the computation of $\overline{F}(O_1,O_2,\omega)$ might not be straightforward. On the one hand, the classical simulation of some quantum mechanical dynamical process is generally  expected to be exponentially expensive in classical computers. Such scaling can be mitigated by using a quantum computer. Here, there are several schemes capable of computing  fixed-time quantities of the form $\langle O_1(t) O_2 \rangle_{\rho}$ ~\cite{pedernales2014efficient,bauer2016hybrid,kreula2016non,sakurai2022hybrid}. Still, the issue remains that one needs to perform the time average. In practice, this can be achieved via the discrete-time approximation
\begin{align}\label{eq:discrete-time}
    \widetilde{F}(O_1,O_2,\omega)
    &=\frac{1}{N}\sum_{t=0}^{N-1} e^{-i\omega \varepsilon t} \langle O_1(\varepsilon t) O_2 \rangle_{\rho}\,,
\end{align}
where we have $\varepsilon=T/N$ (for simplicity, we will henceforth assume that $N$ is a power of 2). That is, for  a given (finite) time window $T$, we are computing the average over $N$ points separated by a spacing $\varepsilon$. As shown in Fig.~\ref{fig:tradeoff}, the spacing $\varepsilon$ determines the level of accuracy in the approximation, as a smaller $\varepsilon$ leads to a more precise discretization of the integral and a better approximation of the true infinite-time average. On the other hand, the final time $T$ determines the resolution of the approximation, as a larger $T$ allows for a longer time interval to be averaged over, capturing more information about the system's behavior over time. One can see that both the resolution and the accuracy  can be improved by a larger number of discrete time steps $N$.

\begin{figure}[t!]
\centering
\includegraphics[width=1\linewidth]{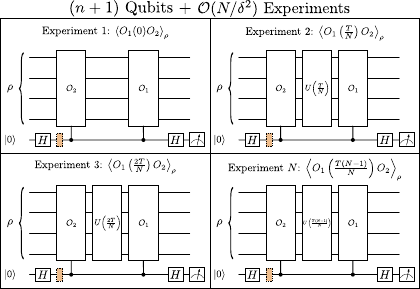}
\caption{\textbf{Algorithm for sequential-in-time estimation of Eq.~\eqref{eq:discrete-time}.} The algorithm  show can be used to individually compute each term in the summation. That is, the circuits can be used to  estimate quantities of the form  $\langle O_1( t) O_2 \rangle_{\rho}$. Then, one can combine those expectation values classically (as well as add the appropriate phases $e^{-i\omega \varepsilon t}$) to estimate the quantity $\widetilde{F}(O_1,O_2,\omega)$ up to precision $\delta$. The colored dashed gate is replaced with an identity (an $S\ad$ gate) to compute the real (imaginary) part of $\langle O_1( t) O_2 \rangle_{\rho}$. This approach requires a quantum device with $(n+1)$-qubits and $\OC(N/\delta^2)$ different experiments.   }
\label{fig:circuit-F-sequential}
\end{figure}

\begin{figure}[th!]
\centering
\includegraphics[width=1\linewidth]{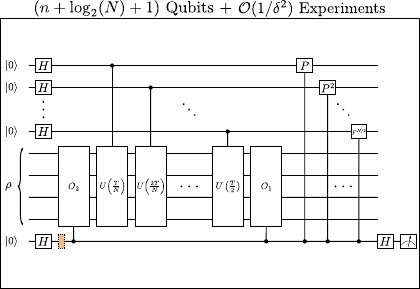}
\caption{\textbf{Algorithm for parallel-in-time estimation of Eq.~\eqref{eq:discrete-time}.} The algorithm shown  can be used to directly estimate the quantity $\widetilde{F}(O_1,O_2,\omega)$ up to precision $\delta$. In the figure, $P$ denotes a $ \omega \varepsilon$ phase gate. It is clear that this algorithm contains as a sub-routine the circuit for preparing the history state of Fig.~\ref{fig:hist-state}. This approach requires a quantum device with $(n+\log(N)+1)$-qubits and $\OC(1/\delta^2)$ experiments.  The colored dashed gate is replaced with an identity (an $S\ad$ gate) to compute the real (imaginary) part of $\widetilde{F}(O_1,O_2,\omega)$.   }
\label{fig:circuit-F-parallel}
\end{figure}

\subsection{Sequential and parallel-in-time protocols}

Let us now consider the task of estimating $\widetilde{F}(O_1,O_2,\omega)$ when $O_1$ and $O_2$ are Pauli operators by either sequential- or parallel-in-time simulations. Here, by sequential, we mean that each term in the sum in Eq.~\eqref{eq:discrete-time} is estimated on a quantum device by running some finite number of ``experiment''. For instance, consider the circuit in Fig.~\ref{fig:circuit-F-sequential}, as explicitly shown in the Supplemental Information, it can be used to estimate an expectation value of the form $\langle O_1(\varepsilon t) O_2 \rangle_{\rho}$. Thus, we have that the following proposition holds. 

\begin{proposition}\label{prop:sequential}
    The circuit in Fig.~\ref{fig:circuit-F-sequential}, which requires $(n+1)$-qubits, can be used to estimate the quantity $\widetilde{F}(O_1,O_2,\omega)$ of Eq.~\eqref{eq:discrete-time} up to $\delta$ accuracy  with $\OC(N/\delta^2)$ experiments.
\end{proposition}

The proof of Proposition~\ref{prop:sequential}, as well as that of all other main results, is presented in Appendix \ref{app:circ}. 

Clearly, the fact that we need to sequentially estimate $\langle O_1(\varepsilon t) O_2 \rangle_{\rho}$ for each $t=0,\ldots,N-1$, leads to a complexity in the number of experiments (i.e., number of calls to the quantum computer) that scales as $\OC(N)$. As we now show, this complexity can be reduced by using a scheme  based on the discrete history state formalism, which allows us to directly estimate the whole sum of Eq.~\eqref{eq:discrete-time}. That is, the following result holds.
\begin{theorem}\label{theo:parallel1}
    The circuit in Fig.~\ref{fig:circuit-F-parallel}, which requires $(n+\log(N)+1)$-qubits, can be used to estimate the quantity $\widetilde{F}(O_1,O_2,\omega)$ of Eq.~\eqref{eq:discrete-time} up to $\delta$ accuracy  with $\OC(1/\delta^2)$ experiments. 
\end{theorem}

Comparing Proposition~\ref{prop:sequential} and Theorem~\ref{theo:parallel1} reveals that by leveraging the discrete history state formalism we can trade the $\OC(N)$ experiment-complexity  for $\OC( \log(N))$-ancillary qubits. That is, the parallel-in-time algorithm of Fig.~\ref{fig:circuit-F-parallel} allows an exponential temporal-to-qubit resource trade-off.  Here we remark that one can see from Fig.~\ref{fig:circuit-F-parallel} that the key step behind the algorithm to compute $\widetilde{F}(O_1,O_2,\omega)$ is the discrete history state. In fact the circuit in Fig.~\ref{fig:hist-state} used to create the history state is a sub-routine in Fig.~\ref{fig:circuit-F-parallel}. Thus, by leveraging $\log(N)$ ancillas, one can simultaneously implement all $N$ time evolution operators $U(\varepsilon t)$ for $t=0,\ldots,N-1$, and concomitantly compute all the terms in the summation leading to $\widetilde{F}(O_1,O_2,\omega)$.

Note that while Proposition~\ref{prop:sequential} and Theorem~\ref{theo:parallel1} are derived and proved for the case of $O_1$ and $O_2$ being unitary operators, one can readily generalize the previous results for the case when they are instead expressed as a linear combination of Pauli operators. In particular, if
\begin{equation}
    O_i=\sum_{\mu=1}^{M_1} c^{(i)}_\mu U_\mu\,,
\end{equation}
for $U_\mu$ being a Pauli operator, then the experiment complexities in Proposition~\ref{prop:sequential} and Theorem~\ref{theo:parallel1} respectively change as $\OC(NM_1M_2/\delta^2)$ and $\OC(M_1M_2/\delta^2)$. Here, we again recover an  exponential temporal-to-qubit resource trade-off by using the parallel-in-time algorithm.

\begin{figure}[t!]
\centering
\includegraphics[width=1\linewidth]{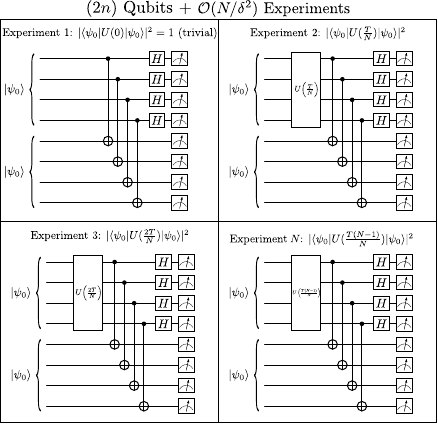}
\caption{\textbf{Algorithm for sequential-in-time estimation of the Loschmidt echo of Eq.~\eqref{eq:temp-averages-discrete}.}  We show an algorithm which computes, up to precision $\delta$, the overlap between $\ket{\psi_0}$ and $U(\varepsilon t)\ket{\psi_0}$ for $t=0,\ldots,(N-1)$. The algorithm is based on Bell-basis measurements as described in~\cite{cincio2018learning}.   Once these overlaps are estimating, we can average them classically  to estimate discrete-time temporal average of the Loschmidt echo $\widetilde{\LC}(\psi_0)$. This approach requires a quantum device with $(2n)$-qubits and $\OC(N/\delta^2)$ different experiments.   }
\label{fig:circuit-loschmidt-sequential}
\end{figure}

\begin{figure}[t!]
\centering
\includegraphics[width=1\linewidth]{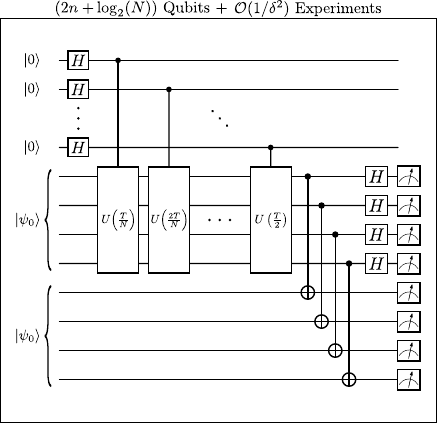}
\caption{\textbf{Algorithm for parallel-in-time estimation of the Loschmidt echo of Eq.~\eqref{eq:temp-averages-discrete}.} 
 We show an algorithm which computes, up to precision $\delta$, the overlap between the discrete history state $\dya{\Psi}$ and $ \id\otimes\dya{\psi_0}$. As shown in Eq.~\eqref{eq:meanvd}, the overlap between these two states is equal to $\widetilde{\LC}(\psi_0)$. The algorithm is based on Bell-basis measurements as described in~\cite{cincio2018learning}.   This approach requires a quantum device with  $(2n+\log(N))$--qubits and $\OC(1/\delta^2)$ different experiments. }
\label{fig:circuit-loschmidt-parallel}
\end{figure}

Next, let us consider 
$\rho=O_1=\dya{\psi_0}$, 
$O_2=\id$ and $\omega=0$. In this special case,
\begin{align}\label{eq:infinite-loschmidt}
   \overline{F}(\dya{\psi_0},\id,0)
   &=\lim_{T\rightarrow \infty}\int_{0}^T \frac{dt}{T}  |\langle \psi_0|U( t)|\psi_0\rangle |^2\,.
\end{align}
The quantity on the right hand side is  the infinite-time Loschmidt echo average ~\cite{venuti2010universality,venuti2011exact,yang2017many,zhou2019signature},  which we denote as $\bar{\LC}(\psi_0)$.  We see that
\begin{equation}
    \bar{\LC}(\psi_0)\equiv \overline{F}(\dya{\psi_0},\id,0)\,.
\end{equation}
Similarly, for its discrete-time approximation $\widetilde{\LC}(\psi_0)$, we can write
\small
\begin{align}\label{eq:temp-averages-discrete}
 \widetilde{\LC}(\psi_0)\equiv \widetilde{F}(\dya{\psi_0},\id,0)
    =\frac{1}{N}\sum_{t=0}^{N-1} |\langle \psi_0|U(\varepsilon t)|\psi_0\rangle |^2\,.
\end{align}
\normalsize

It is clear that while $\widetilde{\LC}(\psi_0)$ can technically be computed with the circuits in Figs.~\eqref{fig:circuit-F-sequential} and~\eqref{fig:circuit-F-parallel}, this requires expanding $O_1=\dya{\psi_0}$ into a linear combination of unitaries, and such summation will generally contain exponentially many terms. To mitigate this issue, we also present two results which allow us to estimate Eq.~\eqref{eq:temp-averages-discrete} by either sequential-, or parallel-in-time simulations.

First, let us consider the following proposition.
\begin{proposition}\label{prop:sequential-losch}
    The circuit in Fig.~\ref{fig:circuit-loschmidt-sequential}, which requires $(2n)$-qubits, can be used to estimate the quantity $\widetilde{\LC}(\psi_0)$ of Eq.~\eqref{eq:temp-averages-discrete} up to $\delta$ accuracy  with $\OC(N/\delta^2)$ experiments.
\end{proposition}
Proposition~\eqref{prop:sequential-losch} simply follows from applying a SWAP  test~\cite{buhrman2001quantum,harrow2013testing,gutoski2015quantum,cincio2018learning} (or more specifically, the state overlap algorithm of~\cite{cincio2018learning}) between $U(\varepsilon t)\ket{\psi_0}$ and $\ket{\psi_0}$ for $t=1,\ldots, (N-1)$. We note that the case $t=0$ is trivial as $|\bra{\psi_0}U(0)\ket{\psi_0}|^2=1$. When using the discrete history state we can prove the following theorem.

\begin{theorem}\label{theo:parallel-losch}
    The circuit in Fig.~\ref{fig:circuit-loschmidt-parallel}, which requires     $(2n+\log(N))$--qubits, can be used to estimate the quantity $\widetilde{\LC}(\psi_0)$ of Eq.~\eqref{eq:discrete-time} up to $\delta$ accuracy  with $\OC(1/\delta^2)$ experiments. 
\end{theorem}

Again, we can see from Theorem~\ref{theo:parallel-losch} that performing a parallel-in-time simulation allows us to exponentially reduce the experiment complexity (from linear in $N$ to being $N$-independent) at the cost of $\log(N)$ ancillas. Similarly to Proposition~\ref{prop:sequential-losch}, the proof of Theorem~\ref{theo:parallel-losch} simply spans from computing the overlap between the discrete history state $\dya{\Psi}$ and $\id_T\otimes \dya{\psi_0}$. Explicitly, we have
\begin{align}
  \Tr[\dya{\Psi}(\id_T\otimes\dya{\psi_0})]&=\langle \Psi| (\id_T\otimes\dya{\psi_0})|\Psi\rangle\nonumber\\
  &=\frac{1}{N}\sum_{t=0}^N \langle \psi(\varepsilon t) \dya{\psi_0}\psi(\varepsilon t)\rangle\nonumber\\
  &=\widetilde{\LC}(\psi_0)\,.\label{eq:meanvd}
\end{align}

\section{Accessing dynamical information via system-time entanglement}\label{sec:stent}

Thus far, we have seen that using the history state allows us to push  the complexity of running multiple experiments onto ancillary clock-qubit requirements. However, as we will now show, the entanglement present between the time and system qubits in the history state   has operational meaning and contains  information that we can use to learn about the system's dynamics. Moreover, we will unveil a rigorous and explicit connection between these correlations and the equilibration problem. Protocols for obtaining these quantities from variations of the previous circuits are also provided in this section. 

\subsection{Properties, relation to the  problem of equilibration and to temporal fluctuations of observables} 

First, let us again recall that the discrete history state is a bipartite state between the system Hilbert space $\mathcal{H}_S$ and the time, or clock Hilbert space $\mathcal{H}_T$. That is, $\ket{\Psi}\in\mathcal{H}_T\otimes\mathcal{H}_S$. Moreover, it is apparent from Fig.~\ref{fig:hist-state} and Eq.~\eqref{eq:phystate}  that history states are in general entangled across the system-time partition. We will henceforth refer to the  correlations between the system qubits and the clock qubits as \emph{system-time entanglement} (following \cite{boette2016system}).

It is important to note that, in general, \eqref{eq:phystate}  is not in the Schmidt's decomposition~\cite{nielsen2000quantum} of $|\Psi\rangle$ (as the states $\ket{\psi(t)}$ are not necessarily orthogonal). However, there exists a basis in which  we can write the history state as 
\begin{equation}\label{eq:schmidt}
    |\Psi\rangle=\sum_l \sqrt{p_l}|l\rangle_T |l\rangle_S\,,
\end{equation}
where $\sqrt{p_l}$ are the so-called Schmidt coefficients, and  $\{|l\rangle_S\}$, $\{|l\rangle_T\}$ are orthonormal sets of states in $\HC_S$ and $\HC_T$, respectively. A simple way to quantify the system-time entanglement is through the linear entropy, defined   as
\begin{equation}\label{eq:linear-entropy}
    E_2=1-\Tr[\rho_T^2]=1-\Tr[\rho_S^2]=1-\sum_l p_l^2\,,
\end{equation}
where $\rho_{T(S)}=\Tr_{S(T)}[\dya{\Psi}]$ is the reduced state of the history state in the clock (system) qubits. Here, we denote as $\Tr_{S(T)}$ the partial trace over the system (clock) qubits. In principle, one can also consider other entropies such as the von Neumann entropy. However,  the linear entropy has the desirable property of being efficiently computable in a quantum device (see below).

There is a deep connection between the system-time entanglement and dynamical properties of the system, in particular to the problem of its equilibration: Let us recall first that given an arbitrary (for simplicity) pure state $|\psi\rangle=\sum_k c_k|k\rangle$ the infinite-time average of the associated density matrix is
\begin{equation}\label{eq:averagedst}
    \bar{\rho}=\int \frac{dt}{T} \sum_{k,k'} c_k c_{k'}^\ast e^{-i(E_k-E_{k'})t}|k\rangle \langle k'|\to\sum_k |c_k|^2 |k\rangle \langle k|\,,
\end{equation}
where one assumes large (infinite) $T$ and with $H|k\rangle=E_k|k\rangle$. In other words, if the state of the system is averaged over large enough times it loses all coherences in the energy basis. Under experimentally realistic conditions it is feasible to identify this state with the stationary equilibrium state \cite{reimann2008foundation}. For ``most''  observables this actually holds for short times $T$ \cite{malabarba2014quantum}, meaning that a finite time window average of observables is also an interesting quantity in general. The quantum time formalism gives a new interpretation to the loss of coherences induced by a time average: since the system is ``entangled with time'', we lose information  by ignoring the ``clock qubits''. This loss induces precisely the (dephasing) quantum channel $\rho\to \bar{\rho}$ in the large $T$ and small $\varepsilon$ limit, a result that can be derived directly from a continuum quantum time formalism  \footnote{for infinite $T$ one has to consider subtleties related to the normalization of states; see \cite{giovannetti2015quantum, diaz2019history, diaz2019historystate}.}. For discrete time the following result holds.

\begin{theorem}\label{theo:deph}
Let $\ket{\Psi}$ be the discrete history-state in Eq.~\eqref{eq:phystate}. The partial trace over the clock induces a quantum channel which in the large time limit implies $\rho_S\to \bar{\rho}$. Moreover, for any $\varepsilon$ and $N$ the following majorization relation holds: 
\label{eq:major}
  \begin{equation}
    \bar{\rho}\prec \rho_S= \tilde{\rho}\,,
\end{equation}
with $\tilde{\rho}$ a discretization of Eq.~\eqref{eq:averagedst}. Furthermore,  for a periodic evolution with period $\tau$ generated by a Hamiltonian with $M$ distinct eigenvalues (i.e., $e^{-iH\tau}=\id$) and given a history state with $\log(M)$ clock qubits and time window $T=\tau$, we have
\begin{equation}
    \rho_S= \bar{\rho}\,.
\end{equation}

\end{theorem}
While phrased in a rather abstract way, this result has many interesting corollaries with clear operational meaning. The reason for this is that roughly speaking the history state is providing a way to prepare the equilibrated state of a quantum system: One simply needs to prepare the history state and ignore the clock-qubits.  In fact, this is the reason why the previous for evaluating time averages work. Moreover, the system time entanglement entropies are in fact a lower bound to the entropies of the state in equilibrium, as it follows directly from Theorem \ref{theo:deph} and basic majorization properties. Furthermore, we show in the Appendix \ref{sec:equilibr} that one can \emph{rediscover} the quantum time formalism from the natural purification of this approximate dephasing channel: The history state arises from a simple isometric extension $U[K_t]$ of the channel as $|\Psi\rangle=U[K_t]|\psi_0\rangle$ with $K_t$ the Krauss operators $K_t=e^{-iHt\epsilon}/\sqrt{N}$. The interested reader can refer to Appendix \ref{sec:equilibr} where the proof of Theorem~\ref{theo:deph} is provided together with a more detailed discussion.

With the previous in mind, let us consider again the task of estimating the  infinite-time Loschmidt echo average in Eq.~\eqref{eq:infinite-loschmidt}.  We recall that $\bar{\LC}(\psi_0)$ quantifies the degree of reversibility of the time evolution and is an indicator of the stability of the quantum system. Moreover, it is easy to see that
\begin{equation}
    \bar{\LC}(\psi_0)=\Tr [\bar{\rho}^2]\,,
\end{equation}
i.e., the infinite-time average of the Loschmidt echo is the purity of the dephased state $\bar{\rho}$. We can now use these considerations and Theorem \ref{theo:deph} to obtain the following result.  
\begin{corollary}\label{theo:ent-loschmidt}
Let $\ket{\Psi}$ be the discrete history-state in Eq.~\eqref{eq:phystate}, and let $E_2$ be the linear entropy of the system-time partition. Then, for any $T$ and $N$ we have
\begin{equation}\label{eq:entineq}
    E_2\leq (1-\bar{\LC}(\psi_0))\,.
\end{equation}
\end{corollary}

Corollary~\ref{theo:ent-loschmidt} has several important implications. First, it bounds the amount of entanglement between the system and the clock qubits. In particular, it shows that the  system-time entanglement can only be large if the  infinite-time average of the  Loschmidt echo value is small. Conversely, if $\bar{\LC}(\psi_0)$ is large, $E_2$ has to be small. Second, let us remark that Eq.~\eqref{eq:entineq} is valid for all values of $T$, but most notably, also for all values of $N$. For large $N$ and $T$ the equality is reached asymptotically, and we have that Eq.~\eqref{eq:entineq} becomes $\Tr[\rho^2_T]\equiv\bar{\LC}(\psi_0)$. Moreover, as we will see below, our numerical  analysis shows that $\Tr[\rho^2_T]$ can provide a better approximation to $\bar{\LC}(\psi_0)$ than $\widetilde{\LC}(\psi_0)$, implying 
 that there exists  no simple general relation between $E_2$ and $\widetilde{\LC}(\psi_0)$. 

 We can understand the intuition behind Corollary~\ref{theo:ent-loschmidt} as follows. Let $\ket{\psi_0}$ be a stationary state of the unitary evolution. For instance, let $\ket{\psi_0}$ be an eigenstate of $H$ with eigenenergy $E_0$, so that $U(\varepsilon t)\ket{\psi_0}=e^{-i \varepsilon t E_0 }\ket{\psi_0}$. Then, the discrete history state becomes 
 \begin{equation}\label{eq:separable}
     \ket{\Psi}=\frac{1}{\sqrt{N}}\sum_{t=1}^{N-1}e^{-i \varepsilon t E_0 }\ket{t}\otimes \ket{\psi_0}\,.
 \end{equation}
Equation~\eqref{eq:separable} reveals that $ \ket{\Psi}$ is separable. It is also not hard to verify that in this case  $\widetilde{\LC}(\psi_0)=1$. On the other hand, if $\ket{\psi_0}$ evolves through $N$ orthogonal states $ \bra{\psi(\varepsilon t)}\ket{\psi(\varepsilon t')}=\delta_{tt'}$ then Eq.~\eqref{eq:phystate} is already the Schmidt decomposition of  $ \ket{\Psi}$ and the state is maximally entangled. The previous toy model shows  that if the state is quasi stationary (i.e., large Loschmidt echo), we can expect small values of entanglement. Similarly, if the state is significantly changing during the evolution (e.g., small Loschmidt echo value), then the history state will likely possess large amounts of entanglement. We note that the  relation between the distinguishability of the evolved state and the system time-entanglement was first reported in~\cite{boette2016system}. However, the connection with the Loschmidt echo was not explored therein.

 The result in Corollary~\ref{theo:ent-loschmidt} can be further strengthened for the special case where the  time evolution is periodic. That is, when 
 \begin{equation}\label{eq:periodic-condition}
    e^{-iH\tau}=\id\,,
 \end{equation}
 for some $\tau$, and where we assume that $H$ has $M$ distinct eigenvalues, for $M$ being a power of  two. Now,  we find that the following result holds.

 \begin{corollary}
\label{theo:ent-periodic}
     For a periodic evolution with period $\tau$ generated by a Hamiltonian with $M$ distinct eigenvalues, as in Eq.~\eqref{eq:periodic-condition}, then for a history state with $\log(M)$ clock qubits and time window $T=\tau$, we have
\begin{equation}
    E_2=(1-\bar{\LC}(\psi_0))=(1-\widetilde{\LC}(\psi_0))\,.
\end{equation} 
 \end{corollary}

Corollary~\ref{theo:ent-periodic} shows that for periodic Hamiltonians the system-time entanglement is exactly the same as the infinite-time average of the Loschmidt echo $\bar{\LC}(\psi_0)$, as well as the  discrete-time approximation $\widetilde{\LC}(\psi_0)$. As shown in Appendix~\ref{sec:equilibr}, tracing out induces now a completely dephasing channel in the energy eigenbasis so that $\rho_S=\tilde{\rho}=\bar{\rho}$.

The previous results connecting the system-time entanglement with the Loschmidt echo allow us to derive even more operational meaning to $E_2$ as a bound for  temporal fluctuations of observable. In Ref.~\cite{reimann2008foundation}, it was shown that given an observable $O$, $\bar{\LC}(\psi_0)$ provides a bound on temporal fluctuations of observables as 
\begin{equation}\label{eq:var}
    \sigma^2_O\leq \Delta_O^2 \bar{\LC}(\psi_0)\,,
\end{equation}
with $\Delta_A^2=\lambda_{\max}[O]-\lambda_{\min}[O]$ (the difference between the largest and smallest  eigenvalues of $O$ in the subspace of states satisfying $\langle n|\psi\rangle\neq 0$), and where   $\sigma^2_O$ denotes the temporal variance
\begin{align}
    \sigma^2_O:=&\overline{F(O)^2}-\overline{F}(O)^2\label{eq:bound-loschmidt-variance}\\
    =&\lim_{T\rightarrow \infty}\int_{0}^T \frac{dt}{T}  \langle O(t) \rangle^2_{\psi_0}-\left(\lim_{T\rightarrow \infty}\int_{0}^T \frac{dt}{T}  \langle O(t) \rangle_{\psi_0}\right)^2\nonumber\,.
\end{align}
Here we have used  the notation defined in Eq.~\eqref{eq:temp-averages} with $F(O)\equiv \langle O\rangle$ (at a given time) while the ``overline'' denotes temporal-average. Eq.~\eqref{eq:var} shows that small temporal Loschmidt echo averages imply a small temporal variance of the observable $O$, and vice versa. In other words, a system with a small $\bar{\LC}(\psi_0)$ can only  exhibit smaller temporal fluctuations in its observables compared to a system with a large Loschmidt echo.

It should be  clear to see that Theorem~\ref{theo:ent-loschmidt} readily implies the following corollary.

\begin{corollary}\label{cor:1}
Let $O$ be an observable, and let $ \sigma^2_O$ denote its temporal variance as in Eq.~\eqref{eq:bound-loschmidt-variance}. The system-clock entanglement provides  bound on temporal fluctuations as
\begin{equation}\label{eq:cor-tempfluct-bound}
      \sigma^2_O\leq \Delta_O^2\left(1-E_2\right)=  \Delta_O^2{\rm Tr}[\rho^2_S]\,.
\end{equation}
\end{corollary}

Corollary~\ref{cor:1} shows a clear physical meaning of the system-time entanglement. Namely, if $E_2$ is small, then the system is stable and predictable. This follows from the fact that the temporal variances of expectation values will be small. Conversely, if the system-time entanglement is large, then the system can be unstable and unpredictable, as evidenced by potentially large observable fluctuations.

\begin{figure}[t!]
\centering
\includegraphics[width=1\linewidth]{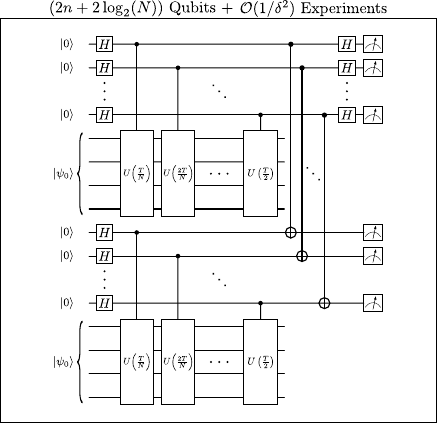}
\caption{\textbf{Algorithm for estimating $E_2$ via state-overlap.} Here we consider the task of evaluating Eq.~\eqref{eq:linear-entropy}. By taking two copies of the history state, we can estimate $\Tr[\rho_T^2]$ up to $\delta$ precision via the state-overlap circuit in~\cite{cincio2018learning}. This approach requires a quantum device with $(2n+2\log(N))$-qubits and $\OC(1/\delta^2)$ different experiments.   }
\label{fig:circuit-Ent_Overlap}
\end{figure}

\subsection{Protocols for computing the system-time entanglement}

The previous theorems and corollary shed light on the exciting possibility of understanding the dynamics of the system through the system-time entanglement. However, in order for these results to be truly useful, one needs to be able to measure  $E_2$ from the history state.  As we can see in  Eq.~\eqref{eq:linear-entropy}, we  need to estimate $\Tr[\rho_S^2]$ or $\Tr[\rho_T^2]$.  While mathematically, it makes no difference whatsoever which subsystem we focus on, as their purity is the same  (see Eq.~\eqref{eq:schmidt}), in practice it can be substantially easier to work with one system or the other. 

As heuristically evidenced by our numerics (see below), the discrete history state with a number of clock qubits $\log(N)$ much smaller than the system size $n$ produces results which accurately reproduce the infinity time average properties  of the system dynamics. Thus, we will henceforth assume that $\log(N)\ll n$. This assumption implies that we can compute $E_2$, and therefore learn about the system, by just looking at the clock qubits. We now present two methods for estimating $\Tr[\rho_T^2]$.

\begin{figure}[t!]
\centering
\includegraphics[width=1\linewidth]{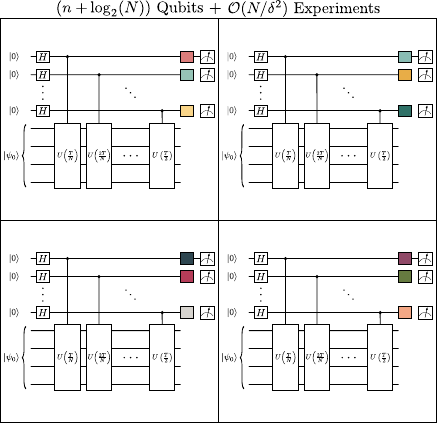}
\caption{\textbf{Algorithm for estimating $E_2$ via randomized measurements.} Here we consider the task of evaluating Eq.~\eqref{eq:linear-entropy}. We start with a copy of the history state, and we apply a random unitary (indicated by a colored gate) to each qubit. Then we measure each qubit in the computational basis and record the measurement outcome. These constitute the so-called ``classical shadows'' of  $\rho_T$. As shown in~\cite{huang2020predicting}, this procedure allows us to  estimate $\Tr[\rho_T^2]$ up to $\delta$ precision with a quantum device with $(n+\log(N))$-qubits and $\OC(N/\delta^2)$ different experiments.   }
\label{fig:circuit-Ent_Rand}
\end{figure}

\begin{theorem}\label{theo:ent}
    The quantity $E_2$ of Eq.~\eqref{eq:linear-entropy} can be estimated up to $\delta$ accuracy with the circuit in Fig.~\ref{fig:circuit-Ent_Overlap}, which requires $(2n+2\log(N))$-qubits with $\OC(1/\delta^2)$ experiments. Similarly, it can also be estimated with  the circuit in Fig.~\ref{fig:circuit-Ent_Rand}, which requires $(n+\log(N))$-qubits  with $\OC(N/\delta^2)$ experiments.
\end{theorem}

When using the circuit in Fig.~\ref{fig:circuit-Ent_Overlap} one prepares two copies of the history state $\ket{\Psi}$ and then performs the  state overlap circuit of Ref.~\cite{cincio2018learning}. On the other hand, when using the circuit in Fig.~\ref{fig:circuit-Ent_Rand} one can estimate $E_2$ with a single copy of $\ket{\Psi}$ by using classical shadows, or randomized measurements~\cite{huang2020predicting,brydges2019probing}. For instance, one can prepare the history state and performs a random unitary on each qubit, followed by a measurement on the computational basis. The measurement outcomes are stored and then combined classically to estimate $\Tr[\rho_T^2]$. 

To finish this section, we note that by comparing  Proposition~\ref{prop:sequential-losch}, Theorem~\ref{theo:parallel-losch}, and Theorem~\ref{theo:ent}, the method  to  estimate either $\widetilde{\LC}(\psi_0)$ or $E_2$ with  the least computational requirement (assuming $\log(N)\ll n$) is that of Fig.~\ref{fig:circuit-Ent_Rand}. Namely, here we can compute $E_2$ up to $\delta$ precision with a quantum computer with    $(n+\log(N))<<2n$ qubits and with $\OC(N/\delta)$ experiments. This result then showcases the power of using the history state as it allows us to study physical properties of the system (such as bounding $\bar{\LC}(\psi_0)$ or the temporal variances $\Delta O^2$) with less requirements than we would otherwise need.

\section{Depth-Estimation and parallel-in-time advantages}

In the previous sections we have presented several methods where we used the  history state  to  study temporal averages of  quantities of the form of $\widetilde{F}(O_1,O_2,\omega)$ in Eq.~\eqref{eq:discrete-time}. At the same time, we have shown how to compute  the system-time entanglement, a new quantity with many interesting applications.
Crucially, these techniques require being able to implement the phase estimation-like circuit for preparing history states in Fig.~\ref{fig:hist-state} as a sub-routine. 
As it is well known, under the assumption of black boxes implementing the controlled unitaries only $\log(N)$ steps are necessary \cite{nielsen2000quantum} for such protocol, which seems to indicate an exponential advantage over sequential approaches. The most famous example of such an advantage is provided by Shor's alghoritm \cite{shor1999polynomial} where the controlled unitaries can be efficiently prepared via modular exponentiation. In this section, we discuss the depth scaling under the more realistic scenario were the controlled unitaries associated with quantum evolution need to be explicitly implemented. We provide two different strategies: a direct
Lie-Trotter product formula \cite{lloyd1996universal} approach and a Hamiltonian diagonalization scheme which we implemented variationally ~\cite{commeau2020variational}. The second strategy, when successful, maintains the black box scaling leading to an exponential advantage. Instead, the Trotterization scheme leads to a more modest quadratic advantage for large enough $N$.

\subsection{Direct Trotterization approach}

We start by recalling that $U(\varepsilon t)=e^{-i H \varepsilon t}$ for some Hamiltonian of interest $H$. Usually, if one wishes to implement $U(\varepsilon t)$, the standard approach is to 
break the evolution into a smaller, easier  to implement evolutions $U(\varepsilon)$, and then repeat it $t$ times. That is, one has  $U(\varepsilon t)=U(\varepsilon)^t$. 
In this way, the depth of the circuit needed to implement  $U(\varepsilon t)$ grows  
with $t$.

To be more specific, let us assume that one is employing a Lie-Trotter decomposition-based product formula. This was the first example of quantum advantage for quantum simulations \cite{lloyd1996universal} and remains a relevant and straightforward technique to this day \cite{childs2021theory}. The basic idea is to decompose a given Hamiltonian
\begin{equation}
    H=\sum_{j=1}^l h_j
\end{equation}
as $U(\varepsilon)\approx \prod_j e^{-ih_j \varepsilon}$. We will further assume that each $h_j$ is local, i.e., it acts non-trivially on at most $\OC(1)$ qubits (similar considerations hold if $l$ grows polynomically with $n$). Importantly, note that the  locality condition implies that $l$ is linear in the system size. 
The evolution operator up to time $\tau=t \varepsilon$ can be approximated with $t$ copies of these gates as $U(\tau)\approx \Big[\prod_j e^{-ih_j \varepsilon}\Big]^t$ which is also logarithmic in the system size.

A rough estimation of the error involved was provided  in \cite{lloyd1996universal} by assuming that the main contribution to the error comes from the second-order term in the Lie-Trotter formula. Under this assumption, the number of times steps required for guaranteeing a fixed precision grows as $t \propto \tau^2$. Then, the total number of gates, denoted as $\#_g(\tau)$, scales as  $\#_g(\tau)=\gamma l \tau^2$ for $\gamma$ a constant dependent on the precision and the particular Hamiltonian. More general bounds were found in \cite{berry2007efficient} which allows us to write $\#_g(\tau)=\gamma l \tau^\alpha$. 
This is the estimation we will be using. Under certain scenarios, this bound can be improved e.g. by using the Lie-algebraic structure associated with the given Hamiltonian \cite{somma2016trotter}, and in general, the ``actual'' error scaling of such
product formulas remains poorly understood \cite{childs2021theory}. For our purpose,  the previous bound will suffice: we want to compare the total number of gates required in the sequential (Figures \ref{fig:circuit-F-sequential} and \ref{fig:circuit-loschmidt-sequential}) versus the parallel-in-time approach (Figures \ref{fig:circuit-F-parallel} and \ref{fig:circuit-loschmidt-parallel}) assuming the same Trotterization scheme is applied to both. The interest in this quantity relies on the fact that the \emph{total} number of gates employed in each simulation protocol is what determines the total time span required to complete the computation \footnote{ If the duration  of time required to implement the gates depends on the interval $\varepsilon$ the reasoning can be adapted by changing $\alpha$. For example, following \cite{lloyd1996universal} one may assume that implementing $e^{-ih_j \varepsilon}$ takes $\varepsilon=\tau/t$ time and we need $\propto t$ gates  meaning a total duration proportional to $\tau$, rather than quadratic as the number of gates.}. By estimating this quantity in each protocol we can establish whether it is more convenient to use a sequential or parallel-in-time approach.

\begin{theorem}\label{theo:scaling}
Consider the total number of gates required for implementing the evolution in the sequential approaches $\#^{\text{seq}}$, and the total number for the parallel in time approaches $\#^{\text{par}}$. They scale as $\#^{\text{seq}}\in \mathcal{O}(l N^{\alpha+1})$, $\#^{\text{par}}\in \mathcal{O}(l^2 N^{\alpha})$ yielding 
    \begin{equation}
\#^{\text{par}}\sim\frac{\beta l}{N} \#^{\text{seq}}\,,
\end{equation}
for $\beta\in \mathcal{O}(1)$ a constant independent both of the system and clock size.
\end{theorem}
Theorem~\ref{theo:scaling} is a consequence of the fact that $\#^{\text{seq}}$ is given by the sum over the amount of
gates of each run so that $\#^{\text{seq}}\sim N^{\alpha+1}$. Instead, in the parallel approach the total number gates only involves a sum over the $log(N)$ gates of the same run, thus giving $\#^{\text{seq}}\sim  N^{\alpha} \beta l$. The extra factor $\beta l$ comes from the fact that those gates need to be controlled. Remarkably, the depth scaling with the number of times of the parallel-in-time approach is the same as the one of a single Trotter evolution up to time $\tau\equiv \varepsilon N$. See Appendix~\ref{app:proof-theo-5} for the details and the proof of the previous theorem.

 Something really interesting has happened: in the parallel approach we have an increase in depth which is \emph{linear} in the system size, but we have reduced the total number of gates \emph{exponentially} in the number of clock qubits with respect to a sequential approach (equivalent to a quadratic improvement in the total number of times $N$). We can then state the following:
 
\begin{proposition}\label{prop:adv}
Given $\log(N)$ clock-qubits and system of Hilbert space dimension $d$, the parallel-in-time approach outperforms the computational times of the sequential approach for 
   \begin{equation}\label{eq:advant}
\log(N)\gtrsim \log(\beta l)\sim \log(\beta\log(d))\,. 
\end{equation}
\end{proposition}
Remarkably, the condition for a convenient clock size is doubly logarithmic in the system's Hilbert space dimension $d$. Typically, a modest number of qubits for the clock, much smaller than the system size,  is sufficient to improve computational times. As an example for $n=200$ system qubits and for $\beta=5$ the inequality $\log(N)>\log(\beta \log(d))$ is achieved when $m\geq 10$, while for $n=400$ one needs $m=11$ clock qubits. Let us also remark that in most applications it is natural to scale $N$ with the system size to properly capture the system behavior. In this sense, a quadratic speed up in $N$ could indicate a substantial improvement when measured by the dimension of the system.

Finally, let us remark that Proposition \ref{prop:adv} will hold under rather general conditions, since it is based on the fact that a sum over $N$ terms is involved in the estimation of $\#^{\text{seq}}$, while a sum over $\log(N)$ terms is required for $\#^{\text{par}}$ (see proof in Appendix \ref{app:proofdepths}). However,
the scaling of the number of gates $\#^{\text{par}}$ we provided in Theorem \ref{theo:scaling} is still based on a pessimistic bound and linked to product formulae: the generic bounds we used for Trotterization can overestimate by far \cite{childs2018toward, heyl2019quantum} the actual errors, which depend on the specific initial states and observables involved in the complete protocols. This means that actual implementations of the parallel-in-time protocols, whether based on product formulae or more advanced methods, might be much more efficient. The important message is that the advantages over sequential-in-time protocols, as stated in Proposition \ref{prop:adv}, hold more generally (see also below and the discussion).  

\subsection{Hamiltonian diagonalization  and Cartan decomposition approach}

In this section, we repeat the circuit depth analysis in another relevant scheme, namely assuming one has access to a diagonalization of the Hamiltonian. In particular, we will also discuss how one can obtain such diagonalization variationally via the algorithm presented in~\cite{commeau2020variational}. 

Let us recall that there always exists a unitary $W$ (whose columns are the eigenvectors of $H$) and a diagonal matrix $D$ (whose entries are the eigenvalues of $H$) such that
\begin{equation}\label{eq:diagonal}
    H=WDW\ad\,.
\end{equation}
Without loss of generality, we can expand $D$ is some basis of mutually commuting operators
\begin{equation}\label{eq:basis-comm}
    D=\sum_\mu c_\mu h_\mu\,,
\end{equation}
where $[h_\mu,h_{\mu'}]=0$ for all $\mu,\mu'$. If one has access to $W$ and $D$, then the unitary evolution can be expressed as
\begin{align}
    U(\varepsilon t)&=U(\varepsilon)^t\nonumber\\
    &=We^{- i D \varepsilon t}W\ad\nonumber\\
    &=We^{- i \sum_\mu c_\mu \varepsilon t h_\mu  }W\ad\nonumber\\
    &=W\left(\prod_\mu e^{- i c_\mu \varepsilon t h_\mu }\right)W\ad\,,\label{eq:diagonal-exp}
\end{align}

\begin{figure}[t!]
\centering
\includegraphics[width=1\linewidth]{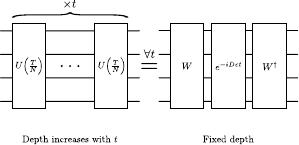}
\caption{\textbf{Depth-reduction via Hamiltonian diagonalization.} On the left, we implement $U(\varepsilon t)$ by expanding the evolution into shorter-time evolutions $U(\varepsilon)$ (which we can then implement via Trotterization), and then perform $U(\varepsilon t)=U(\varepsilon)^t$. This comes at the cost of increasing the depth as $t$ is increased. On the right, we use the diagonalization of $H$ as in Eq.~\eqref{eq:diagonal} to express $U(\varepsilon t)=We^{- i D \varepsilon t}W\ad$. Moreover, it can be seen by expanding $D$ in a basis of mutually commuting operators (as in Eq.~\eqref{eq:basis-comm}) that  the circuit implementation of $e^{- i D \varepsilon t}$ has the same depth for any $t$ (see also Eq.~\eqref{eq:diagonal-exp}).    }
\label{fig:depth-red}
\end{figure}

The power of Eq.~\eqref{eq:diagonal-exp} can be seen from Fig.\ ~\ref{fig:depth-red}, where it is shown that we can use the diagonalization of $H$ to implement $U(\varepsilon t)$ at fixed depth. Namely, the circuit depth for $\prod_\mu e^{- i c_\mu \varepsilon t h_\mu }$ and for $\prod_\mu e^{- i c_\mu \varepsilon t h_\mu  t'}$ is exactly the same, we just change the parameters associated to each time evolution generated by $h_\mu$. 
This means that in a sequential approach, each run requires $\mathcal{O}(n)$ gates independently of the evolution time, and assuming for simplicity that each $h_\mu$ acts as a one-body operator. The total number of gates is then $\#^{\text{seq}}=\mathcal{O}(n N)$.

\begin{figure*}[t!]
\centering
\includegraphics[width=1\linewidth]{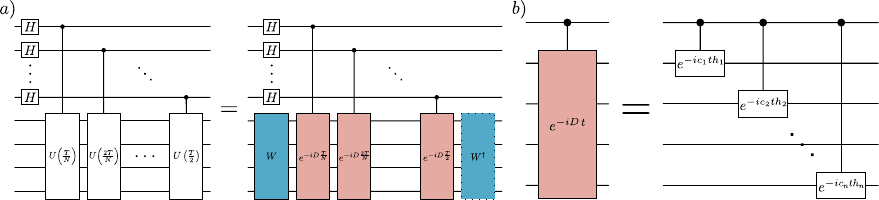}
\caption{\textbf{History state via Hamiltonian diagonalization.} a) Here we show how the circuit in Fig.~\ref{fig:hist-state} for preparing the history state changes when using the Hamiltonian diagonalization. In particular, we see that instead of controlling the $\log{N}$  gates $U(2^{j-1}\varepsilon)$ for $j=1,\ldots,\log{N}$ we now need to implement non-controlled unitaries $W$ and $W\ad$, and simply control the fixed-depth $\log{N}$  gates $e^{-i D 2^{j-1}\varepsilon}$ for $j=1,\ldots,\log{N}$. We note that if we only care about the entanglement in $\rho_T$ as in Figs.~\ref{fig:circuit-Ent_Overlap} and~\ref{fig:circuit-Ent_Rand}, then the final unitary $W\ad$ can be omitted (the entanglement is invariant under local unitaries~\cite{horodecki2009quantum}). b) For the special case when  the diagonal Hamiltonian $D$ is expressed as a sum of single-qubit operators acting on each qubits (i.e., $h_\mu$ in Eq.~\eqref{eq:basis-comm} is a  one-body operator acting on the $\mu$-th qubits), then controlling the gates $e^{-i D 2^{j-1}\varepsilon}$ simply requires controlling single-qubit gates.  }
\label{fig:diag-history}
\end{figure*}

Moreover, the benefits of diagonalizing the Hamiltonian $H$ are amplified when using this technique in the circuit for preparing the history state.  As shown in Fig.~\ref{fig:diag-history}(a), we can see that instead of controlling $\log{N}$  gates $U(2^{j-1}\varepsilon)$ (where for $j=1,\ldots,\log{N}$), the history state can be prepared by first acting on the system qubits with the non-controlled unitary $W$, followed by $\log{N}$ controlled gates $e^{-i D 2^{j-1}\varepsilon}$, and finally by implementing a non-controlled unitary $W\ad$. This further reduces  the depth required to prepare the history state. First, we do not need to control $W$, nor $W\ad$. Second, we note that controlling $e^{-i D \varepsilon t}$ (for any $t$) is equivalent to controlling each term $e^{- i c_\mu \varepsilon t h_\mu}$ (since the  $h_\mu$ are mutually commuting). For instance, we can see in Fig.~\ref{fig:diag-history}(b) that if the $h_\mu$ are single-qubit Pauli operators acting on each qubit, then implementing a controlled $e^{-i D \varepsilon t}$ gate, just requires controlling $n$ single qubit rotations.
This gives the following Theorem.
\begin{theorem}\label{theo:depths-new}
By replacing the history state preparation subroutine with the diagonalized Hamiltonian as in Eq.~\eqref{eq:diagonal} and Fig.~\ref{fig:diag-history}, we can prepare the history state with a total number of gates
\begin{equation}
    \#^{\text{par}}\in \OC(\log(N)n)
\end{equation}
 i.e.  logarithmic in both the number of times and system size.
\end{theorem}

 The parallel-in-time advantage over the sequential approach condition now becomes $N>\beta \log(N)$, which is independent of the system size and is virtually always reached. Of course, one could argue that if we have classical access to the diagonalized Hamiltonian, then we could just expand the initial state state and the measured operators in the energy eigenbasis to compute any expectation value. However, this kind of expansion will not be tractable for large problem sizes. Instead, we will show below that if $W$ is accessible in a quantum computer, then one can still leverage the diagonalization for depth reduction.

Finally, we note that if we want to study the entanglement in $\rho_T$ as in Theorems~\ref{theo:ent-loschmidt} and~\ref{theo:ent} (see also Figs.~\ref{fig:circuit-Ent_Overlap} and~\ref{fig:circuit-Ent_Rand}), then the final unitary $W\ad$ in Fig.~\ref{fig:diag-history}(a) can be omitted. This is due to the fact that the entanglement is invariant under local unitaries~\cite{horodecki2009quantum}, and hence $W\ad$ cannot change the entanglement nor the spectral properties of $\rho_T$.

It is worth highlighting the fact that the main challenge for using Eq.~\eqref{eq:diagonal-exp} is that it requires access to the decomposition in Eq.~\eqref{eq:diagonal}, and that $W$ and $D$ might not be readily accessible. However, one can still attempt to variationally learn them~\cite{cerezo2020variationalreview}. For example, one can use the Variational Hamiltonian Diagonalization algorithm in Ref.~\cite{commeau2020variational} which is aimed at training a parametrized ansatz for the diagonalization of $H$. The ansatz is composed of two parts: 1) A parametrized unitary $W(\vec{\alpha})$, and 2) A diagonal Hamiltonian $D(\vec{\beta})$ such that
\begin{equation}\label{eq:ansatz}
    \widehat{H}(\vec{\alpha},\vec{\beta})=W(\vec{\alpha})D(\vec{\beta})W\ad(\vec{\alpha})\,.
\end{equation}
 One can quantify how much $\widehat{H}(\vec{\alpha},\vec{\beta})$ approximates the target Hamiltonian $H$ by defining the cost function
\begin{equation}\label{eq:vhd_cost_function}
    C(\vec{\alpha},\vec{\beta})=\frac{\norm{H-\widehat{H}(\vec{\alpha},\vec{\beta})}_{HS}^2}{2^n}\,,
\end{equation}
where $\norm{X}_{HS}=\sqrt{\Tr[X\ad X]}$ is the Hilbert-Schmidt norm. Clearly, the cost is equal to zero if $H=\widehat{H}(\vec{\alpha},\vec{\beta})$. Thus, the parameters $\vec{\beta}$ and $\vec{\alpha}$ are trained by solving the optimization task
\begin{equation}\label{eq:opt}
    \argmin_{\vec{\beta},\vec{\alpha}} C(\vec{\alpha},\vec{\beta})\,.
\end{equation}
Here, where a quantum computer is used to estimate the term in $C(\vec{\alpha},\vec{\beta})$~\cite{commeau2020variational}, while classical optimizers are used to train the parameters. 

In this variational setting, it is extremely important to pick an ansatz (i.e., a given unitary $W(\vec{\alpha})$, and diagonal Hamiltonian $D(\vec{\beta})$) which do not lead to trainability issues such as barren plateaus~\cite{mcclean2018barren,cerezo2020cost,holmes2021connecting,cerezo2022challenges,larocca2024review}, where the cost function gradients are exponentially suppressed with the problem size. One of the leading strategies to mitigate such issues is to use the so-called problem-inspired ansatzes, where one creates ansatzes with strong inductive biases~\cite{kubler2021inductive,larocca2021diagnosing,larocca2022group} based on the problem at hand. Recently, one such method was developed which is exploits the Cartan decomposition of the Lie algebra generated by the target Hamiltonian $H$~\cite{kokcu2022fixed,camps2022algebraic,kokcu2022algebraic}. Below we explain such method.

Consider the Hamiltonian of interest $H$. Then, without loss of generality we assume that it can be expressed as a sum of Hermitian traceless operators $\{H_i\}$ as 
\begin{equation}
H = \sum_i a_i H_i\,,
\end{equation}
where $a_i \in \mbb{R}$. Then, let $\mf{g}=\langle \{ i H_i\} \rangle_{Lie}$ be the Lie closure of the set of operators~\cite{zeier2011symmetry}. Note that, by definition, $iH\in \mf{g}$. The result in  Ref.~\cite{kokcu2021fixed},  provides an efficient-in-$\dim(\mf{g})$ circuit for the simulation of any $e^{-i Ht}$ for any set of coefficients $\{a_i\}$. To understand the technique of Ref.~\cite{kokcu2021fixed}, we recall that a Cartan decomposition of the Lie algebra $\mf{g}$ refers to the decomposition of $\mf{g}$ into two orthogonal subspaces $\mf{g}=\mf{k}\oplus \mf{m}$, where $\mf{k}$ is a Lie subalgebra, i.e., $[\mf{k},\mf{k}]\subseteq \mf{k}$, whereas $\mf{m}$ is not: $[\mf{m},\mf{m}]\subseteq \mf{k}$. Moreover, these two orthogonal subspaces satisfy $[\mf{k},\mf{m}]=\mf{m}$, and $\mf{m}$ contains the maximal commutative subalgebra, also known as the Cartan subalgebra,  $\mf{h}$ of $\mf{g}$. Note that for any pair of element $i h_1$ and $i h_2$ in $\mf{h}$, we have $[h_1,h_2]=0$.

The Cartan decomposition provides us an ansatz to diagonalize $H$ as follows. First we note that $H$ always admits a decomposition of the form
\begin{equation}\label{eq:KAK}
H = W h K\ad\,,
\end{equation}
with $K \in e^{\mf{k}}$ and $h\in\mf{h}$. Reference~\cite{kokcu2021fixed}  provides us with an ansatz for Eq.~\eqref{eq:ansatz} as we can now parametrize and optimize over the Lie group $e^{\mf{k}}$ and the algebra $\mf{h}$. That is, we simply pick
\begin{equation}\label{eq:ans-W}
W(\vec{\alpha}) = \prod_\nu e^{i \alpha_\nu B_\nu}\,,
\end{equation}
where $ i B_\nu$ belongs to  a basis of $\mf{k}$, and
\begin{equation}\label{eq:ans-h}
D(\vec{\beta}) = \sum_\mu \beta_\mu h_\mu \,,
\end{equation}
with $i h_\mu$ belonging to  a basis of $\mf{h}$. Taken together, Eqs.~\eqref{eq:ans-W} and~\eqref{eq:ans-h} provide a problem-inspired ansatz for the diagonalization of $H$ which we can use to solve Eq.~\eqref{eq:opt}.

\subsubsection{Example: $XY$ model}\label{sec:cartan_XY}

Let us here exemplify the Cartan decomposition-based method. Consider a general $XY$  Hamiltonian of the form
\begin{equation}\label{eq:hxx}
H=\sum_{j=1}^{n-1} a_j^x X_j X_{j+1}+a_j^y Y_j Y_{j+1}+ \sum_j a_j^z Z_j\,.
\end{equation}
Here, one can prove that  $\mf{g}=\langle i\{ X_j X_{j+1},Y_j Y_{j+1},Z_j\} \rangle_{\Lie} \cong \mf{so}(2n)$. Thus, the ensuing  Cartan decomposition is 
\small
\begin{align}
\mf{k} =& \spn\{ \widehat{X_i Y_j},\widehat{Y_i X_j}\,|\, 1\leq i <j\leq n\}\cong \mf{so}(n) \oplus \mf{so}(n)\label{eq:algebra-k} \\
\mf{m} =& \spn\{ Z_j, \widehat{X_i X_j},\widehat{Y_i Y_j}\,|\, 1\leq i <j\leq n\} \\
\mf{h} =& \spn\{ Z_j \,|\, 1\leq i \leq n\} \cong \oplus_{i=1}^n \mf{u}(1)
\end{align}
\normalsize
where we used the notation $\widehat{A_iB_j}=A_i Z_{i+1} \cdots Z_{j-1} B_j$. Let us mention that we can map all the elements in $\mf{g}$ to operators quadratic in fermionic operators. In fact, all the gates obtained via exponentiation of this algebra are matchgates \cite{jozsa2008matchgates, wan2022matchgate, diaz2023showcasing}.

Here, we can see that the ansatz for the diagonal part of the ansatz in Eq.~\eqref{eq:ansatz} is simply
\begin{equation}\label{eq:ans-h-XY}
D(\vec{\beta}) = \sum_{j=1}^n \beta_\mu Z_j \,.
\end{equation}
On the other hand, it is clear from Eqs.~\eqref{eq:ans-W} and~\eqref{eq:algebra-k} that a drawback in the proposal  of Ref.~\cite{kokcu2021fixed} is that it requires us to implement gates which are obtained by exponentiation of highly non-local operators (e.g. $e^{-i \alpha \widehat{X_1Y_n}}$), a task which can be hard to implement and lead to deep circuits. Hence,  we propose a different parametrization for $W(\vec{\alpha})$. Consider the following set of local operators. 
\begin{equation}
\GC=\{ X_j Y_{j+1},  Y_j X_{j+1} \}_{j=1}^{n-1} \subset \mf{k}\,.
\end{equation}
We can prove that the following proposition holds.
\begin{proposition}\label{prop:generating-set}
 $i\GC$ is a generating set of the algebra $\mf{k}$. That is, $\langle i \GC \rangle_{Lie}=\mf{k}$\,.
\end{proposition}

\begin{figure}[t!]
\centering
\includegraphics[width=1\linewidth]{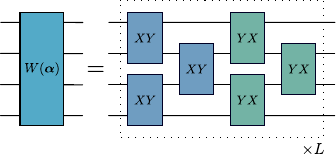}
\caption{\textbf{Ansatz for $W(\vec{\alpha})$.}  By using Proposition~\ref{prop:generating-set} we propose an ansatz for the diagonalizing unitary $W(\vec{\alpha})$ which only uses local gates acting on neighbouring qubits. We shown here a single ``layer'' of an $n=4$ ansatz which is repeated $L$ times.   }
\label{fig:dansatz-W}
\end{figure}

The key implication of Proposition~\ref{prop:generating-set} is that we can generate any unitary in the unitary subgroup $e^{\mf{k}}$ by exponentiating only the local operators in $\GC$. This means, that one can diagonalize $H$ using an ansatz of the form 
\begin{equation}\label{eq:local-gen}
W(\vec{\alpha}) = \prod_{l=1}^L \prod_{j=1}^{n-1} e^{i \alpha_{j,l} X_j Y_{j+1} }\prod_{i=1}^{n-1} e^{i \alpha_{i,l} Y_i X_{i+1} }\,.
\end{equation}
We explicitly show the form of this ansatz in Fig.~\ref{fig:dansatz-W}. Note that since the operators in Eq.~\eqref{eq:local-gen} are  two-body, then the circuit for $W(\vec{\alpha})$ only requires local two-qubit gates. Hence, such construction significantly reduces the circuit requirements over that in Ref.~\cite{kokcu2021fixed}.

The question still remains to how large $L$ needs to be. Here, we can leverage recent results from the quantum machine learning literature which state  that by taking $L\geq \dim(\mf{k})/ (2n-2)$ it is generally sufficient to  guarantee that any $K\in e^{i\mf{k}}$ will be expressible~\cite{larocca2021diagnosing}. Moreover, in this regime the ansatz is said to be overparametrized. In this overparametrization regime, the optimization of Eq.~\eqref{eq:opt} becomes much easier to solve as many spurious local minima disappear~\cite{larocca2021diagnosing,you2022convergence}.

Putting the previous results together, and assuming we can  efficiently solve Eq.~\eqref{eq:opt}, we can derive the following theorem.
\begin{theorem}\label{theo:depths}
    Let $H$ be an $XY$  Hamiltonian of the form in Eq.~\eqref{eq:hxx}. Then, let $D(\vec{\beta})$ be a diagonal operator as in Eq.~\eqref{eq:ans-h-XY}, and let $W(\vec{\alpha})$ be a unitary as in \eqref{eq:local-gen}. By replacing the history state preparation subroutine with the trained diagonalized Hamiltonian as in Eq.~\eqref{eq:ansatz} and Fig.~\ref{fig:diag-history}, we can implement the circuits used in Theorems~\ref{theo:parallel1},~\ref{theo:parallel-losch}, and~\ref{theo:ent} with circuit depths in $\OC(\log(N)n)$.
\end{theorem}

The results in Theorem~\ref{theo:depths} showcase the extreme power of diagonalizing $H$ via its Cartan decomposition as we can implement all the circuits in Figs.~\ref{fig:circuit-F-parallel},~\ref{fig:circuit-loschmidt-parallel},~\ref{fig:circuit-Ent_Overlap}, and~\ref{fig:circuit-Ent_Rand} with a depth that only scales as the product of the number of system and clock qubits.

Let us remark that one key aspect of our example is the polynomial size of the Lie algebra $\mf{g}$, which allows us to efficiently diagonalize the considered Hamiltonian. More complicated Hamiltonians might lead to exponentially large algebras which for deep circuits (as those needed in the overparameterized regime) naturally induce BPs.  We refer the reader to \cite{diaz2023showcasing,larocca2024review} and \cite{cerezo2023does} for a recent discussion of these type of challenges faced by quantum variational protocols.  Importantly, for our the purpose of the present work, the use of VHD is not central, as the idea of parallelization and the benefits offered by using our proposed methods do not rely on solving a variational optimization problem. As previously mentioned, one can use parallel-in-time techniques with Troterrization, VHD, or any compilation method available.

\section{Numerical simulations}\label{sec:numerics}

In this section we first provide numerical simulations that showcase how the discrete-time approximations (computable via our algorithms) can capture the behaviour of their continuum time counterparts. Similarly, we also show numerically that the system-time entanglement provides a new way to understand dynamical properties of the system. Next, we will demonstrate how the  variational Hamiltonian diagonalization (VHD) algorithms can be used to reduce the depth of the history state preparation circuit,  as discussed in Section~\ref{sec:cartan_XY}.

  \begin{figure}[t!]
\centering
\includegraphics[width=1\columnwidth]{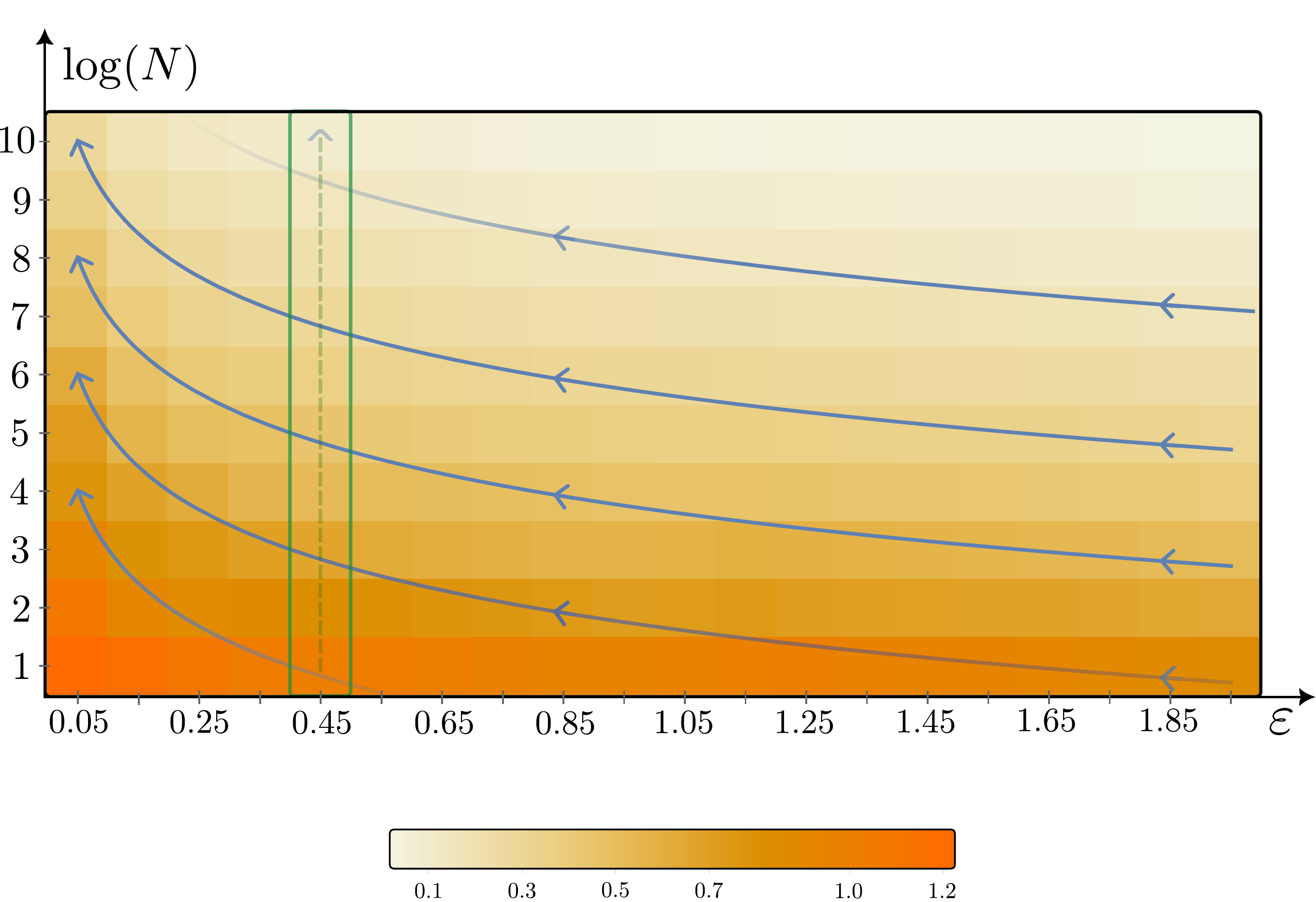}
\caption{\textbf{Average approximation error $|\bar{\LC}(\psi_0)-\widetilde{\LC}(\psi_0)|$ for a chain of $n=200$ sites.} The vertical indicates the number of clock qubits while the horizontal axis is the window size $\varepsilon$. The blue (solid) curves represent constant values of $T$ with the arrows pointing toward the direction of greater resolution. Green arrows indicate a cross cut of fixed $\varepsilon$ an increasing $\log(N)$ that is shown in further detail in Fig.~\ref{fig:loshmidt}.}
\label{fig:2dplot}
\end{figure}

In all of our experiments, we consider a system of $n$-qubits evolving by a unitary generate by the time-independent non-uniform $XXZ$ Heisenberg model, whose  Hamiltonian reads
\begin{equation}\label{eq:hxxnum}
     H=\sum_{j=1}^n \Big[\frac{J}{4}(X_j X_{j+1}+Y_jY_{j+1})+ \Delta  Z_j Z_{j+1}+ h_j(Z_j+\id)\Big]\,,
\end{equation}
where we define either periodic boundary conditions as $h_{n+1}\equiv h_1$ (for $h=X,Y,Z$) or open boundary conditions $h_{n+1}\equiv 0$, and $h_j=\frac{\lambda}{2}\cos(2\pi \alpha j)$.  For $\alpha=\frac{\sqrt{5}-1}{2}$ and $\Delta=0$, one can use the Jordan-Wigner transformation~\cite{aubry1980analyticity} (see details in Appendix \ref{App:methods}) to show that in the thermodynamic limit this model exhibits a delocalization-localization transition at the critical point $\lambda=J$.  Indeed, it is well known that such transition  induces sharp changes in long-time dynamical properties such as the  Loschmidt echo average~\cite{zhou2019signature}. Our goal is then use this paradigmatic model as a test-bed to show that our proposed discrete-time average of the Loschmidt echo can capture the behavior of their continuum time counterparts. Analogously, for $\Delta\neq 0$ it is expected that this model exhibits a many-body localization (MBL) transition with localization surviving the presence of interactions \cite{iyer2013many, vstrkalj2021many}.

\subsection{Single particle localization, Discrete time averages and system-time entanglement}

\begin{figure*}[t]
\centering
\includegraphics[width=1\linewidth]{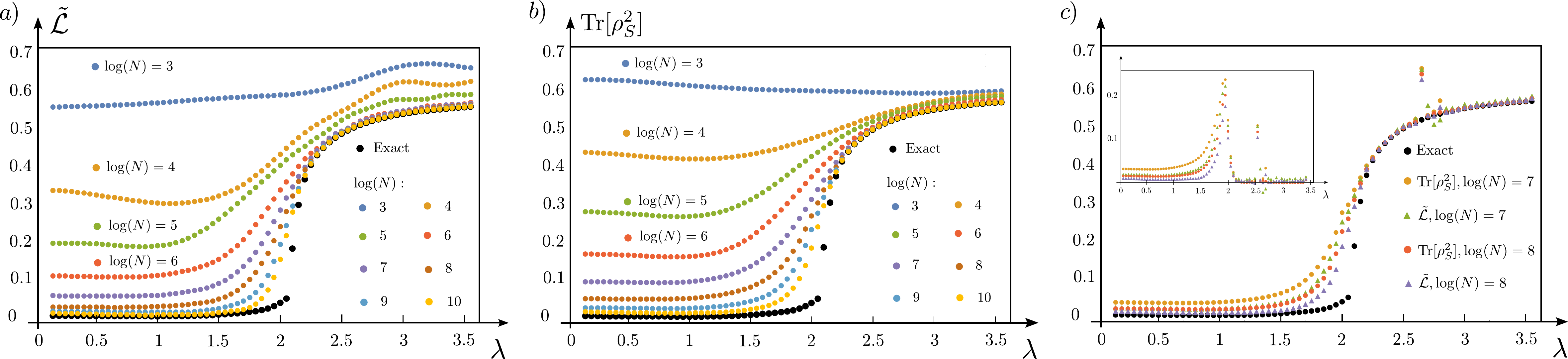}
\caption{\textbf{Discrete-time approximations to the Loschmidt echo $\bar{\mathcal{L}}(\psi_0)$ (black dots) as a function of $\lambda$ for a chain of $n=200$ sites.} In panel $a)$ we depict the exact infinite-time average , as well as the discrete-time approximation $\widetilde{\mathcal{L}}(\psi_0)$ for different values of clock qubits $\log(N)$ and for  $\varepsilon=0.45$. In $b)$ we depict the exact infinite-time average $\bar{\mathcal{L}}(\psi_0)$ (black dots), as well as the purity of the reduced state  $\rho_{S}=\Tr_{T}[\dya{\Psi}]$  for different values of clock qubits $\log(N)$ and for  $\varepsilon=0.45$. In $c) $ we depict the exact infinite-time average $\bar{\mathcal{L}}(\psi_0)$ (black dots), its discrete-time approximation $\widetilde{\mathcal{L}}(\psi_0)$, as well as the purity of the reduced state  $\rho_{S}=\Tr_{T}[\dya{\Psi}]$  for different values of clock qubits $\log(N)$and for $\varepsilon=1.25$. The inset corresponds to the difference between each approximation and $\bar{\mathcal{L}}(\psi_0)$.}
\label{fig:loshmidt}
\end{figure*}

Let us focus in the $\Delta=0$ case first. 
To study the discrete-time average of the Loschmidt's echo we have considered a chain of $n=200$ sites with $J=2$, $\alpha=\frac{\sqrt{5}-1}{2}$, and a number of clock qubits ranging from $1$ to $10$, corresponding to a maximum number of $N=1024$ times. Note that  with this choice, the system dimension is equal to  $2^{200}$ and hence, much larger than the clock Hilbert space dimension, $N$. To study the effects of the window size, we have also considered  values of $\varepsilon$ spanning from $0.05$ up to $1.95$ with spacing $0.1$ (see Fig.~\ref{fig:tradeoff}). The initial state of our simulations is $|\psi_0\rangle=(s^+_{99}+s^+_{100}+s^+_{101})|\downarrow\downarrow\dots \downarrow\rangle/\sqrt{3}$, where $s^+_j$ denotes the rising spin operator at site $j$. As such, at $t=0$, the state is only  partially delocalized in the middle of the chain. All simulations, including the computation of the exact infinite-time average $\bar{\LC}(\psi_0)$, where performed via Jordan-Wigner  diagonalization, and we refer the reader to  Appendix \ref{App:methods} for additional details.

In Fig.~\ref{fig:2dplot} we first present a two-dimensional plot of the error between the infinite-time average of the Loschmidt echo $\bar{\LC}(\psi_0)$ and its discrete-time approximation $\widetilde{\LC}(\psi_0)$ ($|\bar{\LC}(\psi_0)-\widetilde{\LC}(\psi_0)|$) averaged over $\lambda\in (0.1,3.5)$ (with spacing $\Delta \lambda=0.05$), for  different values of  $\varepsilon$ and $n$. Here, we can see that, as expected, the error is reduced by  increasing the number of clock qubits. The improvement follows two tendencies. First, there is an overall improvement when increasing $T$ (i.e., when moving up in the $\log(N)$ axis for fixed $\varepsilon$), as this  corresponds to better accuracy. On the other hand, for constant $T$ it is beneficial to reduce $\varepsilon$ (i.e., increase resolution), as shown by the blue solid curves.

We further explore the effect of fixing $\varepsilon$ and increasing $\log(N)$ in Fig.~\ref{fig:loshmidt} $a)$. Therein we show  $\bar{\mathcal{L}}(\psi_0)$, as well as its discrete-time approximation $\widetilde{\mathcal{L}}(\psi_0)$ for different number of clock qubits  as a function of $\lambda$ for fixed resolution $\varepsilon=0.45$ (vertical dashed line in Fig.~\ref{fig:2dplot}). First, we note that the infinite-time Loschmidt echo captures the delocalization-localization transition occurring at $\lambda=J=2$. In particular, for $\lambda<2$ we see that $\bar{\mathcal{L}}(\psi_0)$ is small, indicating a delocalized phase. On the other hand, for $\lambda>2$ the evolved state is localized as $\bar{\mathcal{L}}(\psi_0)$ is large. Next, let us note that as $\log(N)$ increases, $\widetilde{\mathcal{L}}(\psi_0)$ quickly becomes a good approximation for its infinite-time counterpart (as expected from  Fig.~\ref{fig:2dplot}). However, Fig.~\ref{fig:loshmidt} also reveals that $\widetilde{\mathcal{L}}(\psi_0)$ capture the delocalization-localization transition even for a small number of clock qubits. Already for $\log(N)=6$ the inflection point of $\widetilde{\mathcal{L}}(\psi_0)$ approaches the critical value $\lambda=2$.

Next, we study how the system-time entanglement, as measured through the subsystem purity $\Tr[\rho_{S}^2]$ for $\rho_{S}=\Tr_{T}[\dya{\Psi}]$, approximates the infinite-time average Loschmidt echo (see Corollary~\ref{theo:ent-loschmidt}). In Fig.~\ref{fig:loshmidt} $b)$ we  plot $\bar{\mathcal{L}}(\psi_0)$, as well as $\Tr[\rho_{S}^2]$, for different number of clock qubits  as a function of $\lambda$. Again, we see a clear convergence towards $\bar{\mathcal{L}}(\psi_0)$ as the number of clock qubits is increased. This result shows that the subsystem purity provides an excellent approximation of $\bar{\mathcal{L}}(\psi_0)$. Moreover, one can also observe that the system-time entanglement clearly captures the delocalization-localization transition. This fact can be readily understood from the fact that  in the localized phase the state does not change considerably with time, and hence a small amount of entanglement is expected. This example perfectly exemplifies the fact that the system-time entanglement in the history state carries valuable information about the system dynamics. Moreover, since we know that $\Tr[\rho_{S}^2]=\Tr[\rho_{T}^2]$, then one can estimate the reduced state tomography by studying only the reduced state on the $\log(N) (\ll n)$ clock qubits.

Figures~\ref{fig:loshmidt} $a)$ and $b)$ show that both the discrete-time Loschmidt echo and the subsystem purity provide good approximations of $\bar{\mathcal{L}}(\psi_0)$. To better compare their performance, we show in Fig.~\ref{fig:loshmidt} $c)$ curves for $\bar{\mathcal{L}}(\psi_0)$, $\widetilde{\mathcal{L}}(\psi_0)$  and $\Tr[\rho_{S}^2]$ for the same chain of $n=200$ spins, but for $\varepsilon=1.25$, i.e., for less accuracy (see Fig.~\ref{fig:tradeoff}).  In this regime, one can see that  while $\widetilde{\mathcal{L}}(\psi_0)$ suffers from undesired oscillations, $\Tr[\rho_S^2]$ can still provide a good approximation for the same number of qubits. In particular, Fig.~\ref{fig:loshmidt} $c)$ shows that $\widetilde{\mathcal{L}}(\psi_0)$ can be smaller than $\bar{\mathcal{L}}(\psi_0)$ in unpredictable ways (due to insufficient resolution), meaning that  $\widetilde{\mathcal{L}}$ cannot be strictly used to provide strict bounds such as the one in Corollary~\ref{theo:ent-loschmidt}.  While ${\rm Tr}[\rho_S^2]$ oscillates as well, this quantity never crosses the black points, in agreement with our bounds. Here we also observe that the system-time entanglement provides a better convergence in the localized region. On the other hand, the entanglement curves are above the $\widetilde{\mathcal{L}}$ curves in the delocalized sector. Notice however that this discrepancy can be mitigated by increasing the number of qubits. Finally, we note that in Fig.~\ref{fig:loshmidt} $c)$ we also  depict the differences $\left(\widetilde{\mathcal{L}}(\psi_0)-\bar{\mathcal{L}}(\psi_0)\right)$ and $\left(\Tr[\rho_{S}^2]-\bar{\mathcal{L}}(\psi_0)\right)$, which confirm that $\Tr[\rho_{S}^2]$ is always strictly larger than $\bar{\mathcal{L}}(\psi_0)$, whereas $\widetilde{\mathcal{L}}(\psi_0)$ can indeed be smaller that the infinite-time average.

\begin{figure}[t]
\centering
\includegraphics[width=1\columnwidth]{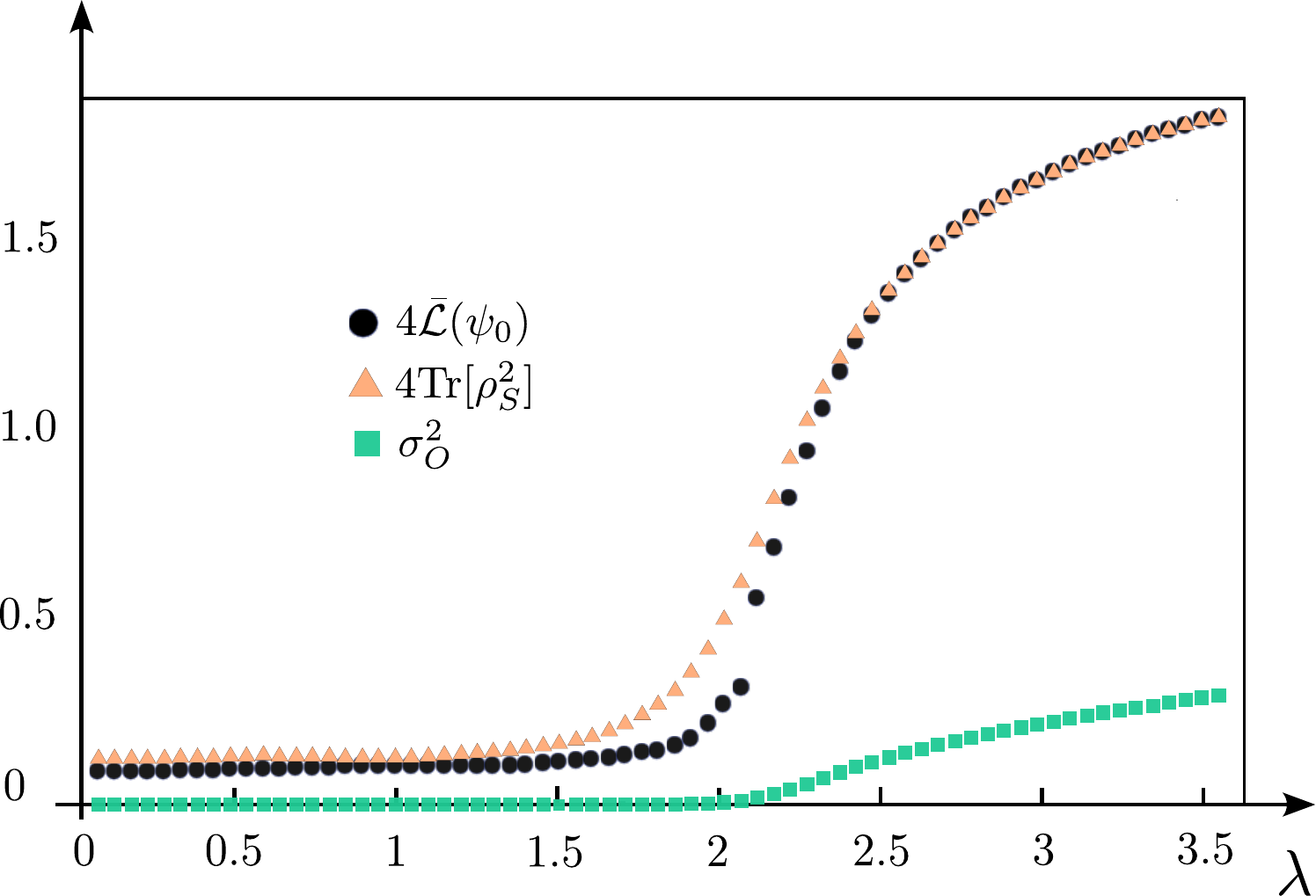}
\caption{\textbf{Observable fluctuations as a function of $\lambda$ for a chain of $n=100$ sites.}  We depict the observable fluctuations $\sigma_O^2$, the infinity-time average of the Loschmidt echo $\bar{\mathcal{L}}(\psi_0)$ and the reduced subsystem purity $\Tr[\rho_S^2]$. We consider $\log(N)=9$ qubit clocks, and we take $\varepsilon=0.5$   }
\label{fig:fluctuations}
\end{figure}

Finally, as an example of Corollary \ref{cor:1} we also numerically show  how the system-time entanglement provides a bound for the fluctuation of observables. We use as an example the observable $O= s^+_{L/2}s^-_{L/2+1}+s^+_{L/2+1}s^-_{L/2}$ and as the initial state $|\psi_0\rangle=(s^+_{L/2}+s^+_{L/2+1})|\downarrow \downarrow \dots \downarrow\rangle/\sqrt{2}$. In this case, the bounds of Eq.~\eqref{eq:cor-tempfluct-bound} becomes 
\begin{equation}
    \sigma_O^2\leq 4\bar{\mathcal{L}}(\psi_0)\leq 4\Tr[\rho_S^2]
\end{equation}
since $\Delta_O=2$. In Fig.~\ref{fig:fluctuations} we plot the numerical results for a chain of $n=100$ sites and $\log(N)=9$ qubit clocks (i.e., $512$ times). We see that while the bound is not tight, both $\bar{\mathcal{L}}(\psi_0)$  and $\Tr[\rho_S^2]$ are capable of clearly separating the different phases. As expected from our bounds, the system-time entanglement provides a less tight but strict bound. However, given that one can experimentally compute the system-time entanglement efficiently in quantum computers, this bound is still useful for practical purposes. Moreover, it is important to highlight again the fact that the system-time entanglement is obtained from a  discrete-time formalism (in contrast to $\bar{\mathcal{L}}(\psi_0)$ which requires infinite time averages). As such, our new notion of system-time entanglement provides valuable and strict information about the system's observable dynamics and its eventual equilibration (a feature not available  for the discrete time Loschmidt echo $\widetilde{\mathcal{L}}(\psi_0)$).

\subsection{Diagonalization via Cartan decomposition}

In this section we show how one can use the variational Hamiltonian diagonalization (to diagonalize the Hamiltonian in Eq.~\eqref{eq:hxxnum} and thus reduce the depth of the history state preparation circuit. We will  take the ansatz for $D(\vec{\beta})$ and $W(\vec{\alpha})$ as appearing in Eqs.~\eqref{eq:ans-h-XY} and~\eqref{eq:local-gen}. Thus, as depicted in Fig.~\ref{fig:dansatz-W}, $W(\vec{\alpha})$ consists of $L$ layers of two qubit gates generate by $XY$ and $YX$ arranged in a brick wall fashion, whereas the diagonal part $D(\vec{\beta})$ is just a sum of Pauli $Z$ operators on each qubit.  To train the parameters $\vec{\alpha}$ and $\vec{\beta})$, we will optimize the  Hilbert-Schmidt  cost function defined in Eq.~\eqref{eq:vhd_cost_function}.   Details of the simulation can be found in Appendix~\ref{App:methods}.

We considered a chain of $n=6$ sites, and set $J=2$, and $\alpha=\frac{\sqrt{5}-1}{2}$. Moreover, we diagonalized the Hamiltonian for  $\lambda=1,2,3$, thus allowing us to show the success of the algorithm in each important region of the phase diagram.

\begin{figure}[h]
    \centering
    \includegraphics[width=1\columnwidth]{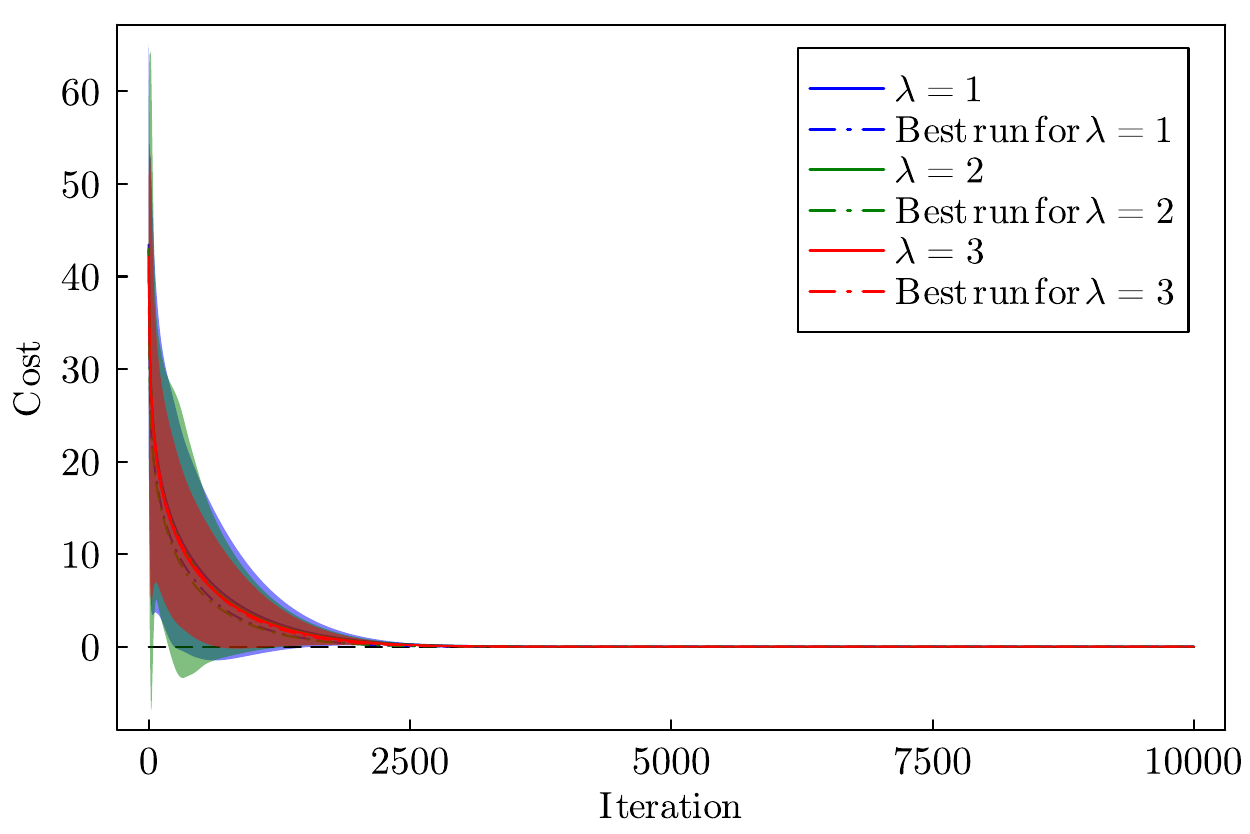}
    \caption{\textbf{Variational Hamiltonian diagonalization training curves.} The plot shows the decay of the loss function in Eq.~\eqref{eq:vhd_cost_function} versus the number of  training iterations. Solid lines correspond to the average over $10$ different initializations for each value of $\lambda \in \{1,2,3\}$. Shaded regions indicate the variance over the runs sample, while dash-dot lines show the best run for each batch of training runs, i.e., the one that reached the stopping value first.}
    \label{fig:vhd_training}
\end{figure}

\begin{figure}[t]
    \centering
    \includegraphics[width=0.5\textwidth]{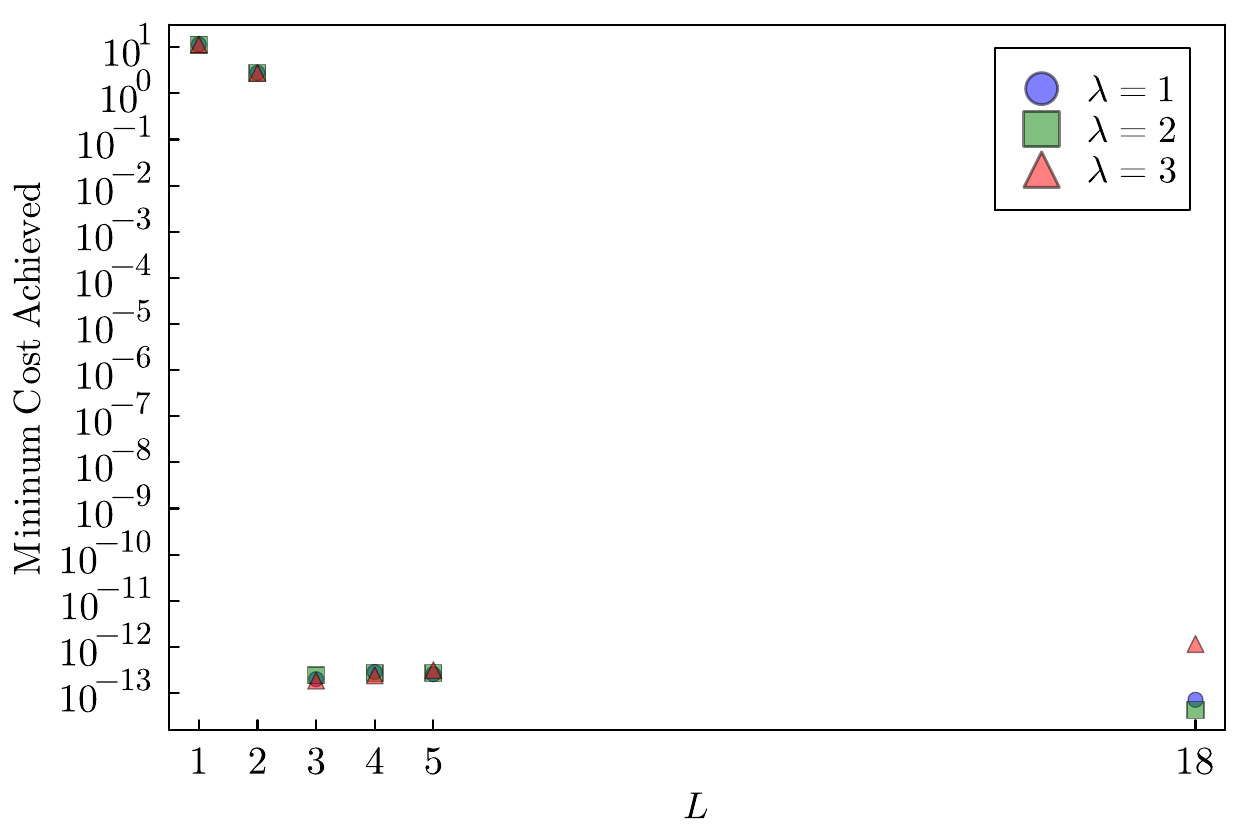}
    \caption{\textbf{Overparametrization transition.} We show the minimum cost achieved when training the Variational Hamiltonian Diagonalization algorithm for a problem system of  $n=6$ sites. The overparametrization regime is reached with a number of layers $L\geq 3$. }
    \label{fig:vhd_overparm_transition}
\end{figure}

\begin{figure}[t]
    \centering
    \includegraphics[width=0.5\textwidth]{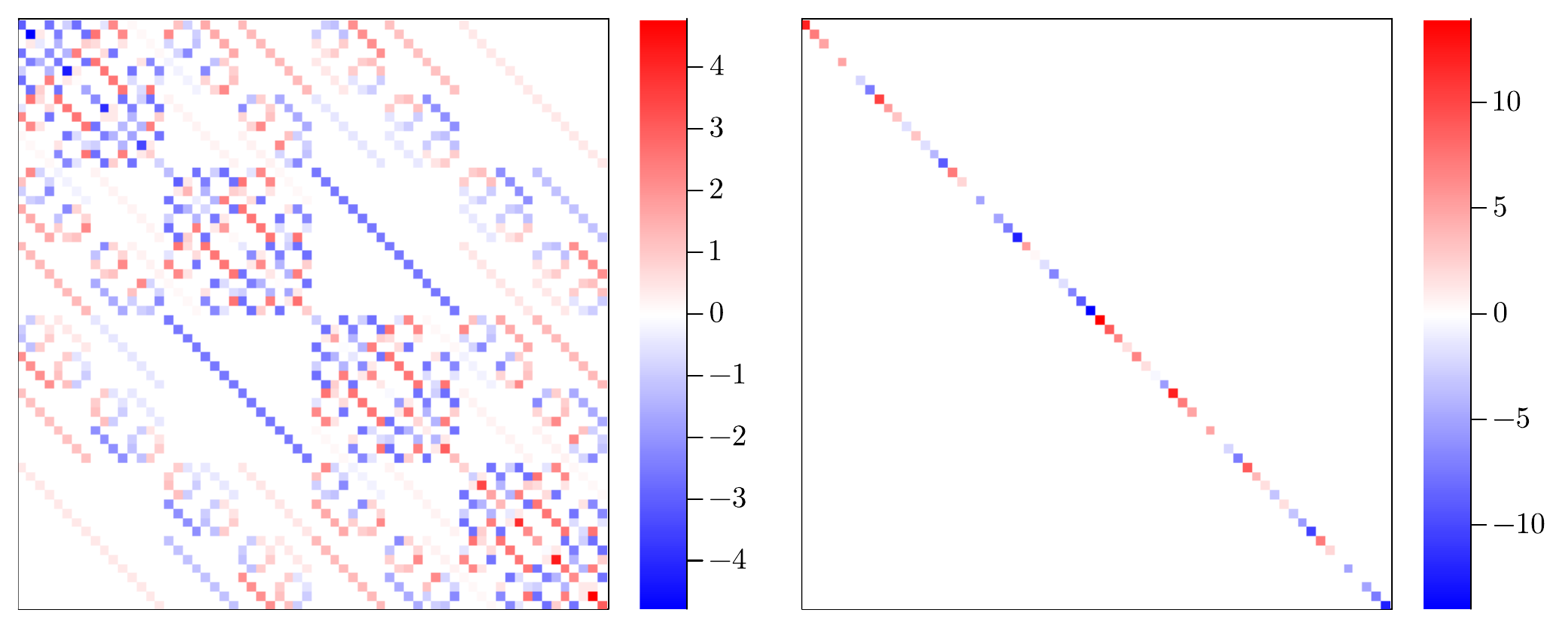}
    \caption{\textbf{Testing the success of the Hamiltonian diagonalization algorithm.}  The left panel shows the initial ``diagonalized'' Hamiltonian $\hat{D}(\vec{\alpha})=W(\vec{\alpha})H W\ad(\vec{\alpha})$ where $\vec{\alpha}$ is a vector of initial random parameters $\alpha_i \in [0, 2\pi)$. The right panel instead features $\hat{D}(\vec{\alpha}^*)$ obtained applying the trained diagonalizing unitary to its target  Hamiltonian $H$. Here $\vec{\alpha}^*$ are the optimal parameters of the random successful run at hand.}
    \label{fig:vhd_diag_hamiltonian}
\end{figure}

In Fig.~\ref{fig:vhd_training} we show the training curves (loss function versus iteration step) for an ansatz with $L=18$ layers the three different values of $\lambda$ that we considered. Here we can clearly see that as the number of iterations increases, the cost function value goes to zero, indicating that we can accurately diagonalize the target Hamiltonian. Notably, we can see that all the trained curves converged to the solution, meaning that the optimizer did not get stuck in a local minima. Such extremely high optimization success rate can be understood from the fact that the circuit is overparametrized~\cite{larocca2021theory}, i.e., it contains enough parameters to explore all relevant directions. In fact, using the results from~\cite{larocca2021theory} we know that a circuit with a set of generators $\GC$ will be overparametrized if the number of parameters is $~\dim(\langle i\GC \rangle_{{\rm Lie}})$. Importantly, we can use Proposition~\ref{prop:generating-set} to know that $\dim(\langle i\GC \rangle_{{\rm Lie}})=n(n-1)=30$. Since the ansatz $W(\vec{\alpha})$ contains $2(n-1)$ parameters per layer (see Fig.~\ref{fig:dansatz-W}), then we can overparametrize it with $L=\left\lceil n/2 \right\rceil$. Indeed, in Fig.~\ref{fig:vhd_overparm_transition} we show the minimal cost achieved versus the number of layers $L$, and as expected we see a computational phase transition at $L=3$: For smaller number of layers, the ansatz is underparametrized and can get stuck in local minima, but for $L\geq 3$ it is overparametrized and training becomes easier. As the plot shows, once the model is overparametrized,  further increasing the number of layers does not lead to any improvement in the minimum loss value achievable. Lastly, in Fig.~\ref{fig:vhd_diag_hamiltonian} we show the success of the diagonalization by applying a successfully trained diagonalizing unitary $W(\vec{\alpha})$ to its target Hamiltonian $H$, obtaining a perfectly diagonal matrix and verifying the success of the algorithm.

\subsection{Many-body localization and system-time entanglement}

\begin{figure*}[t!]
  \centering
  \includegraphics[width=\linewidth]{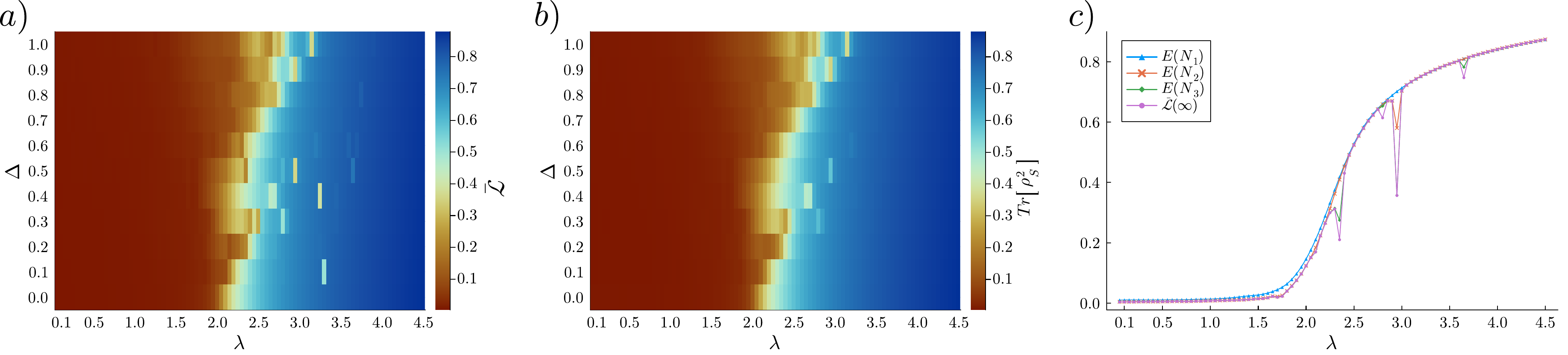}
  \caption{\textbf{System-time entanglement and Loschmidt echo for Hubbard-like interacting model.}
  In all three panels we consider $|\psi_0\rangle=s^+_{n/2}s^+_{n/2+1}|\downarrow\downarrow\dots \downarrow\rangle$ as the initial state of a chain of $n=54$ spins. 
  Panel $a)$: Contour plot of the discrete-time approximation of $\bar{\LC}$ as a function of the ``disorder'' strength $\lambda\in (0.05,4.5)$ with spacing $0.05$ and the interaction coupling $\Delta\in (0,1)$ with spacing $0.1$. Here the number of clock qubits is $15$ ($N=2^{15}$), while the time step is taken as $\epsilon=0.05$.  
  Panel $b)$: Contour plot of the discrete-time approximation of $\Tr[\rho_S^2]$ under the same conditions of panel a).  
  Panel $c)$: System-time entanglement ($E=\Tr[\rho_S^2]$) vs exact Loschmidt echo ($\bar{\LC}(\infty)$) for fixed $\Delta=0.05,\epsilon=0.05$ and time registers of sizes $N_1=2^{12},N_2=2^{15},N_3=2^{18}$. 
  }
  \label{fig:test}
\end{figure*}

Consider now the full $\Delta\neq 0$  model in an open chain. As shown in the Supplemental material, the Hamiltonian is essentially mapped to a Hubbard-like interacting model via the JW transformation, with $\Delta$ the interacting strength.
It has been argued \cite{iyer2013many} that the localization phenomenon in this system could survive the interaction, at least for small $\Delta$, with the transition point shifting to higher values of the critical $\lambda$ as $\Delta$ increases (see the proposed phase diagram in \cite{iyer2013many}). 

In this section we numerically show how our parallel-in-time algorithms can be used to capture this behavior. We considered a chain of $n=54$ spins and an initial state $|\psi_0\rangle=s^+_{n/2}s^+_{n/2+1}|\downarrow\downarrow\dots \downarrow\rangle$, corresponding to two particles localized in the middle of the chain. In Figure \ref{fig:test} a) we show a contour plot of the time average of the Loschmidt's echo for $N=2^{15}$ corresponding to $15$ clock qubits, and with $\epsilon=0.05$. One can qualitatively appreciate the tendency of the localization sector $\lambda$ moving to higher values of $\lambda$ as the interaction strength increases. Similarly, the system-time entanglement, measured by the purity $\Tr[\rho_S^2]$, is depicted in Figure  \ref{fig:test} b), capturing the same behavior. Let us notice that many ``fast oscillations'' are present. While these may be smoothed out by considering averages over an additional offset added to the argument of the cosine function defining $h_j$ and over the initial state \cite{iyer2013many}, a single run of our algorithms would correspond to  have these quantities as fixed, as in our example. At the same time, these oscillations are clearly present, as shown in Figure \ref{fig:test} c), through the exact infinite time average $\bar{\LC}(\psi_0)$. As shown therein, in order to fully capture them via entanglement, a sufficiently large number of click qubits is required, with $18$ clock qubits already capturing most of this subtle behavior. We  remark also how the bound of Theorem  \ref{theo:ent-loschmidt} is strictly satisfied, with the entanglement approaching the ``dips'' always from above as $N$ increases.

Let us now briefly discuss our choice of the initial state. Notice that we are considering ``low-energy'' states. This allows us to consider large temporal averages and modestly large chains (see details in the Supplemental material for a description of our methods). The more usual assumption made in the literature \cite{iyer2013many,vstrkalj2021many} of an initial half-filled state renders the time-averages we considered practically inaccessible for classical computations except for small chains. At the same time, even for this simple state, the system exhibits an interesting behavior as shown in Figure \ref{fig:test}, thus providing us with a non-trivial example of the ability of our algorithms to capture many-body properties, even for a relatively small number of clock-qubits. 
In the Appendix \ref{App:methods}, we discuss how to simulate the interacting model both via exact diagonalization after mapping it to a fermionic model, and via standard Tensor Network techniques. Particularly, since the system at hand is 1-D, in the latter case we resort to Matrix Product State (MPS) algorithms \cite{orus2014practical}, for which we provide scaling of complexity and deviation from the exact results. The analysis of bond dimension confirms the non-trivial behavior of the system for our choice of initial state. The MPS perspective also allows us to discuss the effects of Trotter error (see Appendix  \ref{App:methods} for details).

\section{Discussions}

The simulation of quantum systems has widely been considered the most important application of quantum computation since its conception \cite{feynman1982simulating}.
Traditionally, the focus of quantum simulations has revolved around computing quantum states and physical quantities at a given time, harnessing the exponential growth of the Hilbert space of qubits to mimic the behavior of many-body systems.
However,  many fundamental quantities, such as correlation functions or the equilibrium state of a quantum system,  are associated with large temporal sums of the previous. In this manuscript
we have shown that by treating time itself quantum mechanically, which in a computational scheme corresponds to using clock-qubits, those quantities become readily accessible. 

This result arises from a fruitful analogy between the recent quantum time discussions in quantum foundations and quantum gravity fields, and the fields of quantum information and computation. More importantly, by developing  quantum-time inspired algorithms, we have disclosed new important connections between the  correlations contained in 
history states
and the problem of equilibration of an isolated quantum system, thus unveiling a link to statistical mechanics as well.  
In particular, we have shown that the system-time entanglement is a good measure of equilibration, and as a by product, how the formalism provides a way to prepare approximate equilibrium states. Whether under proper conditions this can provide a useful scheme for studying thermalization as well is left for future investigations. 
These considerations show that, in addition to the practical applications of the various proposed algorithms, the framework we presented offers new insights that can be applied to diverse areas of many-body physics, and potentially to other scenarios where one needs to simulate non-unitary quantum processes (e.g., via linear combination of unitaries).

A key advantage of our framework is that whatever compilation method one would use to transform $e^{-i Ht}$ into a sequence of gates for sequential deployment, can also be readily adapted to parallel deployment. While here we discuss Trotterization-approaches and the VHD algorithm, we note that parallel-in-time can be also combined with any form of state-of-the for compilation and compression. For instance, one could instead use  tensor-network compilation techniques~\cite{gibbs2024deep,zhang2024scalable,gibbs2025learning} based on out-of-distribution generalization~\cite{caro2021generalization,caro2022outofdistribution}, and further reduce the depth of the circuit via parallelization. Such improvement, while quadratic in the circuit depth, should not be readily disregarded as quantum circuit runs are expected to be (1) expensive, and (2) slow in the foreseeable future, meaning that the reduction in the number of circuit runs our techniques offers could lead to significant financial and time savings.

Next, let us note that the system-time entanglement is also indicative of the need for entangling gates to prepare history states. In this work we have considered a fixed architecture which even for a direct Trotterization approach provides a running time advantage over sequential in time schemes. We have also considered a further simplification via a variational Hamiltonian diagonalization approach. Interestingly, the fact that the entanglement is state and evolution dependent suggests that one can  consider more adaptive preparation schemes, where the entanglement is used as a measure of ``compressibility'' (or complexity) of the history state. While we leave the study of these possibilities for future investigations, we should mention that a global  variational scheme has been proposed recently \cite{barison2022variational} based on the Feynman-Kitaev Hamiltonian \cite{mcclean2013feynman}. As with any variational protocol, knowledge of the structure of the solution is valuable for choosing proper ansatzes, thus rendering our efficient protocol particularly relevant for any near-term implementation as well. 
At the same time, all of our protocols can be easily updated by replacing the history state preparation subroutine. The newly disclosed parallel-in-time advantages clearly hold and can be readily extended to any related proposal including \cite{barison2022variational}.

In all the protocols we have considered, the circuits give information about the history of a closed system. However, these protocols can be extended to consider the history of states which at some points in time are being subject to measurements. This becomes feasible by following the recent treatment of Ref.~\cite{giovannetti2015quantum} that incorporates  ancillary memories (following von Neumann) and describe the history of the whole (history of the system+ancillas). This framework opens many new interesting possibilities for novel quantum algorithms. A similar treatment may be applied to quantum evolutions associated to non-unitary channels (open systems).

On the other hand,  even if we focus on closed systems there is  much yet to explore: Most of the parallel protocols in the paper involve measurements on the system side. We also know that adding a projective measurement in the time basis of the clock qubits yields predictions at a given time. However, the most characteristic features of quantum mechanics arise when one is considering measurements in different bases, meaning that the full potential of quantum time effects remains unexplored.  In addition, multiple copies of history states may be used to study higher momenta of mean values, thus opening even more possibilities. Protocols exploiting time-reversed evolution \cite{yoshida2023reversing,chen2024quantum} might also be considered.
Going even further, it has been recently discussed~\cite{diaz2019history,diaz2019historystate,diaz2021spacetime,diaz2021path, giovannetti2023geometric, diaz2023spacetime, diaz2025spacetime} that the PaW mechanism alone is not enough for achieving a fully symmetric version of QM. This has led to nontrivial extensions of the original formalism of Page and Wootters (see also Appendix \ref{sec:literat}). We can speculate that in the near future, motivated by the current results, these new extensions may provide further informational and computational insights related to the time domain.

\section*{Acknowledgements}

The authors would like to thank Luis Pedro Garcia Pintos, Lodovico Scarpa, Lucia Vilchez, Ivan Chernyshev, and Hasan Sayginel for fruitful discussions. N.L.D,  P.B. and M.C. were supported by the Laboratory Directed Research and Development (LDRD) program of Los Alamos National Laboratory (LANL) under project numbers 20210116DR, 20230049DR and 20230527ECR, respectively. P.B. was also  supported by the U.S. Department of Energy (DOE) through a quantum computing program sponsored by the LANL Information Science \& Technology Institute.  N.L.D. was also supported by  the Center for Nonlinear Studies (CNLS) at LANL. M.L. was initially supported by the Center for Nonlinear Studies at LANL and by the U.S. DOE, Office of Science, Office of Advanced Scientific Computing Research, under the Accelerated Research in Quantum Computing (ARQC) program. M.C. was also supported by LANL's ASC Beyond Moore’s Law project. N.L.D., J.M.M acknowledge support from CONICET, and R.R. from CIC Argentina. 
This research used resources provided by the Los Alamos National Laboratory Institutional Computing Program, and by the Darwin testbed at Los Alamos National Laboratory (LANL) which is funded by the Computational Systems and Software Environments subprogram of LANL’s Advanced Simulation and Computing program (NNSA/DOE).

%


\begin{thebibliography}{146}%
\makeatletter
\providecommand \@ifxundefined [1]{%
 \@ifx{#1\undefined}
}%
\providecommand \@ifnum [1]{%
 \ifnum #1\expandafter \@firstoftwo
 \else \expandafter \@secondoftwo
 \fi
}%
\providecommand \@ifx [1]{%
 \ifx #1\expandafter \@firstoftwo
 \else \expandafter \@secondoftwo
 \fi
}%
\providecommand \natexlab [1]{#1}%
\providecommand \enquote  [1]{``#1''}%
\providecommand \bibnamefont  [1]{#1}%
\providecommand \bibfnamefont [1]{#1}%
\providecommand \citenamefont [1]{#1}%
\providecommand \href@noop [0]{\@secondoftwo}%
\providecommand \href [0]{\begingroup \@sanitize@url \@href}%
\providecommand \@href[1]{\@@startlink{#1}\@@href}%
\providecommand \@@href[1]{\endgroup#1\@@endlink}%
\providecommand \@sanitize@url [0]{\catcode `\\12\catcode `\$12\catcode
  `\&12\catcode `\#12\catcode `\^12\catcode `\_12\catcode `\%12\relax}%
\providecommand \@@startlink[1]{}%
\providecommand \@@endlink[0]{}%
\providecommand \url  [0]{\begingroup\@sanitize@url \@url }%
\providecommand \@url [1]{\endgroup\@href {#1}{\urlprefix }}%
\providecommand \urlprefix  [0]{URL }%
\providecommand \Eprint [0]{\href }%
\providecommand \doibase [0]{https://doi.org/}%
\providecommand \selectlanguage [0]{\@gobble}%
\providecommand \bibinfo  [0]{\@secondoftwo}%
\providecommand \bibfield  [0]{\@secondoftwo}%
\providecommand \translation [1]{[#1]}%
\providecommand \BibitemOpen [0]{}%
\providecommand \bibitemStop [0]{}%
\providecommand \bibitemNoStop [0]{.\EOS\space}%
\providecommand \EOS [0]{\spacefactor3000\relax}%
\providecommand \BibitemShut  [1]{\csname bibitem#1\endcsname}%
\let\auto@bib@innerbib\@empty
\bibitem [{\citenamefont {Auletta}\ and\ \citenamefont
  {Parisi}(2001)}]{auletta2001foundations}%
  \BibitemOpen
  \bibfield  {author} {\bibinfo {author} {\bibfnamefont {G.}~\bibnamefont
  {Auletta}}\ and\ \bibinfo {author} {\bibfnamefont {G.}~\bibnamefont
  {Parisi}},\ }\href@noop {} {\emph {\bibinfo {title} {Foundations and
  Interpretation of Quantum Mechanics: In the Light of a Critical-Historical
  Analysis of the Problems and of a Synthesis of the Results}}}\ (\bibinfo
  {publisher} {World Scientific},\ \bibinfo {year} {2001})\BibitemShut
  {NoStop}%
\bibitem [{\citenamefont {Zurek}(2003)}]{zurek2003decoherence}%
  \BibitemOpen
  \bibfield  {author} {\bibinfo {author} {\bibfnamefont {W.~H.}\ \bibnamefont
  {Zurek}},\ }\bibfield  {title} {\bibinfo {title} {Decoherence, einselection,
  and the quantum origins of the classical},\ }\href
  {https://doi.org/10.1103/RevModPhys.75.715} {\bibfield  {journal} {\bibinfo
  {journal} {Reviews of modern physics}\ }\textbf {\bibinfo {volume} {75}},\
  \bibinfo {pages} {715} (\bibinfo {year} {2003})}\BibitemShut {NoStop}%
\bibitem [{\citenamefont {Schlosshauer}(2005)}]{schlosshauer2005decoherence}%
  \BibitemOpen
  \bibfield  {author} {\bibinfo {author} {\bibfnamefont {M.}~\bibnamefont
  {Schlosshauer}},\ }\bibfield  {title} {\bibinfo {title} {Decoherence, the
  measurement problem, and interpretations of quantum mechanics},\ }\href
  {https://doi.org/10.1103/RevModPhys.76.1267} {\bibfield  {journal} {\bibinfo
  {journal} {Reviews of Modern physics}\ }\textbf {\bibinfo {volume} {76}},\
  \bibinfo {pages} {1267} (\bibinfo {year} {2005})}\BibitemShut {NoStop}%
\bibitem [{\citenamefont {Reimann}(2008)}]{reimann2008foundation}%
  \BibitemOpen
  \bibfield  {author} {\bibinfo {author} {\bibfnamefont {P.}~\bibnamefont
  {Reimann}},\ }\bibfield  {title} {\bibinfo {title} {Foundation of statistical
  mechanics under experimentally realistic conditions},\ }\href
  {https://doi.org/10.1103/PhysRevLett.101.190403} {\bibfield  {journal}
  {\bibinfo  {journal} {Physical review letters}\ }\textbf {\bibinfo {volume}
  {101}},\ \bibinfo {pages} {190403} (\bibinfo {year} {2008})}\BibitemShut
  {NoStop}%
\bibitem [{\citenamefont {Wheeler}\ and\ \citenamefont
  {Zurek}(2014)}]{wheeler2014quantum}%
  \BibitemOpen
  \bibfield  {author} {\bibinfo {author} {\bibfnamefont {J.~A.}\ \bibnamefont
  {Wheeler}}\ and\ \bibinfo {author} {\bibfnamefont {W.~H.}\ \bibnamefont
  {Zurek}},\ }\href {https://www.jstor.org/stable/j.ctt7ztxn5} {\emph {\bibinfo
  {title} {Quantum theory and measurement}}},\ Vol.~\bibinfo {volume} {53}\
  (\bibinfo  {publisher} {Princeton University Press},\ \bibinfo {year}
  {2014})\BibitemShut {NoStop}%
\bibitem [{\citenamefont {McClean}\ \emph {et~al.}(2013)\citenamefont
  {McClean}, \citenamefont {Parkhill},\ and\ \citenamefont
  {Aspuru-Guzik}}]{mcclean2013feynman}%
  \BibitemOpen
  \bibfield  {author} {\bibinfo {author} {\bibfnamefont {J.~R.}\ \bibnamefont
  {McClean}}, \bibinfo {author} {\bibfnamefont {J.~A.}\ \bibnamefont
  {Parkhill}},\ and\ \bibinfo {author} {\bibfnamefont {A.}~\bibnamefont
  {Aspuru-Guzik}},\ }\bibfield  {title} {\bibinfo {title} {Feynman’s clock, a
  new variational principle, and parallel-in-time quantum dynamics},\ }\href
  {https://doi.org/10.1073/pnas.1308069110} {\bibfield  {journal} {\bibinfo
  {journal} {Proceedings of the National Academy of Sciences}\ }\textbf
  {\bibinfo {volume} {110}},\ \bibinfo {pages} {E3901} (\bibinfo {year}
  {2013})}\BibitemShut {NoStop}%
\bibitem [{\citenamefont {Giovannetti}\ \emph {et~al.}(2015)\citenamefont
  {Giovannetti}, \citenamefont {Lloyd},\ and\ \citenamefont
  {Maccone}}]{giovannetti2015quantum}%
  \BibitemOpen
  \bibfield  {author} {\bibinfo {author} {\bibfnamefont {V.}~\bibnamefont
  {Giovannetti}}, \bibinfo {author} {\bibfnamefont {S.}~\bibnamefont {Lloyd}},\
  and\ \bibinfo {author} {\bibfnamefont {L.}~\bibnamefont {Maccone}},\
  }\bibfield  {title} {\bibinfo {title} {Quantum time},\ }\href
  {https://doi.org/10.1103/PhysRevD.92.045033} {\bibfield  {journal} {\bibinfo
  {journal} {Physical Review D}\ }\textbf {\bibinfo {volume} {92}},\ \bibinfo
  {pages} {045033} (\bibinfo {year} {2015})}\BibitemShut {NoStop}%
\bibitem [{\citenamefont {Boette}\ \emph {et~al.}(2016)\citenamefont {Boette},
  \citenamefont {Rossignoli}, \citenamefont {Gigena},\ and\ \citenamefont
  {Cerezo}}]{boette2016system}%
  \BibitemOpen
  \bibfield  {author} {\bibinfo {author} {\bibfnamefont {A.}~\bibnamefont
  {Boette}}, \bibinfo {author} {\bibfnamefont {R.}~\bibnamefont {Rossignoli}},
  \bibinfo {author} {\bibfnamefont {N.}~\bibnamefont {Gigena}},\ and\ \bibinfo
  {author} {\bibfnamefont {M.}~\bibnamefont {Cerezo}},\ }\bibfield  {title}
  {\bibinfo {title} {System-time entanglement in a discrete-time model},\
  }\href {https://doi.org/10.1103/PhysRevA.93.062127} {\bibfield  {journal}
  {\bibinfo  {journal} {Physical Review A}\ }\textbf {\bibinfo {volume} {93}},\
  \bibinfo {pages} {062127} (\bibinfo {year} {2016})}\BibitemShut {NoStop}%
\bibitem [{\citenamefont {Boette}\ and\ \citenamefont
  {Rossignoli}(2018)}]{boette2018history}%
  \BibitemOpen
  \bibfield  {author} {\bibinfo {author} {\bibfnamefont {A.}~\bibnamefont
  {Boette}}\ and\ \bibinfo {author} {\bibfnamefont {R.}~\bibnamefont
  {Rossignoli}},\ }\bibfield  {title} {\bibinfo {title} {History states of
  systems and operators},\ }\href {https://doi.org/10.1103/PhysRevA.98.032108}
  {\bibfield  {journal} {\bibinfo  {journal} {Physical Review A}\ }\textbf
  {\bibinfo {volume} {98}},\ \bibinfo {pages} {032108} (\bibinfo {year}
  {2018})}\BibitemShut {NoStop}%
\bibitem [{\citenamefont {Horsman}\ \emph {et~al.}(2017)\citenamefont
  {Horsman}, \citenamefont {Heunen}, \citenamefont {Pusey}, \citenamefont
  {Barrett},\ and\ \citenamefont {Spekkens}}]{horsman2017can}%
  \BibitemOpen
  \bibfield  {author} {\bibinfo {author} {\bibfnamefont {D.}~\bibnamefont
  {Horsman}}, \bibinfo {author} {\bibfnamefont {C.}~\bibnamefont {Heunen}},
  \bibinfo {author} {\bibfnamefont {M.~F.}\ \bibnamefont {Pusey}}, \bibinfo
  {author} {\bibfnamefont {J.}~\bibnamefont {Barrett}},\ and\ \bibinfo {author}
  {\bibfnamefont {R.~W.}\ \bibnamefont {Spekkens}},\ }\bibfield  {title}
  {\bibinfo {title} {Can a quantum state over time resemble a quantum state at
  a single time?},\ }\href {https://doi.org/10.1098/rspa.2017.0395} {\bibfield
  {journal} {\bibinfo  {journal} {Proceedings of the Royal Society A:
  Mathematical, Physical and Engineering Sciences}\ }\textbf {\bibinfo {volume}
  {473}},\ \bibinfo {pages} {20170395} (\bibinfo {year} {2017})}\BibitemShut
  {NoStop}%
\bibitem [{\citenamefont {Diaz}\ \emph {et~al.}(2019)\citenamefont {Diaz},
  \citenamefont {Matera},\ and\ \citenamefont {Rossignoli}}]{diaz2019history}%
  \BibitemOpen
  \bibfield  {author} {\bibinfo {author} {\bibfnamefont {N.~L.}\ \bibnamefont
  {Diaz}}, \bibinfo {author} {\bibfnamefont {J.~M.}\ \bibnamefont {Matera}},\
  and\ \bibinfo {author} {\bibfnamefont {R.}~\bibnamefont {Rossignoli}},\
  }\bibfield  {title} {\bibinfo {title} {History state formalism for scalar
  particles},\ }\href {https://doi.org/10.1103/PhysRevD.100.125020} {\bibfield
  {journal} {\bibinfo  {journal} {Physical Review D}\ }\textbf {\bibinfo
  {volume} {100}},\ \bibinfo {pages} {125020} (\bibinfo {year}
  {2019})}\BibitemShut {NoStop}%
\bibitem [{\citenamefont {Diaz}\ and\ \citenamefont
  {Rossignoli}(2019)}]{diaz2019historystate}%
  \BibitemOpen
  \bibfield  {author} {\bibinfo {author} {\bibfnamefont {N.~L.}\ \bibnamefont
  {Diaz}}\ and\ \bibinfo {author} {\bibfnamefont {R.}~\bibnamefont
  {Rossignoli}},\ }\bibfield  {title} {\bibinfo {title} {History state
  formalism for dirac’s theory},\ }\href
  {https://doi.org/10.1103/PhysRevD.99.045008} {\bibfield  {journal} {\bibinfo
  {journal} {Physical Review D}\ }\textbf {\bibinfo {volume} {99}},\ \bibinfo
  {pages} {045008} (\bibinfo {year} {2019})}\BibitemShut {NoStop}%
\bibitem [{\citenamefont {Diaz}\ \emph {et~al.}(2021)\citenamefont {Diaz},
  \citenamefont {Matera},\ and\ \citenamefont
  {Rossignoli}}]{diaz2021spacetime}%
  \BibitemOpen
  \bibfield  {author} {\bibinfo {author} {\bibfnamefont {N.~L.}\ \bibnamefont
  {Diaz}}, \bibinfo {author} {\bibfnamefont {J.~M.}\ \bibnamefont {Matera}},\
  and\ \bibinfo {author} {\bibfnamefont {R.}~\bibnamefont {Rossignoli}},\
  }\bibfield  {title} {\bibinfo {title} {Spacetime quantum actions},\ }\href
  {https://doi.org/10.1103/PhysRevD.103.065011} {\bibfield  {journal} {\bibinfo
   {journal} {Physical Review D}\ }\textbf {\bibinfo {volume} {103}},\ \bibinfo
  {pages} {065011} (\bibinfo {year} {2021})}\BibitemShut {NoStop}%
\bibitem [{\citenamefont {Diaz}\ \emph {et~al.}(2025)\citenamefont {Diaz},
  \citenamefont {Matera},\ and\ \citenamefont {Rossignoli}}]{diaz2021path}%
  \BibitemOpen
  \bibfield  {author} {\bibinfo {author} {\bibfnamefont {N.~L.}\ \bibnamefont
  {Diaz}}, \bibinfo {author} {\bibfnamefont {J.~M.}\ \bibnamefont {Matera}},\
  and\ \bibinfo {author} {\bibfnamefont {R.}~\bibnamefont {Rossignoli}},\
  }\bibfield  {title} {\bibinfo {title} {Path integrals from spacetime quantum
  actions},\ }\href {https://doi.org/10.1016/j.aop.2025.170052} {\bibfield
  {journal} {\bibinfo  {journal} {Annals of Physics}\ ,\ \bibinfo {pages}
  {170052}} (\bibinfo {year} {2025})}\BibitemShut {NoStop}%
\bibitem [{\citenamefont {Pab{\'o}n}\ \emph {et~al.}(2019)\citenamefont
  {Pab{\'o}n}, \citenamefont {Reb{\'o}n}, \citenamefont {Bordakevich},
  \citenamefont {Gigena}, \citenamefont {Boette}, \citenamefont {Iemmi},
  \citenamefont {Rossignoli},\ and\ \citenamefont
  {Ledesma}}]{pabon2019parallel}%
  \BibitemOpen
  \bibfield  {author} {\bibinfo {author} {\bibfnamefont {D.}~\bibnamefont
  {Pab{\'o}n}}, \bibinfo {author} {\bibfnamefont {L.}~\bibnamefont
  {Reb{\'o}n}}, \bibinfo {author} {\bibfnamefont {S.}~\bibnamefont
  {Bordakevich}}, \bibinfo {author} {\bibfnamefont {N.}~\bibnamefont {Gigena}},
  \bibinfo {author} {\bibfnamefont {A.}~\bibnamefont {Boette}}, \bibinfo
  {author} {\bibfnamefont {C.}~\bibnamefont {Iemmi}}, \bibinfo {author}
  {\bibfnamefont {R.}~\bibnamefont {Rossignoli}},\ and\ \bibinfo {author}
  {\bibfnamefont {S.}~\bibnamefont {Ledesma}},\ }\bibfield  {title} {\bibinfo
  {title} {Parallel-in-time optical simulation of history states},\ }\href
  {https://doi.org/10.1103/PhysRevA.99.062333} {\bibfield  {journal} {\bibinfo
  {journal} {Physical Review A}\ }\textbf {\bibinfo {volume} {99}},\ \bibinfo
  {pages} {062333} (\bibinfo {year} {2019})}\BibitemShut {NoStop}%
\bibitem [{\citenamefont {Mendes}\ and\ \citenamefont
  {Soares-Pinto}(2019)}]{mendes2019time}%
  \BibitemOpen
  \bibfield  {author} {\bibinfo {author} {\bibfnamefont {L.~R.}\ \bibnamefont
  {Mendes}}\ and\ \bibinfo {author} {\bibfnamefont {D.~O.}\ \bibnamefont
  {Soares-Pinto}},\ }\bibfield  {title} {\bibinfo {title} {Time as a
  consequence of internal coherence},\ }\href
  {https://doi.org/10.1098/rspa.2019.0470} {\bibfield  {journal} {\bibinfo
  {journal} {Proceedings of the Royal Society A}\ }\textbf {\bibinfo {volume}
  {475}},\ \bibinfo {pages} {20190470} (\bibinfo {year} {2019})}\BibitemShut
  {NoStop}%
\bibitem [{\citenamefont {Favalli}\ and\ \citenamefont
  {Smerzi}(2020)}]{favalli2020time}%
  \BibitemOpen
  \bibfield  {author} {\bibinfo {author} {\bibfnamefont {T.}~\bibnamefont
  {Favalli}}\ and\ \bibinfo {author} {\bibfnamefont {A.}~\bibnamefont
  {Smerzi}},\ }\bibfield  {title} {\bibinfo {title} {Time observables in a
  timeless universe},\ }\href {https://doi.org/10.22331/q-2020-10-29-354}
  {\bibfield  {journal} {\bibinfo  {journal} {Quantum}\ }\textbf {\bibinfo
  {volume} {4}},\ \bibinfo {pages} {354} (\bibinfo {year} {2020})}\BibitemShut
  {NoStop}%
\bibitem [{\citenamefont {Vald{\'e}s-Hern{\'a}ndez}\ \emph
  {et~al.}(2020)\citenamefont {Vald{\'e}s-Hern{\'a}ndez}, \citenamefont
  {Maglione}, \citenamefont {Majtey},\ and\ \citenamefont
  {Plastino}}]{valdes2020emergent}%
  \BibitemOpen
  \bibfield  {author} {\bibinfo {author} {\bibfnamefont {A.}~\bibnamefont
  {Vald{\'e}s-Hern{\'a}ndez}}, \bibinfo {author} {\bibfnamefont {C.~G.}\
  \bibnamefont {Maglione}}, \bibinfo {author} {\bibfnamefont {A.~P.}\
  \bibnamefont {Majtey}},\ and\ \bibinfo {author} {\bibfnamefont
  {A.}~\bibnamefont {Plastino}},\ }\bibfield  {title} {\bibinfo {title}
  {Emergent dynamics from entangled mixed states},\ }\href
  {https://doi.org/10.1103/PhysRevA.102.052417} {\bibfield  {journal} {\bibinfo
   {journal} {Physical Review A}\ }\textbf {\bibinfo {volume} {102}},\ \bibinfo
  {pages} {052417} (\bibinfo {year} {2020})}\BibitemShut {NoStop}%
\bibitem [{\citenamefont {Castro-Ruiz}\ \emph {et~al.}(2020)\citenamefont
  {Castro-Ruiz}, \citenamefont {Giacomini}, \citenamefont {Belenchia},\ and\
  \citenamefont {Brukner}}]{castro2020quantum}%
  \BibitemOpen
  \bibfield  {author} {\bibinfo {author} {\bibfnamefont {E.}~\bibnamefont
  {Castro-Ruiz}}, \bibinfo {author} {\bibfnamefont {F.}~\bibnamefont
  {Giacomini}}, \bibinfo {author} {\bibfnamefont {A.}~\bibnamefont
  {Belenchia}},\ and\ \bibinfo {author} {\bibfnamefont {{\v{C}}.}~\bibnamefont
  {Brukner}},\ }\bibfield  {title} {\bibinfo {title} {Quantum clocks and the
  temporal localisability of events in the presence of gravitating quantum
  systems},\ }\href {https://doi.org/10.1038/s41467-020-16013-1} {\bibfield
  {journal} {\bibinfo  {journal} {Nature Communications}\ }\textbf {\bibinfo
  {volume} {11}},\ \bibinfo {pages} {2672} (\bibinfo {year}
  {2020})}\BibitemShut {NoStop}%
\bibitem [{\citenamefont {Mendes}\ \emph {et~al.}(2021)\citenamefont {Mendes},
  \citenamefont {Brito},\ and\ \citenamefont {Soares-Pinto}}]{mendes2021non}%
  \BibitemOpen
  \bibfield  {author} {\bibinfo {author} {\bibfnamefont {L.~R.}\ \bibnamefont
  {Mendes}}, \bibinfo {author} {\bibfnamefont {F.}~\bibnamefont {Brito}},\ and\
  \bibinfo {author} {\bibfnamefont {D.~O.}\ \bibnamefont {Soares-Pinto}},\
  }\bibfield  {title} {\bibinfo {title} {Non-linear equation of motion for
  page-wootters mechanism with interaction and quasi-ideal clocks},\ }\href
  {https://arxiv.org/abs/2107.11452} {\bibfield  {journal} {\bibinfo  {journal}
  {arXiv preprint arXiv:2107.11452}\ } (\bibinfo {year} {2021})}\BibitemShut
  {NoStop}%
\bibitem [{\citenamefont {Foti}\ \emph {et~al.}(2021)\citenamefont {Foti},
  \citenamefont {Coppo}, \citenamefont {Barni}, \citenamefont {Cuccoli},\ and\
  \citenamefont {Verrucchi}}]{foti2021time}%
  \BibitemOpen
  \bibfield  {author} {\bibinfo {author} {\bibfnamefont {C.}~\bibnamefont
  {Foti}}, \bibinfo {author} {\bibfnamefont {A.}~\bibnamefont {Coppo}},
  \bibinfo {author} {\bibfnamefont {G.}~\bibnamefont {Barni}}, \bibinfo
  {author} {\bibfnamefont {A.}~\bibnamefont {Cuccoli}},\ and\ \bibinfo {author}
  {\bibfnamefont {P.}~\bibnamefont {Verrucchi}},\ }\bibfield  {title} {\bibinfo
  {title} {Time and classical equations of motion from quantum entanglement via
  the page and wootters mechanism with generalized coherent states},\ }\href
  {https://doi.org/10.1038/s41467-021-21782-4} {\bibfield  {journal} {\bibinfo
  {journal} {Nature Communications}\ }\textbf {\bibinfo {volume} {12}},\
  \bibinfo {pages} {1787} (\bibinfo {year} {2021})}\BibitemShut {NoStop}%
\bibitem [{\citenamefont {H{\"o}hn}\ \emph
  {et~al.}(2021{\natexlab{a}})\citenamefont {H{\"o}hn}, \citenamefont {Smith},\
  and\ \citenamefont {Lock}}]{hohn2021trinity}%
  \BibitemOpen
  \bibfield  {author} {\bibinfo {author} {\bibfnamefont {P.~A.}\ \bibnamefont
  {H{\"o}hn}}, \bibinfo {author} {\bibfnamefont {A.~R.}\ \bibnamefont
  {Smith}},\ and\ \bibinfo {author} {\bibfnamefont {M.~P.}\ \bibnamefont
  {Lock}},\ }\bibfield  {title} {\bibinfo {title} {Trinity of relational
  quantum dynamics},\ }\href
  {https://journals.aps.org/prd/abstract/10.1103/PhysRevD.104.066001}
  {\bibfield  {journal} {\bibinfo  {journal} {Physical Review D}\ }\textbf
  {\bibinfo {volume} {104}},\ \bibinfo {pages} {066001} (\bibinfo {year}
  {2021}{\natexlab{a}})}\BibitemShut {NoStop}%
\bibitem [{\citenamefont {H{\"o}hn}\ \emph
  {et~al.}(2021{\natexlab{b}})\citenamefont {H{\"o}hn}, \citenamefont {Smith},\
  and\ \citenamefont {Lock}}]{hohn2021equivalence}%
  \BibitemOpen
  \bibfield  {author} {\bibinfo {author} {\bibfnamefont {P.~A.}\ \bibnamefont
  {H{\"o}hn}}, \bibinfo {author} {\bibfnamefont {A.~R.}\ \bibnamefont
  {Smith}},\ and\ \bibinfo {author} {\bibfnamefont {M.~P.}\ \bibnamefont
  {Lock}},\ }\bibfield  {title} {\bibinfo {title} {Equivalence of approaches to
  relational quantum dynamics in relativistic settings},\ }\href
  {https://www.frontiersin.org/articles/10.3389/fphy.2021.587083/full}
  {\bibfield  {journal} {\bibinfo  {journal} {Frontiers in Physics}\ }\textbf
  {\bibinfo {volume} {9}},\ \bibinfo {pages} {587083} (\bibinfo {year}
  {2021}{\natexlab{b}})}\BibitemShut {NoStop}%
\bibitem [{\citenamefont {Lomoc}\ \emph {et~al.}(2022)\citenamefont {Lomoc},
  \citenamefont {Boette}, \citenamefont {Canosa},\ and\ \citenamefont
  {Rossignoli}}]{lomoc2022history}%
  \BibitemOpen
  \bibfield  {author} {\bibinfo {author} {\bibfnamefont {F.}~\bibnamefont
  {Lomoc}}, \bibinfo {author} {\bibfnamefont {A.}~\bibnamefont {Boette}},
  \bibinfo {author} {\bibfnamefont {N.}~\bibnamefont {Canosa}},\ and\ \bibinfo
  {author} {\bibfnamefont {R.}~\bibnamefont {Rossignoli}},\ }\bibfield  {title}
  {\bibinfo {title} {History states of one-dimensional quantum walks},\ }\href
  {https://doi.org/10.1103/PhysRevA.106.062215} {\bibfield  {journal} {\bibinfo
   {journal} {Physical Review A}\ }\textbf {\bibinfo {volume} {106}},\ \bibinfo
  {pages} {062215} (\bibinfo {year} {2022})}\BibitemShut {NoStop}%
\bibitem [{\citenamefont {Loc}(2022)}]{loc2022time}%
  \BibitemOpen
  \bibfield  {author} {\bibinfo {author} {\bibfnamefont {N.~P.~D.}\
  \bibnamefont {Loc}},\ }\bibfield  {title} {\bibinfo {title} {Time-system
  entanglement and special relativity},\ }\href
  {https://arxiv.org/abs/2212.13348} {\bibfield  {journal} {\bibinfo  {journal}
  {arXiv preprint arXiv:2212.13348}\ } (\bibinfo {year} {2022})}\BibitemShut
  {NoStop}%
\bibitem [{\citenamefont {Paiva}\ \emph
  {et~al.}(2022{\natexlab{a}})\citenamefont {Paiva}, \citenamefont {Te’eni},
  \citenamefont {Peled}, \citenamefont {Cohen},\ and\ \citenamefont
  {Aharonov}}]{paiva2022non}%
  \BibitemOpen
  \bibfield  {author} {\bibinfo {author} {\bibfnamefont {I.~L.}\ \bibnamefont
  {Paiva}}, \bibinfo {author} {\bibfnamefont {A.}~\bibnamefont {Te’eni}},
  \bibinfo {author} {\bibfnamefont {B.~Y.}\ \bibnamefont {Peled}}, \bibinfo
  {author} {\bibfnamefont {E.}~\bibnamefont {Cohen}},\ and\ \bibinfo {author}
  {\bibfnamefont {Y.}~\bibnamefont {Aharonov}},\ }\bibfield  {title} {\bibinfo
  {title} {Non-inertial quantum clock frames lead to non-hermitian dynamics},\
  }\href {https://doi.org/10.1038/s42005-022-01081-0} {\bibfield  {journal}
  {\bibinfo  {journal} {Communications Physics}\ }\textbf {\bibinfo {volume}
  {5}},\ \bibinfo {pages} {298} (\bibinfo {year}
  {2022}{\natexlab{a}})}\BibitemShut {NoStop}%
\bibitem [{\citenamefont {Paiva}\ \emph
  {et~al.}(2022{\natexlab{b}})\citenamefont {Paiva}, \citenamefont {Lobo},\
  and\ \citenamefont {Cohen}}]{paiva2022flow}%
  \BibitemOpen
  \bibfield  {author} {\bibinfo {author} {\bibfnamefont {I.~L.}\ \bibnamefont
  {Paiva}}, \bibinfo {author} {\bibfnamefont {A.~C.}\ \bibnamefont {Lobo}},\
  and\ \bibinfo {author} {\bibfnamefont {E.}~\bibnamefont {Cohen}},\ }\bibfield
   {title} {\bibinfo {title} {Flow of time during energy measurements and the
  resulting time-energy uncertainty relations},\ }\href
  {https://doi.org/10.22331/q-2022-04-07-683} {\bibfield  {journal} {\bibinfo
  {journal} {Quantum}\ }\textbf {\bibinfo {volume} {6}},\ \bibinfo {pages}
  {683} (\bibinfo {year} {2022}{\natexlab{b}})}\BibitemShut {NoStop}%
\bibitem [{\citenamefont {Favalli}\ and\ \citenamefont
  {Smerzi}(2022)}]{favalli2022peaceful}%
  \BibitemOpen
  \bibfield  {author} {\bibinfo {author} {\bibfnamefont {T.}~\bibnamefont
  {Favalli}}\ and\ \bibinfo {author} {\bibfnamefont {A.}~\bibnamefont
  {Smerzi}},\ }\bibfield  {title} {\bibinfo {title} {Peaceful coexistence of
  thermal equilibrium and the emergence of time},\ }\href
  {https://doi.org/10.1103/PhysRevD.105.023525} {\bibfield  {journal} {\bibinfo
   {journal} {Physical Review D}\ }\textbf {\bibinfo {volume} {105}},\ \bibinfo
  {pages} {023525} (\bibinfo {year} {2022})}\BibitemShut {NoStop}%
\bibitem [{\citenamefont {Rijavec}(2022)}]{rijavec2022heisenberg}%
  \BibitemOpen
  \bibfield  {author} {\bibinfo {author} {\bibfnamefont {S.}~\bibnamefont
  {Rijavec}},\ }\bibfield  {title} {\bibinfo {title} {Heisenberg-picture
  evolution without evolution},\ }\href {arxiv.org/abs/2204.11740} {\bibfield
  {journal} {\bibinfo  {journal} {arXiv preprint arXiv:2204.11740}\ } (\bibinfo
  {year} {2022})}\BibitemShut {NoStop}%
\bibitem [{\citenamefont {Baumann}\ \emph {et~al.}(2022)\citenamefont
  {Baumann}, \citenamefont {Krumm}, \citenamefont {Gu{\'e}rin},\ and\
  \citenamefont {Brukner}}]{baumann2022noncausal}%
  \BibitemOpen
  \bibfield  {author} {\bibinfo {author} {\bibfnamefont {V.}~\bibnamefont
  {Baumann}}, \bibinfo {author} {\bibfnamefont {M.}~\bibnamefont {Krumm}},
  \bibinfo {author} {\bibfnamefont {P.~A.}\ \bibnamefont {Gu{\'e}rin}},\ and\
  \bibinfo {author} {\bibfnamefont {{\v{C}}.}~\bibnamefont {Brukner}},\
  }\bibfield  {title} {\bibinfo {title} {Noncausal page-wootters circuits},\
  }\href {https://doi.org/10.1103/PhysRevResearch.4.013180} {\bibfield
  {journal} {\bibinfo  {journal} {Physical Review Research}\ }\textbf {\bibinfo
  {volume} {4}},\ \bibinfo {pages} {013180} (\bibinfo {year}
  {2022})}\BibitemShut {NoStop}%
\bibitem [{\citenamefont {Apadula}\ \emph {et~al.}(2022)\citenamefont
  {Apadula}, \citenamefont {Castro-Ruiz},\ and\ \citenamefont
  {Brukner}}]{apadula2022quantum}%
  \BibitemOpen
  \bibfield  {author} {\bibinfo {author} {\bibfnamefont {L.}~\bibnamefont
  {Apadula}}, \bibinfo {author} {\bibfnamefont {E.}~\bibnamefont
  {Castro-Ruiz}},\ and\ \bibinfo {author} {\bibfnamefont
  {{\v{C}}.}~\bibnamefont {Brukner}},\ }\bibfield  {title} {\bibinfo {title}
  {Quantum reference frames for lorentz symmetry},\ }\href
  {https://arxiv.org/abs/2212.14081} {\bibfield  {journal} {\bibinfo  {journal}
  {arXiv preprint arXiv:2212.14081}\ } (\bibinfo {year} {2022})}\BibitemShut
  {NoStop}%
\bibitem [{\citenamefont {Barison}\ \emph {et~al.}(2022)\citenamefont
  {Barison}, \citenamefont {Vicentini}, \citenamefont {Cirac},\ and\
  \citenamefont {Carleo}}]{barison2022variational}%
  \BibitemOpen
  \bibfield  {author} {\bibinfo {author} {\bibfnamefont {S.}~\bibnamefont
  {Barison}}, \bibinfo {author} {\bibfnamefont {F.}~\bibnamefont {Vicentini}},
  \bibinfo {author} {\bibfnamefont {I.}~\bibnamefont {Cirac}},\ and\ \bibinfo
  {author} {\bibfnamefont {G.}~\bibnamefont {Carleo}},\ }\bibfield  {title}
  {\bibinfo {title} {Variational dynamics as a ground-state problem on a
  quantum computer},\ }\href {https://doi.org/10.1103/PhysRevResearch.4.043161}
  {\bibfield  {journal} {\bibinfo  {journal} {Physical Review Research}\
  }\textbf {\bibinfo {volume} {4}},\ \bibinfo {pages} {043161} (\bibinfo {year}
  {2022})}\BibitemShut {NoStop}%
\bibitem [{\citenamefont {Giovannetti}\ \emph {et~al.}(2023)\citenamefont
  {Giovannetti}, \citenamefont {Lloyd},\ and\ \citenamefont
  {Maccone}}]{giovannetti2023geometric}%
  \BibitemOpen
  \bibfield  {author} {\bibinfo {author} {\bibfnamefont {V.}~\bibnamefont
  {Giovannetti}}, \bibinfo {author} {\bibfnamefont {S.}~\bibnamefont {Lloyd}},\
  and\ \bibinfo {author} {\bibfnamefont {L.}~\bibnamefont {Maccone}},\
  }\bibfield  {title} {\bibinfo {title} {Geometric event-based quantum
  mechanics},\ }\href {https://doi.org/10.1088/1367-2630/acb793} {\bibfield
  {journal} {\bibinfo  {journal} {New Journal of Physics}\ }\textbf {\bibinfo
  {volume} {25}},\ \bibinfo {pages} {023027} (\bibinfo {year}
  {2023})}\BibitemShut {NoStop}%
\bibitem [{\citenamefont {Chester}\ \emph {et~al.}(2023)\citenamefont
  {Chester}, \citenamefont {Arsiwalla}, \citenamefont {Kauffman}, \citenamefont
  {Planat},\ and\ \citenamefont {Irwin}}]{chester2023covariant}%
  \BibitemOpen
  \bibfield  {author} {\bibinfo {author} {\bibfnamefont {D.}~\bibnamefont
  {Chester}}, \bibinfo {author} {\bibfnamefont {X.~D.}\ \bibnamefont
  {Arsiwalla}}, \bibinfo {author} {\bibfnamefont {L.}~\bibnamefont {Kauffman}},
  \bibinfo {author} {\bibfnamefont {M.}~\bibnamefont {Planat}},\ and\ \bibinfo
  {author} {\bibfnamefont {K.}~\bibnamefont {Irwin}},\ }\bibfield  {title}
  {\bibinfo {title} {Covariant, canonical and symplectic quantization of
  relativistic field theories},\ }\href {https://arxiv.org/abs/2305.08864}
  {\bibfield  {journal} {\bibinfo  {journal} {arXiv preprint arXiv:2305.08864}\
  } (\bibinfo {year} {2023})}\BibitemShut {NoStop}%
\bibitem [{\citenamefont {Sakurai}\ and\ \citenamefont
  {Commins}(1995)}]{sakurai1995modern}%
  \BibitemOpen
  \bibfield  {author} {\bibinfo {author} {\bibfnamefont {J.~J.}\ \bibnamefont
  {Sakurai}}\ and\ \bibinfo {author} {\bibfnamefont {E.~D.}\ \bibnamefont
  {Commins}},\ }\href@noop {} {\bibinfo {title} {Modern quantum mechanics,
  revised edition}} (\bibinfo {year} {1995})\BibitemShut {NoStop}%
\bibitem [{\citenamefont {Isham}(1993)}]{isham1993canonical}%
  \BibitemOpen
  \bibfield  {author} {\bibinfo {author} {\bibfnamefont {C.~J.}\ \bibnamefont
  {Isham}},\ }\bibfield  {title} {\bibinfo {title} {Canonical quantum gravity
  and the problem of time},\ }\href
  {https://doi.org/10.1007/978-94-011-1980-1_6} {\bibfield  {journal} {\bibinfo
   {journal} {Integrable systems, quantum groups, and quantum field theories}\
  ,\ \bibinfo {pages} {157}} (\bibinfo {year} {1993})}\BibitemShut {NoStop}%
\bibitem [{\citenamefont {Gambini}\ \emph {et~al.}(2009)\citenamefont
  {Gambini}, \citenamefont {Porto}, \citenamefont {Pullin},\ and\ \citenamefont
  {Torterolo}}]{gambini2009conditional}%
  \BibitemOpen
  \bibfield  {author} {\bibinfo {author} {\bibfnamefont {R.}~\bibnamefont
  {Gambini}}, \bibinfo {author} {\bibfnamefont {R.~A.}\ \bibnamefont {Porto}},
  \bibinfo {author} {\bibfnamefont {J.}~\bibnamefont {Pullin}},\ and\ \bibinfo
  {author} {\bibfnamefont {S.}~\bibnamefont {Torterolo}},\ }\bibfield  {title}
  {\bibinfo {title} {Conditional probabilities with dirac observables and the
  problem of time in quantum gravity},\ }\href
  {https://doi.org/10.1103/PhysRevD.79.04150} {\bibfield  {journal} {\bibinfo
  {journal} {Physical Review D}\ }\textbf {\bibinfo {volume} {79}},\ \bibinfo
  {pages} {041501} (\bibinfo {year} {2009})}\BibitemShut {NoStop}%
\bibitem [{\citenamefont {Kucha{\v{r}}}(2011)}]{kuchavr2011time}%
  \BibitemOpen
  \bibfield  {author} {\bibinfo {author} {\bibfnamefont {K.~V.}\ \bibnamefont
  {Kucha{\v{r}}}},\ }\bibfield  {title} {\bibinfo {title} {Time and
  interpretations of quantum gravity},\ }\href
  {https://doi.org/10.1142/S0218271811019347} {\bibfield  {journal} {\bibinfo
  {journal} {International Journal of Modern Physics D}\ }\textbf {\bibinfo
  {volume} {20}},\ \bibinfo {pages} {3} (\bibinfo {year} {2011})}\BibitemShut
  {NoStop}%
\bibitem [{\citenamefont {Pauli}(1933)}]{pauli1933allgemeinen}%
  \BibitemOpen
  \bibfield  {author} {\bibinfo {author} {\bibfnamefont {W.}~\bibnamefont
  {Pauli}},\ }\href@noop {} {\emph {\bibinfo {title} {Die allgemeinen
  prinzipien der wellenmechanik}}}\ (\bibinfo  {publisher} {Springer},\
  \bibinfo {year} {1933})\BibitemShut {NoStop}%
\bibitem [{\citenamefont {Page}\ and\ \citenamefont
  {Wootters}(1983)}]{page1983evolution}%
  \BibitemOpen
  \bibfield  {author} {\bibinfo {author} {\bibfnamefont {D.~N.}\ \bibnamefont
  {Page}}\ and\ \bibinfo {author} {\bibfnamefont {W.~K.}\ \bibnamefont
  {Wootters}},\ }\bibfield  {title} {\bibinfo {title} {Evolution without
  evolution: Dynamics described by stationary observables},\ }\href
  {https://doi.org/10.1103/PhysRevD.27.2885} {\bibfield  {journal} {\bibinfo
  {journal} {Physical Review D}\ }\textbf {\bibinfo {volume} {27}},\ \bibinfo
  {pages} {2885} (\bibinfo {year} {1983})}\BibitemShut {NoStop}%
\bibitem [{\citenamefont {Connes}\ and\ \citenamefont
  {Rovelli}(1994)}]{connes1994neumann}%
  \BibitemOpen
  \bibfield  {author} {\bibinfo {author} {\bibfnamefont {A.}~\bibnamefont
  {Connes}}\ and\ \bibinfo {author} {\bibfnamefont {C.}~\bibnamefont
  {Rovelli}},\ }\bibfield  {title} {\bibinfo {title} {Von neumann algebra
  automorphisms and time-thermodynamics relation in generally covariant quantum
  theories},\ }\href {https://doi.org/10.1088/0264-9381/11/12/007} {\bibfield
  {journal} {\bibinfo  {journal} {Classical and Quantum Gravity}\ }\textbf
  {\bibinfo {volume} {11}},\ \bibinfo {pages} {2899} (\bibinfo {year}
  {1994})}\BibitemShut {NoStop}%
\bibitem [{\citenamefont {Isham}(1994)}]{isham1994quantum}%
  \BibitemOpen
  \bibfield  {author} {\bibinfo {author} {\bibfnamefont {C.~J.}\ \bibnamefont
  {Isham}},\ }\bibfield  {title} {\bibinfo {title} {Quantum logic and the
  histories approach to quantum theory},\ }\href
  {https://doi.org/10.1063/1.530544} {\bibfield  {journal} {\bibinfo  {journal}
  {Journal of Mathematical Physics}\ }\textbf {\bibinfo {volume} {35}},\
  \bibinfo {pages} {2157} (\bibinfo {year} {1994})}\BibitemShut {NoStop}%
\bibitem [{\citenamefont {Fitzsimons}\ \emph {et~al.}(2015)\citenamefont
  {Fitzsimons}, \citenamefont {Jones},\ and\ \citenamefont
  {Vedral}}]{fitzsimons2015quantum}%
  \BibitemOpen
  \bibfield  {author} {\bibinfo {author} {\bibfnamefont {J.~F.}\ \bibnamefont
  {Fitzsimons}}, \bibinfo {author} {\bibfnamefont {J.~A.}\ \bibnamefont
  {Jones}},\ and\ \bibinfo {author} {\bibfnamefont {V.}~\bibnamefont
  {Vedral}},\ }\bibfield  {title} {\bibinfo {title} {Quantum correlations which
  imply causation},\ }\href {https://doi.org/10.1038/srep18281} {\bibfield
  {journal} {\bibinfo  {journal} {Scientific reports}\ }\textbf {\bibinfo
  {volume} {5}},\ \bibinfo {pages} {18281} (\bibinfo {year}
  {2015})}\BibitemShut {NoStop}%
\bibitem [{\citenamefont {Cotler}\ \emph {et~al.}(2018)\citenamefont {Cotler},
  \citenamefont {Jian}, \citenamefont {Qi},\ and\ \citenamefont
  {Wilczek}}]{cotler2018superdensity}%
  \BibitemOpen
  \bibfield  {author} {\bibinfo {author} {\bibfnamefont {J.}~\bibnamefont
  {Cotler}}, \bibinfo {author} {\bibfnamefont {C.-M.}\ \bibnamefont {Jian}},
  \bibinfo {author} {\bibfnamefont {X.-L.}\ \bibnamefont {Qi}},\ and\ \bibinfo
  {author} {\bibfnamefont {F.}~\bibnamefont {Wilczek}},\ }\bibfield  {title}
  {\bibinfo {title} {Superdensity operators for spacetime quantum mechanics},\
  }\href {https://doi.org/10.1007/JHEP09(2018)093} {\bibfield  {journal}
  {\bibinfo  {journal} {Journal of High Energy Physics}\ }\textbf {\bibinfo
  {volume} {2018}},\ \bibinfo {pages} {1} (\bibinfo {year} {2018})}\BibitemShut
  {NoStop}%
\bibitem [{\citenamefont {Cincio}\ \emph {et~al.}(2018)\citenamefont {Cincio},
  \citenamefont {Suba{\c{s}}{\i}}, \citenamefont {Sornborger},\ and\
  \citenamefont {Coles}}]{cincio2018learning}%
  \BibitemOpen
  \bibfield  {author} {\bibinfo {author} {\bibfnamefont {L.}~\bibnamefont
  {Cincio}}, \bibinfo {author} {\bibfnamefont {Y.}~\bibnamefont
  {Suba{\c{s}}{\i}}}, \bibinfo {author} {\bibfnamefont {A.~T.}\ \bibnamefont
  {Sornborger}},\ and\ \bibinfo {author} {\bibfnamefont {P.~J.}\ \bibnamefont
  {Coles}},\ }\bibfield  {title} {\bibinfo {title} {Learning the quantum
  algorithm for state overlap},\ }\href
  {https://doi.org/10.1088/1367-2630/aae94a} {\bibfield  {journal} {\bibinfo
  {journal} {New Journal of Physics}\ }\textbf {\bibinfo {volume} {20}},\
  \bibinfo {pages} {113022} (\bibinfo {year} {2018})}\BibitemShut {NoStop}%
\bibitem [{\citenamefont {Huang}\ \emph {et~al.}(2020)\citenamefont {Huang},
  \citenamefont {Kueng},\ and\ \citenamefont {Preskill}}]{huang2020predicting}%
  \BibitemOpen
  \bibfield  {author} {\bibinfo {author} {\bibfnamefont {H.-Y.}\ \bibnamefont
  {Huang}}, \bibinfo {author} {\bibfnamefont {R.}~\bibnamefont {Kueng}},\ and\
  \bibinfo {author} {\bibfnamefont {J.}~\bibnamefont {Preskill}},\ }\bibfield
  {title} {\bibinfo {title} {Predicting many properties of a quantum system
  from very few measurements},\ }\href
  {https://doi.org/10.1038/s41567-020-0932-7} {\bibfield  {journal} {\bibinfo
  {journal} {Nature Physics}\ }\textbf {\bibinfo {volume} {16}},\ \bibinfo
  {pages} {1050} (\bibinfo {year} {2020})}\BibitemShut {NoStop}%
\bibitem [{\citenamefont {Brydges}\ \emph {et~al.}(2019)\citenamefont
  {Brydges}, \citenamefont {Elben}, \citenamefont {Jurcevic}, \citenamefont
  {Vermersch}, \citenamefont {Maier}, \citenamefont {Lanyon}, \citenamefont
  {Zoller}, \citenamefont {Blatt},\ and\ \citenamefont
  {Roos}}]{brydges2019probing}%
  \BibitemOpen
  \bibfield  {author} {\bibinfo {author} {\bibfnamefont {T.}~\bibnamefont
  {Brydges}}, \bibinfo {author} {\bibfnamefont {A.}~\bibnamefont {Elben}},
  \bibinfo {author} {\bibfnamefont {P.}~\bibnamefont {Jurcevic}}, \bibinfo
  {author} {\bibfnamefont {B.}~\bibnamefont {Vermersch}}, \bibinfo {author}
  {\bibfnamefont {C.}~\bibnamefont {Maier}}, \bibinfo {author} {\bibfnamefont
  {B.~P.}\ \bibnamefont {Lanyon}}, \bibinfo {author} {\bibfnamefont
  {P.}~\bibnamefont {Zoller}}, \bibinfo {author} {\bibfnamefont
  {R.}~\bibnamefont {Blatt}},\ and\ \bibinfo {author} {\bibfnamefont {C.~F.}\
  \bibnamefont {Roos}},\ }\bibfield  {title} {\bibinfo {title} {Probing
  r{\'e}nyi entanglement entropy via randomized measurements},\ }\href
  {https://doi.org/10.1126/science.aau4963} {\bibfield  {journal} {\bibinfo
  {journal} {Science}\ }\textbf {\bibinfo {volume} {364}},\ \bibinfo {pages}
  {260} (\bibinfo {year} {2019})}\BibitemShut {NoStop}%
\bibitem [{\citenamefont {Commeau}\ \emph {et~al.}(2020)\citenamefont
  {Commeau}, \citenamefont {Cerezo}, \citenamefont {Holmes}, \citenamefont
  {Cincio}, \citenamefont {Coles},\ and\ \citenamefont
  {Sornborger}}]{commeau2020variational}%
  \BibitemOpen
  \bibfield  {author} {\bibinfo {author} {\bibfnamefont {B.}~\bibnamefont
  {Commeau}}, \bibinfo {author} {\bibfnamefont {M.}~\bibnamefont {Cerezo}},
  \bibinfo {author} {\bibfnamefont {Z.}~\bibnamefont {Holmes}}, \bibinfo
  {author} {\bibfnamefont {L.}~\bibnamefont {Cincio}}, \bibinfo {author}
  {\bibfnamefont {P.~J.}\ \bibnamefont {Coles}},\ and\ \bibinfo {author}
  {\bibfnamefont {A.}~\bibnamefont {Sornborger}},\ }\bibfield  {title}
  {\bibinfo {title} {Variational {H}amiltonian diagonalization for dynamical
  quantum simulation},\ }\href {https://arxiv.org/abs/2009.02559} {\bibfield
  {journal} {\bibinfo  {journal} {arXiv preprint arXiv:2009.02559}\ } (\bibinfo
  {year} {2020})}\BibitemShut {NoStop}%
\bibitem [{\citenamefont {K{\"o}kc{\"u}}\ \emph
  {et~al.}(2022{\natexlab{a}})\citenamefont {K{\"o}kc{\"u}}, \citenamefont
  {Steckmann}, \citenamefont {Wang}, \citenamefont {Freericks}, \citenamefont
  {Dumitrescu},\ and\ \citenamefont {Kemper}}]{kokcu2021fixed}%
  \BibitemOpen
  \bibfield  {author} {\bibinfo {author} {\bibfnamefont {E.}~\bibnamefont
  {K{\"o}kc{\"u}}}, \bibinfo {author} {\bibfnamefont {T.}~\bibnamefont
  {Steckmann}}, \bibinfo {author} {\bibfnamefont {Y.}~\bibnamefont {Wang}},
  \bibinfo {author} {\bibfnamefont {J.}~\bibnamefont {Freericks}}, \bibinfo
  {author} {\bibfnamefont {E.~F.}\ \bibnamefont {Dumitrescu}},\ and\ \bibinfo
  {author} {\bibfnamefont {A.~F.}\ \bibnamefont {Kemper}},\ }\bibfield  {title}
  {\bibinfo {title} {Fixed depth hamiltonian simulation via cartan
  decomposition},\ }\href {https://doi.org/10.1103/PhysRevLett.129.070501}
  {\bibfield  {journal} {\bibinfo  {journal} {Physical Review Letters}\
  }\textbf {\bibinfo {volume} {129}},\ \bibinfo {pages} {070501} (\bibinfo
  {year} {2022}{\natexlab{a}})}\BibitemShut {NoStop}%
\bibitem [{\citenamefont {Cerezo}\ \emph
  {et~al.}(2021{\natexlab{a}})\citenamefont {Cerezo}, \citenamefont
  {Arrasmith}, \citenamefont {Babbush}, \citenamefont {Benjamin}, \citenamefont
  {Endo}, \citenamefont {Fujii}, \citenamefont {McClean}, \citenamefont
  {Mitarai}, \citenamefont {Yuan}, \citenamefont {Cincio},\ and\ \citenamefont
  {Coles}}]{cerezo2020variationalreview}%
  \BibitemOpen
  \bibfield  {author} {\bibinfo {author} {\bibfnamefont {M.}~\bibnamefont
  {Cerezo}}, \bibinfo {author} {\bibfnamefont {A.}~\bibnamefont {Arrasmith}},
  \bibinfo {author} {\bibfnamefont {R.}~\bibnamefont {Babbush}}, \bibinfo
  {author} {\bibfnamefont {S.~C.}\ \bibnamefont {Benjamin}}, \bibinfo {author}
  {\bibfnamefont {S.}~\bibnamefont {Endo}}, \bibinfo {author} {\bibfnamefont
  {K.}~\bibnamefont {Fujii}}, \bibinfo {author} {\bibfnamefont {J.~R.}\
  \bibnamefont {McClean}}, \bibinfo {author} {\bibfnamefont {K.}~\bibnamefont
  {Mitarai}}, \bibinfo {author} {\bibfnamefont {X.}~\bibnamefont {Yuan}},
  \bibinfo {author} {\bibfnamefont {L.}~\bibnamefont {Cincio}},\ and\ \bibinfo
  {author} {\bibfnamefont {P.~J.}\ \bibnamefont {Coles}},\ }\bibfield  {title}
  {\bibinfo {title} {Variational quantum algorithms},\ }\href
  {https://doi.org/10.1038/s42254-021-00348-9} {\bibfield  {journal} {\bibinfo
  {journal} {Nature Reviews Physics}\ }\textbf {\bibinfo {volume} {3}},\
  \bibinfo {pages} {625–644} (\bibinfo {year}
  {2021}{\natexlab{a}})}\BibitemShut {NoStop}%
\bibitem [{\citenamefont {Peng}\ \emph {et~al.}(2020)\citenamefont {Peng},
  \citenamefont {Harrow}, \citenamefont {Ozols},\ and\ \citenamefont
  {Wu}}]{peng2020simulating}%
  \BibitemOpen
  \bibfield  {author} {\bibinfo {author} {\bibfnamefont {T.}~\bibnamefont
  {Peng}}, \bibinfo {author} {\bibfnamefont {A.~W.}\ \bibnamefont {Harrow}},
  \bibinfo {author} {\bibfnamefont {M.}~\bibnamefont {Ozols}},\ and\ \bibinfo
  {author} {\bibfnamefont {X.}~\bibnamefont {Wu}},\ }\bibfield  {title}
  {\bibinfo {title} {Simulating large quantum circuits on a small quantum
  computer},\ }\href {https://doi.org/10.1103/PhysRevLett.125.150504}
  {\bibfield  {journal} {\bibinfo  {journal} {Phys. Rev. Lett.}\ }\textbf
  {\bibinfo {volume} {125}},\ \bibinfo {pages} {150504} (\bibinfo {year}
  {2020})}\BibitemShut {NoStop}%
\bibitem [{\citenamefont {Bharti}\ \emph {et~al.}(2022)\citenamefont {Bharti},
  \citenamefont {Cervera-Lierta}, \citenamefont {Kyaw}, \citenamefont {Haug},
  \citenamefont {Alperin-Lea}, \citenamefont {Anand}, \citenamefont {Degroote},
  \citenamefont {Heimonen}, \citenamefont {Kottmann}, \citenamefont {Menke}
  \emph {et~al.}}]{bharti2021noisy}%
  \BibitemOpen
  \bibfield  {author} {\bibinfo {author} {\bibfnamefont {K.}~\bibnamefont
  {Bharti}}, \bibinfo {author} {\bibfnamefont {A.}~\bibnamefont
  {Cervera-Lierta}}, \bibinfo {author} {\bibfnamefont {T.~H.}\ \bibnamefont
  {Kyaw}}, \bibinfo {author} {\bibfnamefont {T.}~\bibnamefont {Haug}}, \bibinfo
  {author} {\bibfnamefont {S.}~\bibnamefont {Alperin-Lea}}, \bibinfo {author}
  {\bibfnamefont {A.}~\bibnamefont {Anand}}, \bibinfo {author} {\bibfnamefont
  {M.}~\bibnamefont {Degroote}}, \bibinfo {author} {\bibfnamefont
  {H.}~\bibnamefont {Heimonen}}, \bibinfo {author} {\bibfnamefont {J.~S.}\
  \bibnamefont {Kottmann}}, \bibinfo {author} {\bibfnamefont {T.}~\bibnamefont
  {Menke}}, \emph {et~al.},\ }\bibfield  {title} {\bibinfo {title} {Noisy
  intermediate-scale quantum algorithms},\ }\href
  {https://doi.org/10.1103/RevModPhys.94.015004} {\bibfield  {journal}
  {\bibinfo  {journal} {Reviews of Modern Physics}\ }\textbf {\bibinfo {volume}
  {94}},\ \bibinfo {pages} {015004} (\bibinfo {year} {2022})}\BibitemShut
  {NoStop}%
\bibitem [{\citenamefont {Ollitrault}\ \emph {et~al.}(2023)\citenamefont
  {Ollitrault}, \citenamefont {Jandura}, \citenamefont {Miessen}, \citenamefont
  {Burghardt}, \citenamefont {Martinazzo}, \citenamefont {Tacchino},\ and\
  \citenamefont {Tavernelli}}]{ollitrault2022quantum}%
  \BibitemOpen
  \bibfield  {author} {\bibinfo {author} {\bibfnamefont {P.~J.}\ \bibnamefont
  {Ollitrault}}, \bibinfo {author} {\bibfnamefont {S.}~\bibnamefont {Jandura}},
  \bibinfo {author} {\bibfnamefont {A.}~\bibnamefont {Miessen}}, \bibinfo
  {author} {\bibfnamefont {I.}~\bibnamefont {Burghardt}}, \bibinfo {author}
  {\bibfnamefont {R.}~\bibnamefont {Martinazzo}}, \bibinfo {author}
  {\bibfnamefont {F.}~\bibnamefont {Tacchino}},\ and\ \bibinfo {author}
  {\bibfnamefont {I.}~\bibnamefont {Tavernelli}},\ }\bibfield  {title}
  {\bibinfo {title} {Quantum algorithms for grid-based variational time
  evolution},\ }\href {https://quantum-journal.org/papers/q-2023-10-12-1139/}
  {\bibfield  {journal} {\bibinfo  {journal} {Quantum}\ }\textbf {\bibinfo
  {volume} {7}},\ \bibinfo {pages} {1139} (\bibinfo {year} {2023})}\BibitemShut
  {NoStop}%
\bibitem [{\citenamefont {Biamonte}\ \emph {et~al.}(2017)\citenamefont
  {Biamonte}, \citenamefont {Wittek}, \citenamefont {Pancotti}, \citenamefont
  {Rebentrost}, \citenamefont {Wiebe},\ and\ \citenamefont
  {Lloyd}}]{biamonte2017quantum}%
  \BibitemOpen
  \bibfield  {author} {\bibinfo {author} {\bibfnamefont {J.}~\bibnamefont
  {Biamonte}}, \bibinfo {author} {\bibfnamefont {P.}~\bibnamefont {Wittek}},
  \bibinfo {author} {\bibfnamefont {N.}~\bibnamefont {Pancotti}}, \bibinfo
  {author} {\bibfnamefont {P.}~\bibnamefont {Rebentrost}}, \bibinfo {author}
  {\bibfnamefont {N.}~\bibnamefont {Wiebe}},\ and\ \bibinfo {author}
  {\bibfnamefont {S.}~\bibnamefont {Lloyd}},\ }\bibfield  {title} {\bibinfo
  {title} {Quantum machine learning},\ }\href
  {https://doi.org/10.1038/nature23474} {\bibfield  {journal} {\bibinfo
  {journal} {Nature}\ }\textbf {\bibinfo {volume} {549}},\ \bibinfo {pages}
  {195} (\bibinfo {year} {2017})}\BibitemShut {NoStop}%
\bibitem [{\citenamefont {Schuld}\ \emph {et~al.}(2015)\citenamefont {Schuld},
  \citenamefont {Sinayskiy},\ and\ \citenamefont
  {Petruccione}}]{schuld2015introduction}%
  \BibitemOpen
  \bibfield  {author} {\bibinfo {author} {\bibfnamefont {M.}~\bibnamefont
  {Schuld}}, \bibinfo {author} {\bibfnamefont {I.}~\bibnamefont {Sinayskiy}},\
  and\ \bibinfo {author} {\bibfnamefont {F.}~\bibnamefont {Petruccione}},\
  }\bibfield  {title} {\bibinfo {title} {An introduction to quantum machine
  learning},\ }\href {https://doi.org/10.1080/00107514.2014.964942} {\bibfield
  {journal} {\bibinfo  {journal} {Contemporary Physics}\ }\textbf {\bibinfo
  {volume} {56}},\ \bibinfo {pages} {172} (\bibinfo {year} {2015})}\BibitemShut
  {NoStop}%
\bibitem [{\citenamefont {Cerezo}\ \emph {et~al.}(2022)\citenamefont {Cerezo},
  \citenamefont {Verdon}, \citenamefont {Huang}, \citenamefont {Cincio},\ and\
  \citenamefont {Coles}}]{cerezo2022challenges}%
  \BibitemOpen
  \bibfield  {author} {\bibinfo {author} {\bibfnamefont {M.}~\bibnamefont
  {Cerezo}}, \bibinfo {author} {\bibfnamefont {G.}~\bibnamefont {Verdon}},
  \bibinfo {author} {\bibfnamefont {H.-Y.}\ \bibnamefont {Huang}}, \bibinfo
  {author} {\bibfnamefont {L.}~\bibnamefont {Cincio}},\ and\ \bibinfo {author}
  {\bibfnamefont {P.~J.}\ \bibnamefont {Coles}},\ }\bibfield  {title} {\bibinfo
  {title} {Challenges and opportunities in quantum machine learning},\
  }\bibfield  {journal} {\bibinfo  {journal} {Nature Computational Science}\
  }\href {https://doi.org/10.1038/s43588-022-00311-3}
  {10.1038/s43588-022-00311-3} (\bibinfo {year} {2022})\BibitemShut {NoStop}%
\bibitem [{\citenamefont {{\v{S}}trkalj}\ \emph {et~al.}(2021)\citenamefont
  {{\v{S}}trkalj}, \citenamefont {Doggen}, \citenamefont {Gornyi},\ and\
  \citenamefont {Zilberberg}}]{vstrkalj2021many}%
  \BibitemOpen
  \bibfield  {author} {\bibinfo {author} {\bibfnamefont {A.}~\bibnamefont
  {{\v{S}}trkalj}}, \bibinfo {author} {\bibfnamefont {E.~V.}\ \bibnamefont
  {Doggen}}, \bibinfo {author} {\bibfnamefont {I.~V.}\ \bibnamefont {Gornyi}},\
  and\ \bibinfo {author} {\bibfnamefont {O.}~\bibnamefont {Zilberberg}},\
  }\bibfield  {title} {\bibinfo {title} {Many-body localization in the
  interpolating aubry-andr{\'e}-fibonacci model},\ }\href
  {https://doi.org/10.1103/PhysRevResearch.3.033257} {\bibfield  {journal}
  {\bibinfo  {journal} {Physical Review Research}\ }\textbf {\bibinfo {volume}
  {3}},\ \bibinfo {pages} {033257} (\bibinfo {year} {2021})}\BibitemShut
  {NoStop}%
\bibitem [{\citenamefont {Linden}\ \emph {et~al.}(2009)\citenamefont {Linden},
  \citenamefont {Popescu}, \citenamefont {Short},\ and\ \citenamefont
  {Winter}}]{linden2009quantum}%
  \BibitemOpen
  \bibfield  {author} {\bibinfo {author} {\bibfnamefont {N.}~\bibnamefont
  {Linden}}, \bibinfo {author} {\bibfnamefont {S.}~\bibnamefont {Popescu}},
  \bibinfo {author} {\bibfnamefont {A.~J.}\ \bibnamefont {Short}},\ and\
  \bibinfo {author} {\bibfnamefont {A.}~\bibnamefont {Winter}},\ }\bibfield
  {title} {\bibinfo {title} {Quantum mechanical evolution towards thermal
  equilibrium},\ }\href {https://doi.org/10.1103/PhysRevE.79.061103} {\bibfield
   {journal} {\bibinfo  {journal} {Physical Review E}\ }\textbf {\bibinfo
  {volume} {79}},\ \bibinfo {pages} {061103} (\bibinfo {year}
  {2009})}\BibitemShut {NoStop}%
\bibitem [{\citenamefont {Malabarba}\ \emph {et~al.}(2014)\citenamefont
  {Malabarba}, \citenamefont {Garc{\'\i}a-Pintos}, \citenamefont {Linden},
  \citenamefont {Farrelly},\ and\ \citenamefont
  {Short}}]{malabarba2014quantum}%
  \BibitemOpen
  \bibfield  {author} {\bibinfo {author} {\bibfnamefont {A.~S.}\ \bibnamefont
  {Malabarba}}, \bibinfo {author} {\bibfnamefont {L.~P.}\ \bibnamefont
  {Garc{\'\i}a-Pintos}}, \bibinfo {author} {\bibfnamefont {N.}~\bibnamefont
  {Linden}}, \bibinfo {author} {\bibfnamefont {T.~C.}\ \bibnamefont
  {Farrelly}},\ and\ \bibinfo {author} {\bibfnamefont {A.~J.}\ \bibnamefont
  {Short}},\ }\bibfield  {title} {\bibinfo {title} {Quantum systems equilibrate
  rapidly for most observables},\ }\href
  {https://doi.org/10.1103/PhysRevE.90.012121} {\bibfield  {journal} {\bibinfo
  {journal} {Physical Review E}\ }\textbf {\bibinfo {volume} {90}},\ \bibinfo
  {pages} {012121} (\bibinfo {year} {2014})}\BibitemShut {NoStop}%
\bibitem [{\citenamefont {Mussardo}(2013)}]{mussardo2013infinite}%
  \BibitemOpen
  \bibfield  {author} {\bibinfo {author} {\bibfnamefont {G.}~\bibnamefont
  {Mussardo}},\ }\bibfield  {title} {\bibinfo {title} {Infinite-time average of
  local fields in an integrable quantum field theory after a quantum quench},\
  }\href {https://doi.org/10.1103/PhysRevLett.111.100401} {\bibfield  {journal}
  {\bibinfo  {journal} {Physical review letters}\ }\textbf {\bibinfo {volume}
  {111}},\ \bibinfo {pages} {100401} (\bibinfo {year} {2013})}\BibitemShut
  {NoStop}%
\bibitem [{\citenamefont {Venuti}\ and\ \citenamefont
  {Zanardi}(2010)}]{venuti2010universality}%
  \BibitemOpen
  \bibfield  {author} {\bibinfo {author} {\bibfnamefont {L.~C.}\ \bibnamefont
  {Venuti}}\ and\ \bibinfo {author} {\bibfnamefont {P.}~\bibnamefont
  {Zanardi}},\ }\bibfield  {title} {\bibinfo {title} {Universality in the
  equilibration of quantum systems after a small quench},\ }\href
  {https://doi.org/10.1103/PhysRevA.81.032113} {\bibfield  {journal} {\bibinfo
  {journal} {Physical Review A}\ }\textbf {\bibinfo {volume} {81}},\ \bibinfo
  {pages} {032113} (\bibinfo {year} {2010})}\BibitemShut {NoStop}%
\bibitem [{\citenamefont {Venuti}\ \emph {et~al.}(2011)\citenamefont {Venuti},
  \citenamefont {Jacobson}, \citenamefont {Santra},\ and\ \citenamefont
  {Zanardi}}]{venuti2011exact}%
  \BibitemOpen
  \bibfield  {author} {\bibinfo {author} {\bibfnamefont {L.~C.}\ \bibnamefont
  {Venuti}}, \bibinfo {author} {\bibfnamefont {N.~T.}\ \bibnamefont
  {Jacobson}}, \bibinfo {author} {\bibfnamefont {S.}~\bibnamefont {Santra}},\
  and\ \bibinfo {author} {\bibfnamefont {P.}~\bibnamefont {Zanardi}},\
  }\bibfield  {title} {\bibinfo {title} {Exact infinite-time statistics of the
  loschmidt echo for a quantum quench},\ }\href
  {https://doi.org/10.1103/PhysRevLett.107.010403} {\bibfield  {journal}
  {\bibinfo  {journal} {Physical Review Letters}\ }\textbf {\bibinfo {volume}
  {107}},\ \bibinfo {pages} {010403} (\bibinfo {year} {2011})}\BibitemShut
  {NoStop}%
\bibitem [{\citenamefont {Goussev}\ \emph {et~al.}(2012)\citenamefont
  {Goussev}, \citenamefont {Jalabert}, \citenamefont {Pastawski},\ and\
  \citenamefont {Wisniacki}}]{goussev2012loschmidt}%
  \BibitemOpen
  \bibfield  {author} {\bibinfo {author} {\bibfnamefont {A.}~\bibnamefont
  {Goussev}}, \bibinfo {author} {\bibfnamefont {R.~A.}\ \bibnamefont
  {Jalabert}}, \bibinfo {author} {\bibfnamefont {H.~M.}\ \bibnamefont
  {Pastawski}},\ and\ \bibinfo {author} {\bibfnamefont {D.~A.}\ \bibnamefont
  {Wisniacki}},\ }\bibfield  {title} {\bibinfo {title} {{L}oschmidt echo},\
  }\href {https://doi.org/10.4249/scholarpedia.11687} {\bibfield  {journal}
  {\bibinfo  {journal} {Scholarpedia}\ }\textbf {\bibinfo {volume} {7}},\
  \bibinfo {pages} {11687} (\bibinfo {year} {2012})},\ \bibinfo {note}
  {revision \#127578}\BibitemShut {NoStop}%
\bibitem [{\citenamefont {Yang}\ and\ \citenamefont
  {Hamma}(2017)}]{yang2017many}%
  \BibitemOpen
  \bibfield  {author} {\bibinfo {author} {\bibfnamefont {J.}~\bibnamefont
  {Yang}}\ and\ \bibinfo {author} {\bibfnamefont {A.}~\bibnamefont {Hamma}},\
  }\bibfield  {title} {\bibinfo {title} {Many-body localization transition,
  temporal fluctuations of the loschmidt echo, and scrambling},\ }\href
  {https://arxiv.org/abs/1702.00445} {\bibfield  {journal} {\bibinfo  {journal}
  {arXiv preprint arXiv:1702.00445}\ } (\bibinfo {year} {2017})}\BibitemShut
  {NoStop}%
\bibitem [{\citenamefont {Zhou}\ \emph {et~al.}(2019)\citenamefont {Zhou},
  \citenamefont {Yang},\ and\ \citenamefont {Chen}}]{zhou2019signature}%
  \BibitemOpen
  \bibfield  {author} {\bibinfo {author} {\bibfnamefont {B.}~\bibnamefont
  {Zhou}}, \bibinfo {author} {\bibfnamefont {C.}~\bibnamefont {Yang}},\ and\
  \bibinfo {author} {\bibfnamefont {S.}~\bibnamefont {Chen}},\ }\bibfield
  {title} {\bibinfo {title} {Signature of a nonequilibrium quantum phase
  transition in the long-time average of the loschmidt echo},\ }\href
  {https://doi.org/10.1103/PhysRevB.100.184313} {\bibfield  {journal} {\bibinfo
   {journal} {Physical Review B}\ }\textbf {\bibinfo {volume} {100}},\ \bibinfo
  {pages} {184313} (\bibinfo {year} {2019})}\BibitemShut {NoStop}%
\bibitem [{\citenamefont {Rickayzen}(2013)}]{rickayzen2013green}%
  \BibitemOpen
  \bibfield  {author} {\bibinfo {author} {\bibfnamefont {G.}~\bibnamefont
  {Rickayzen}},\ }\href@noop {} {\emph {\bibinfo {title} {Green's functions and
  condensed matter}}}\ (\bibinfo  {publisher} {Courier Corporation},\ \bibinfo
  {year} {2013})\BibitemShut {NoStop}%
\bibitem [{\citenamefont {Khatami}\ \emph {et~al.}(2013)\citenamefont
  {Khatami}, \citenamefont {Pupillo}, \citenamefont {Srednicki},\ and\
  \citenamefont {Rigol}}]{khatami2013fluctuation}%
  \BibitemOpen
  \bibfield  {author} {\bibinfo {author} {\bibfnamefont {E.}~\bibnamefont
  {Khatami}}, \bibinfo {author} {\bibfnamefont {G.}~\bibnamefont {Pupillo}},
  \bibinfo {author} {\bibfnamefont {M.}~\bibnamefont {Srednicki}},\ and\
  \bibinfo {author} {\bibfnamefont {M.}~\bibnamefont {Rigol}},\ }\bibfield
  {title} {\bibinfo {title} {Fluctuation-dissipation theorem in an isolated
  system of quantum dipolar bosons after a quench},\ }\href
  {https://doi.org/10.1103/PhysRevLett.111.050403} {\bibfield  {journal}
  {\bibinfo  {journal} {Physical review letters}\ }\textbf {\bibinfo {volume}
  {111}},\ \bibinfo {pages} {050403} (\bibinfo {year} {2013})}\BibitemShut
  {NoStop}%
\bibitem [{\citenamefont {Luitz}\ and\ \citenamefont
  {Lev}(2016)}]{luitz2016anomalous}%
  \BibitemOpen
  \bibfield  {author} {\bibinfo {author} {\bibfnamefont {D.~J.}\ \bibnamefont
  {Luitz}}\ and\ \bibinfo {author} {\bibfnamefont {Y.~B.}\ \bibnamefont
  {Lev}},\ }\bibfield  {title} {\bibinfo {title} {Anomalous thermalization in
  ergodic systems},\ }\href {https://doi.org/10.1103/PhysRevLett.117.170404}
  {\bibfield  {journal} {\bibinfo  {journal} {Physical review letters}\
  }\textbf {\bibinfo {volume} {117}},\ \bibinfo {pages} {170404} (\bibinfo
  {year} {2016})}\BibitemShut {NoStop}%
\bibitem [{\citenamefont {K{\"o}kc{\"u}}\ \emph {et~al.}(2023)\citenamefont
  {K{\"o}kc{\"u}}, \citenamefont {Labib}, \citenamefont {Freericks},\ and\
  \citenamefont {Kemper}}]{kokcu2023linear}%
  \BibitemOpen
  \bibfield  {author} {\bibinfo {author} {\bibfnamefont {E.}~\bibnamefont
  {K{\"o}kc{\"u}}}, \bibinfo {author} {\bibfnamefont {H.~A.}\ \bibnamefont
  {Labib}}, \bibinfo {author} {\bibfnamefont {J.}~\bibnamefont {Freericks}},\
  and\ \bibinfo {author} {\bibfnamefont {A.~F.}\ \bibnamefont {Kemper}},\
  }\bibfield  {title} {\bibinfo {title} {A linear response framework for
  simulating bosonic and fermionic correlation functions illustrated on quantum
  computers},\ }\href {https://arxiv.org/abs/2302.10219} {\bibfield  {journal}
  {\bibinfo  {journal} {arXiv preprint arXiv:2302.10219}\ } (\bibinfo {year}
  {2023})}\BibitemShut {NoStop}%
\bibitem [{\citenamefont {Alhambra}\ \emph {et~al.}(2020)\citenamefont
  {Alhambra}, \citenamefont {Riddell},\ and\ \citenamefont
  {Garc{\'\i}a-Pintos}}]{alhambra2020time}%
  \BibitemOpen
  \bibfield  {author} {\bibinfo {author} {\bibfnamefont {{\'A}.~M.}\
  \bibnamefont {Alhambra}}, \bibinfo {author} {\bibfnamefont {J.}~\bibnamefont
  {Riddell}},\ and\ \bibinfo {author} {\bibfnamefont {L.~P.}\ \bibnamefont
  {Garc{\'\i}a-Pintos}},\ }\bibfield  {title} {\bibinfo {title} {Time evolution
  of correlation functions in quantum many-body systems},\ }\href
  {https://doi.org/10.1103/PhysRevLett.124.110605} {\bibfield  {journal}
  {\bibinfo  {journal} {Physical Review Letters}\ }\textbf {\bibinfo {volume}
  {124}},\ \bibinfo {pages} {110605} (\bibinfo {year} {2020})}\BibitemShut
  {NoStop}%
\bibitem [{\citenamefont {Pedernales}\ \emph {et~al.}(2014)\citenamefont
  {Pedernales}, \citenamefont {Di~Candia}, \citenamefont {Egusquiza},
  \citenamefont {Casanova},\ and\ \citenamefont
  {Solano}}]{pedernales2014efficient}%
  \BibitemOpen
  \bibfield  {author} {\bibinfo {author} {\bibfnamefont {J.}~\bibnamefont
  {Pedernales}}, \bibinfo {author} {\bibfnamefont {R.}~\bibnamefont
  {Di~Candia}}, \bibinfo {author} {\bibfnamefont {I.}~\bibnamefont
  {Egusquiza}}, \bibinfo {author} {\bibfnamefont {J.}~\bibnamefont
  {Casanova}},\ and\ \bibinfo {author} {\bibfnamefont {E.}~\bibnamefont
  {Solano}},\ }\bibfield  {title} {\bibinfo {title} {Efficient quantum
  algorithm for computing n-time correlation functions},\ }\href
  {https://doi.org/10.1103/PhysRevLett.113.020505} {\bibfield  {journal}
  {\bibinfo  {journal} {Physical Review Letters}\ }\textbf {\bibinfo {volume}
  {113}},\ \bibinfo {pages} {020505} (\bibinfo {year} {2014})}\BibitemShut
  {NoStop}%
\bibitem [{\citenamefont {Baez}\ \emph {et~al.}(2020)\citenamefont {Baez},
  \citenamefont {Goihl}, \citenamefont {Haferkamp}, \citenamefont
  {Bermejo-Vega}, \citenamefont {Gluza},\ and\ \citenamefont
  {Eisert}}]{baez2020dynamical}%
  \BibitemOpen
  \bibfield  {author} {\bibinfo {author} {\bibfnamefont {M.~L.}\ \bibnamefont
  {Baez}}, \bibinfo {author} {\bibfnamefont {M.}~\bibnamefont {Goihl}},
  \bibinfo {author} {\bibfnamefont {J.}~\bibnamefont {Haferkamp}}, \bibinfo
  {author} {\bibfnamefont {J.}~\bibnamefont {Bermejo-Vega}}, \bibinfo {author}
  {\bibfnamefont {M.}~\bibnamefont {Gluza}},\ and\ \bibinfo {author}
  {\bibfnamefont {J.}~\bibnamefont {Eisert}},\ }\bibfield  {title} {\bibinfo
  {title} {Dynamical structure factors of dynamical quantum simulators},\
  }\href {https://doi.org/10.1073/pnas.2006103117} {\bibfield  {journal}
  {\bibinfo  {journal} {Proceedings of the National Academy of Sciences}\
  }\textbf {\bibinfo {volume} {117}},\ \bibinfo {pages} {26123} (\bibinfo
  {year} {2020})}\BibitemShut {NoStop}%
\bibitem [{\citenamefont {Coldea}\ \emph {et~al.}(2010)\citenamefont {Coldea},
  \citenamefont {Tennant}, \citenamefont {Wheeler}, \citenamefont {Wawrzynska},
  \citenamefont {Prabhakaran}, \citenamefont {Telling}, \citenamefont
  {Habicht}, \citenamefont {Smeibidl},\ and\ \citenamefont
  {Kiefer}}]{coldea2010quantum}%
  \BibitemOpen
  \bibfield  {author} {\bibinfo {author} {\bibfnamefont {R.}~\bibnamefont
  {Coldea}}, \bibinfo {author} {\bibfnamefont {D.}~\bibnamefont {Tennant}},
  \bibinfo {author} {\bibfnamefont {E.}~\bibnamefont {Wheeler}}, \bibinfo
  {author} {\bibfnamefont {E.}~\bibnamefont {Wawrzynska}}, \bibinfo {author}
  {\bibfnamefont {D.}~\bibnamefont {Prabhakaran}}, \bibinfo {author}
  {\bibfnamefont {M.}~\bibnamefont {Telling}}, \bibinfo {author} {\bibfnamefont
  {K.}~\bibnamefont {Habicht}}, \bibinfo {author} {\bibfnamefont
  {P.}~\bibnamefont {Smeibidl}},\ and\ \bibinfo {author} {\bibfnamefont
  {K.}~\bibnamefont {Kiefer}},\ }\bibfield  {title} {\bibinfo {title} {Quantum
  criticality in an ising chain: experimental evidence for emergent e8
  symmetry},\ }\href {https://doi.org/10.1126/science.1180085} {\bibfield
  {journal} {\bibinfo  {journal} {Science}\ }\textbf {\bibinfo {volume}
  {327}},\ \bibinfo {pages} {177} (\bibinfo {year} {2010})}\BibitemShut
  {NoStop}%
\bibitem [{\citenamefont {Jia}\ \emph {et~al.}(2014)\citenamefont {Jia},
  \citenamefont {Nowadnick}, \citenamefont {Wohlfeld}, \citenamefont {Kung},
  \citenamefont {Chen}, \citenamefont {Johnston}, \citenamefont {Tohyama},
  \citenamefont {Moritz},\ and\ \citenamefont {Devereaux}}]{jia2014persistent}%
  \BibitemOpen
  \bibfield  {author} {\bibinfo {author} {\bibfnamefont {C.}~\bibnamefont
  {Jia}}, \bibinfo {author} {\bibfnamefont {E.}~\bibnamefont {Nowadnick}},
  \bibinfo {author} {\bibfnamefont {K.}~\bibnamefont {Wohlfeld}}, \bibinfo
  {author} {\bibfnamefont {Y.}~\bibnamefont {Kung}}, \bibinfo {author}
  {\bibfnamefont {C.-C.}\ \bibnamefont {Chen}}, \bibinfo {author}
  {\bibfnamefont {S.}~\bibnamefont {Johnston}}, \bibinfo {author}
  {\bibfnamefont {T.}~\bibnamefont {Tohyama}}, \bibinfo {author} {\bibfnamefont
  {B.}~\bibnamefont {Moritz}},\ and\ \bibinfo {author} {\bibfnamefont
  {T.}~\bibnamefont {Devereaux}},\ }\bibfield  {title} {\bibinfo {title}
  {Persistent spin excitations in doped antiferromagnets revealed by resonant
  inelastic light scattering},\ }\href {https://doi.org/10.1038/ncomms4314}
  {\bibfield  {journal} {\bibinfo  {journal} {Nature communications}\ }\textbf
  {\bibinfo {volume} {5}},\ \bibinfo {pages} {3314} (\bibinfo {year}
  {2014})}\BibitemShut {NoStop}%
\bibitem [{\citenamefont {Bauer}\ \emph {et~al.}(2016)\citenamefont {Bauer},
  \citenamefont {Wecker}, \citenamefont {Millis}, \citenamefont {Hastings},\
  and\ \citenamefont {Troyer}}]{bauer2016hybrid}%
  \BibitemOpen
  \bibfield  {author} {\bibinfo {author} {\bibfnamefont {B.}~\bibnamefont
  {Bauer}}, \bibinfo {author} {\bibfnamefont {D.}~\bibnamefont {Wecker}},
  \bibinfo {author} {\bibfnamefont {A.~J.}\ \bibnamefont {Millis}}, \bibinfo
  {author} {\bibfnamefont {M.~B.}\ \bibnamefont {Hastings}},\ and\ \bibinfo
  {author} {\bibfnamefont {M.}~\bibnamefont {Troyer}},\ }\bibfield  {title}
  {\bibinfo {title} {Hybrid quantum-classical approach to correlated
  materials},\ }\href
  {https://journals.aps.org/prx/abstract/10.1103/PhysRevX.6.031045} {\bibfield
  {journal} {\bibinfo  {journal} {Physical Review X}\ }\textbf {\bibinfo
  {volume} {6}},\ \bibinfo {pages} {031045} (\bibinfo {year}
  {2016})}\BibitemShut {NoStop}%
\bibitem [{\citenamefont {Kreula}\ \emph {et~al.}(2016)\citenamefont {Kreula},
  \citenamefont {Clark},\ and\ \citenamefont {Jaksch}}]{kreula2016non}%
  \BibitemOpen
  \bibfield  {author} {\bibinfo {author} {\bibfnamefont {J.}~\bibnamefont
  {Kreula}}, \bibinfo {author} {\bibfnamefont {S.~R.}\ \bibnamefont {Clark}},\
  and\ \bibinfo {author} {\bibfnamefont {D.}~\bibnamefont {Jaksch}},\
  }\bibfield  {title} {\bibinfo {title} {Non-linear quantum-classical scheme to
  simulate non-equilibrium strongly correlated fermionic many-body dynamics},\
  }\href {https://doi.org/10.1038/srep32940} {\bibfield  {journal} {\bibinfo
  {journal} {Scientific reports}\ }\textbf {\bibinfo {volume} {6}},\ \bibinfo
  {pages} {1} (\bibinfo {year} {2016})}\BibitemShut {NoStop}%
\bibitem [{\citenamefont {Sakurai}\ \emph {et~al.}(2022)\citenamefont
  {Sakurai}, \citenamefont {Mizukami},\ and\ \citenamefont
  {Shinaoka}}]{sakurai2022hybrid}%
  \BibitemOpen
  \bibfield  {author} {\bibinfo {author} {\bibfnamefont {R.}~\bibnamefont
  {Sakurai}}, \bibinfo {author} {\bibfnamefont {W.}~\bibnamefont {Mizukami}},\
  and\ \bibinfo {author} {\bibfnamefont {H.}~\bibnamefont {Shinaoka}},\
  }\bibfield  {title} {\bibinfo {title} {Hybrid quantum-classical algorithm for
  computing imaginary-time correlation functions},\ }\href
  {https://doi.org/10.1103/PhysRevResearch.4.023219} {\bibfield  {journal}
  {\bibinfo  {journal} {Physical Review Research}\ }\textbf {\bibinfo {volume}
  {4}},\ \bibinfo {pages} {023219} (\bibinfo {year} {2022})}\BibitemShut
  {NoStop}%
\bibitem [{\citenamefont {Buhrman}\ \emph {et~al.}(2001)\citenamefont
  {Buhrman}, \citenamefont {Cleve}, \citenamefont {Watrous},\ and\
  \citenamefont {De~Wolf}}]{buhrman2001quantum}%
  \BibitemOpen
  \bibfield  {author} {\bibinfo {author} {\bibfnamefont {H.}~\bibnamefont
  {Buhrman}}, \bibinfo {author} {\bibfnamefont {R.}~\bibnamefont {Cleve}},
  \bibinfo {author} {\bibfnamefont {J.}~\bibnamefont {Watrous}},\ and\ \bibinfo
  {author} {\bibfnamefont {R.}~\bibnamefont {De~Wolf}},\ }\bibfield  {title}
  {\bibinfo {title} {Quantum fingerprinting},\ }\href
  {https://doi.org/10.1103/PhysRevLett.87.167902} {\bibfield  {journal}
  {\bibinfo  {journal} {Physical Review Letters}\ }\textbf {\bibinfo {volume}
  {87}},\ \bibinfo {pages} {167902} (\bibinfo {year} {2001})}\BibitemShut
  {NoStop}%
\bibitem [{\citenamefont {Harrow}\ and\ \citenamefont
  {Montanaro}(2013)}]{harrow2013testing}%
  \BibitemOpen
  \bibfield  {author} {\bibinfo {author} {\bibfnamefont {A.~W.}\ \bibnamefont
  {Harrow}}\ and\ \bibinfo {author} {\bibfnamefont {A.}~\bibnamefont
  {Montanaro}},\ }\bibfield  {title} {\bibinfo {title} {Testing product states,
  quantum merlin-arthur games and tensor optimization},\ }\href
  {https://doi.org/10.1109/FOCS.2010.66 10.1145/2432622.2432625} {\bibfield
  {journal} {\bibinfo  {journal} {Journal of the ACM (JACM)}\ }\textbf
  {\bibinfo {volume} {60}},\ \bibinfo {pages} {1} (\bibinfo {year}
  {2013})}\BibitemShut {NoStop}%
\bibitem [{\citenamefont {Gutoski}\ \emph {et~al.}(2015)\citenamefont
  {Gutoski}, \citenamefont {Hayden}, \citenamefont {Milner},\ and\
  \citenamefont {Wilde}}]{gutoski2015quantum}%
  \BibitemOpen
  \bibfield  {author} {\bibinfo {author} {\bibfnamefont {G.}~\bibnamefont
  {Gutoski}}, \bibinfo {author} {\bibfnamefont {P.}~\bibnamefont {Hayden}},
  \bibinfo {author} {\bibfnamefont {K.}~\bibnamefont {Milner}},\ and\ \bibinfo
  {author} {\bibfnamefont {M.~M.}\ \bibnamefont {Wilde}},\ }\bibfield  {title}
  {\bibinfo {title} {Quantum interactive proofs and the complexity of
  separability testing},\ }\href {https://doi.org/10.4086/toc.2015.v011a003}
  {\bibfield  {journal} {\bibinfo  {journal} {Theory of Computing}\ }\textbf
  {\bibinfo {volume} {11}},\ \bibinfo {pages} {59} (\bibinfo {year}
  {2015})}\BibitemShut {NoStop}%
\bibitem [{\citenamefont {Nielsen}\ and\ \citenamefont
  {Chuang}(2000)}]{nielsen2000quantum}%
  \BibitemOpen
  \bibfield  {author} {\bibinfo {author} {\bibfnamefont {M.~A.}\ \bibnamefont
  {Nielsen}}\ and\ \bibinfo {author} {\bibfnamefont {I.~L.}\ \bibnamefont
  {Chuang}},\ }\href@noop {} {\emph {\bibinfo {title} {Quantum Computation and
  Quantum Information}}}\ (\bibinfo  {publisher} {Cambridge University Press},\
  \bibinfo {address} {Cambridge},\ \bibinfo {year} {2000})\BibitemShut
  {NoStop}%
\bibitem [{Note1()}]{Note1}%
  \BibitemOpen
  \bibinfo {note} {For infinite $T$ one has to consider subtleties related to
  the normalization of states; see \cite {giovannetti2015quantum,
  diaz2019history, diaz2019historystate}.}\BibitemShut {Stop}%
\bibitem [{\citenamefont {Shor}(1999)}]{shor1999polynomial}%
  \BibitemOpen
  \bibfield  {author} {\bibinfo {author} {\bibfnamefont {P.~W.}\ \bibnamefont
  {Shor}},\ }\bibfield  {title} {\bibinfo {title} {Polynomial-time algorithms
  for prime factorization and discrete logarithms on a quantum computer},\
  }\href {https://doi.org/10.1137/S0036144598347011} {\bibfield  {journal}
  {\bibinfo  {journal} {SIAM review}\ }\textbf {\bibinfo {volume} {41}},\
  \bibinfo {pages} {303} (\bibinfo {year} {1999})}\BibitemShut {NoStop}%
\bibitem [{\citenamefont {Lloyd}(1996)}]{lloyd1996universal}%
  \BibitemOpen
  \bibfield  {author} {\bibinfo {author} {\bibfnamefont {S.}~\bibnamefont
  {Lloyd}},\ }\bibfield  {title} {\bibinfo {title} {Universal quantum
  simulators},\ }\href {https://doi.org/10.1126/science.273.5278.1073}
  {\bibfield  {journal} {\bibinfo  {journal} {Science}\ ,\ \bibinfo {pages}
  {1073}} (\bibinfo {year} {1996})}\BibitemShut {NoStop}%
\bibitem [{\citenamefont {Childs}\ \emph {et~al.}(2021)\citenamefont {Childs},
  \citenamefont {Su}, \citenamefont {Tran}, \citenamefont {Wiebe},\ and\
  \citenamefont {Zhu}}]{childs2021theory}%
  \BibitemOpen
  \bibfield  {author} {\bibinfo {author} {\bibfnamefont {A.~M.}\ \bibnamefont
  {Childs}}, \bibinfo {author} {\bibfnamefont {Y.}~\bibnamefont {Su}}, \bibinfo
  {author} {\bibfnamefont {M.~C.}\ \bibnamefont {Tran}}, \bibinfo {author}
  {\bibfnamefont {N.}~\bibnamefont {Wiebe}},\ and\ \bibinfo {author}
  {\bibfnamefont {S.}~\bibnamefont {Zhu}},\ }\bibfield  {title} {\bibinfo
  {title} {Theory of trotter error with commutator scaling},\ }\href
  {https://doi.org/https://doi.org/10.1103/PhysRevX.11.011020} {\bibfield
  {journal} {\bibinfo  {journal} {Physical Review X}\ }\textbf {\bibinfo
  {volume} {11}},\ \bibinfo {pages} {011020} (\bibinfo {year}
  {2021})}\BibitemShut {NoStop}%
\bibitem [{\citenamefont {Berry}\ \emph {et~al.}(2007)\citenamefont {Berry},
  \citenamefont {Ahokas}, \citenamefont {Cleve},\ and\ \citenamefont
  {Sanders}}]{berry2007efficient}%
  \BibitemOpen
  \bibfield  {author} {\bibinfo {author} {\bibfnamefont {D.~W.}\ \bibnamefont
  {Berry}}, \bibinfo {author} {\bibfnamefont {G.}~\bibnamefont {Ahokas}},
  \bibinfo {author} {\bibfnamefont {R.}~\bibnamefont {Cleve}},\ and\ \bibinfo
  {author} {\bibfnamefont {B.~C.}\ \bibnamefont {Sanders}},\ }\bibfield
  {title} {\bibinfo {title} {Efficient quantum algorithms for simulating sparse
  hamiltonians},\ }\href {https://doi.org/10.1007/s00220-006-0150-x} {\bibfield
   {journal} {\bibinfo  {journal} {Communications in Mathematical Physics}\
  }\textbf {\bibinfo {volume} {270}},\ \bibinfo {pages} {359} (\bibinfo {year}
  {2007})}\BibitemShut {NoStop}%
\bibitem [{\citenamefont {Somma}(2016)}]{somma2016trotter}%
  \BibitemOpen
  \bibfield  {author} {\bibinfo {author} {\bibfnamefont {R.~D.}\ \bibnamefont
  {Somma}},\ }\bibfield  {title} {\bibinfo {title} {A trotter-suzuki
  approximation for lie groups with applications to hamiltonian simulation},\
  }\href {https://doi.org/10.1063/1.4952761} {\bibfield  {journal} {\bibinfo
  {journal} {Journal of Mathematical Physics}\ }\textbf {\bibinfo {volume}
  {57}},\ \bibinfo {pages} {062202} (\bibinfo {year} {2016})}\BibitemShut
  {NoStop}%
\bibitem [{Note2()}]{Note2}%
  \BibitemOpen
  \bibinfo {note} {If the duration of time required to implement the gates
  depends on the interval $\varepsilon $ the reasoning can be adapted by
  changing $\alpha $. For example, following \cite {lloyd1996universal} one may
  assume that implementing $e^{-ih_j \varepsilon }$ takes $\varepsilon =\tau
  /t$ time and we need $\propto t$ gates meaning a total duration proportional
  to $\tau $, rather than quadratic as the number of gates.}\BibitemShut
  {Stop}%
\bibitem [{\citenamefont {Childs}\ \emph {et~al.}(2018)\citenamefont {Childs},
  \citenamefont {Maslov}, \citenamefont {Nam}, \citenamefont {Ross},\ and\
  \citenamefont {Su}}]{childs2018toward}%
  \BibitemOpen
  \bibfield  {author} {\bibinfo {author} {\bibfnamefont {A.~M.}\ \bibnamefont
  {Childs}}, \bibinfo {author} {\bibfnamefont {D.}~\bibnamefont {Maslov}},
  \bibinfo {author} {\bibfnamefont {Y.}~\bibnamefont {Nam}}, \bibinfo {author}
  {\bibfnamefont {N.~J.}\ \bibnamefont {Ross}},\ and\ \bibinfo {author}
  {\bibfnamefont {Y.}~\bibnamefont {Su}},\ }\bibfield  {title} {\bibinfo
  {title} {Toward the first quantum simulation with quantum speedup},\ }\href
  {https://doi.org/10.1073/pnas.1801723115} {\bibfield  {journal} {\bibinfo
  {journal} {Proceedings of the National Academy of Sciences}\ }\textbf
  {\bibinfo {volume} {115}},\ \bibinfo {pages} {9456} (\bibinfo {year}
  {2018})}\BibitemShut {NoStop}%
\bibitem [{\citenamefont {Heyl}\ \emph {et~al.}(2019)\citenamefont {Heyl},
  \citenamefont {Hauke},\ and\ \citenamefont {Zoller}}]{heyl2019quantum}%
  \BibitemOpen
  \bibfield  {author} {\bibinfo {author} {\bibfnamefont {M.}~\bibnamefont
  {Heyl}}, \bibinfo {author} {\bibfnamefont {P.}~\bibnamefont {Hauke}},\ and\
  \bibinfo {author} {\bibfnamefont {P.}~\bibnamefont {Zoller}},\ }\bibfield
  {title} {\bibinfo {title} {Quantum localization bounds trotter errors in
  digital quantum simulation},\ }\href
  {https://www.science.org/doi/10.1126/sciadv.aau8342} {\bibfield  {journal}
  {\bibinfo  {journal} {Science advances}\ }\textbf {\bibinfo {volume} {5}},\
  \bibinfo {pages} {eaau8342} (\bibinfo {year} {2019})}\BibitemShut {NoStop}%
\bibitem [{\citenamefont {Horodecki}\ \emph {et~al.}(2009)\citenamefont
  {Horodecki}, \citenamefont {Horodecki}, \citenamefont {Horodecki},\ and\
  \citenamefont {Horodecki}}]{horodecki2009quantum}%
  \BibitemOpen
  \bibfield  {author} {\bibinfo {author} {\bibfnamefont {R.}~\bibnamefont
  {Horodecki}}, \bibinfo {author} {\bibfnamefont {P.}~\bibnamefont
  {Horodecki}}, \bibinfo {author} {\bibfnamefont {M.}~\bibnamefont
  {Horodecki}},\ and\ \bibinfo {author} {\bibfnamefont {K.}~\bibnamefont
  {Horodecki}},\ }\bibfield  {title} {\bibinfo {title} {Quantum entanglement},\
  }\href {https://doi.org/10.1103/RevModPhys.81.865} {\bibfield  {journal}
  {\bibinfo  {journal} {Reviews of modern physics}\ }\textbf {\bibinfo {volume}
  {81}},\ \bibinfo {pages} {865} (\bibinfo {year} {2009})}\BibitemShut
  {NoStop}%
\bibitem [{\citenamefont {McClean}\ \emph {et~al.}(2018)\citenamefont
  {McClean}, \citenamefont {Boixo}, \citenamefont {Smelyanskiy}, \citenamefont
  {Babbush},\ and\ \citenamefont {Neven}}]{mcclean2018barren}%
  \BibitemOpen
  \bibfield  {author} {\bibinfo {author} {\bibfnamefont {J.~R.}\ \bibnamefont
  {McClean}}, \bibinfo {author} {\bibfnamefont {S.}~\bibnamefont {Boixo}},
  \bibinfo {author} {\bibfnamefont {V.~N.}\ \bibnamefont {Smelyanskiy}},
  \bibinfo {author} {\bibfnamefont {R.}~\bibnamefont {Babbush}},\ and\ \bibinfo
  {author} {\bibfnamefont {H.}~\bibnamefont {Neven}},\ }\bibfield  {title}
  {\bibinfo {title} {Barren plateaus in quantum neural network training
  landscapes},\ }\href {https://doi.org/10.1038/s41467-018-07090-4} {\bibfield
  {journal} {\bibinfo  {journal} {Nature {C}ommunications}\ }\textbf {\bibinfo
  {volume} {9}},\ \bibinfo {pages} {1} (\bibinfo {year} {2018})}\BibitemShut
  {NoStop}%
\bibitem [{\citenamefont {Cerezo}\ \emph
  {et~al.}(2021{\natexlab{b}})\citenamefont {Cerezo}, \citenamefont {Sone},
  \citenamefont {Volkoff}, \citenamefont {Cincio},\ and\ \citenamefont
  {Coles}}]{cerezo2020cost}%
  \BibitemOpen
  \bibfield  {author} {\bibinfo {author} {\bibfnamefont {M.}~\bibnamefont
  {Cerezo}}, \bibinfo {author} {\bibfnamefont {A.}~\bibnamefont {Sone}},
  \bibinfo {author} {\bibfnamefont {T.}~\bibnamefont {Volkoff}}, \bibinfo
  {author} {\bibfnamefont {L.}~\bibnamefont {Cincio}},\ and\ \bibinfo {author}
  {\bibfnamefont {P.~J.}\ \bibnamefont {Coles}},\ }\bibfield  {title} {\bibinfo
  {title} {Cost function dependent barren plateaus in shallow parametrized
  quantum circuits},\ }\href {https://doi.org/10.1038/s41467-021-21728-w}
  {\bibfield  {journal} {\bibinfo  {journal} {Nature {C}ommunications}\
  }\textbf {\bibinfo {volume} {12}},\ \bibinfo {pages} {1} (\bibinfo {year}
  {2021}{\natexlab{b}})}\BibitemShut {NoStop}%
\bibitem [{\citenamefont {Holmes}\ \emph {et~al.}(2022)\citenamefont {Holmes},
  \citenamefont {Sharma}, \citenamefont {Cerezo},\ and\ \citenamefont
  {Coles}}]{holmes2021connecting}%
  \BibitemOpen
  \bibfield  {author} {\bibinfo {author} {\bibfnamefont {Z.}~\bibnamefont
  {Holmes}}, \bibinfo {author} {\bibfnamefont {K.}~\bibnamefont {Sharma}},
  \bibinfo {author} {\bibfnamefont {M.}~\bibnamefont {Cerezo}},\ and\ \bibinfo
  {author} {\bibfnamefont {P.~J.}\ \bibnamefont {Coles}},\ }\bibfield  {title}
  {\bibinfo {title} {Connecting ansatz expressibility to gradient magnitudes
  and barren plateaus},\ }\href {https://doi.org/10.1103/PRXQuantum.3.010313}
  {\bibfield  {journal} {\bibinfo  {journal} {PRX Quantum}\ }\textbf {\bibinfo
  {volume} {3}},\ \bibinfo {pages} {010313} (\bibinfo {year}
  {2022})}\BibitemShut {NoStop}%
\bibitem [{\citenamefont {Larocca}\ \emph {et~al.}(2025)\citenamefont
  {Larocca}, \citenamefont {Thanasilp}, \citenamefont {Wang}, \citenamefont
  {Sharma}, \citenamefont {Biamonte}, \citenamefont {Coles}, \citenamefont
  {Cincio}, \citenamefont {McClean}, \citenamefont {Holmes},\ and\
  \citenamefont {Cerezo}}]{larocca2024review}%
  \BibitemOpen
  \bibfield  {author} {\bibinfo {author} {\bibfnamefont {M.}~\bibnamefont
  {Larocca}}, \bibinfo {author} {\bibfnamefont {S.}~\bibnamefont {Thanasilp}},
  \bibinfo {author} {\bibfnamefont {S.}~\bibnamefont {Wang}}, \bibinfo {author}
  {\bibfnamefont {K.}~\bibnamefont {Sharma}}, \bibinfo {author} {\bibfnamefont
  {J.}~\bibnamefont {Biamonte}}, \bibinfo {author} {\bibfnamefont {P.~J.}\
  \bibnamefont {Coles}}, \bibinfo {author} {\bibfnamefont {L.}~\bibnamefont
  {Cincio}}, \bibinfo {author} {\bibfnamefont {J.~R.}\ \bibnamefont {McClean}},
  \bibinfo {author} {\bibfnamefont {Z.}~\bibnamefont {Holmes}},\ and\ \bibinfo
  {author} {\bibfnamefont {M.}~\bibnamefont {Cerezo}},\ }\bibfield  {title}
  {\bibinfo {title} {A review of barren plateaus in variational quantum
  computing},\ }\href {https://doi.org/10.1038/s42254-025-00813-9} {\bibfield
  {journal} {\bibinfo  {journal} {Nature Reviews Physics}\ }\textbf {\bibinfo
  {volume} {3}},\ \bibinfo {pages} {625–644} (\bibinfo {year}
  {2025})}\BibitemShut {NoStop}%
\bibitem [{\citenamefont {K{\"u}bler}\ \emph {et~al.}(2021)\citenamefont
  {K{\"u}bler}, \citenamefont {Buchholz},\ and\ \citenamefont
  {Sch{\"o}lkopf}}]{kubler2021inductive}%
  \BibitemOpen
  \bibfield  {author} {\bibinfo {author} {\bibfnamefont {J.}~\bibnamefont
  {K{\"u}bler}}, \bibinfo {author} {\bibfnamefont {S.}~\bibnamefont
  {Buchholz}},\ and\ \bibinfo {author} {\bibfnamefont {B.}~\bibnamefont
  {Sch{\"o}lkopf}},\ }\bibfield  {title} {\bibinfo {title} {The inductive bias
  of quantum kernels},\ }\href
  {https://proceedings.neurips.cc/paper/2021/hash/69adc1e107f7f7d035d7baf04342e1ca-Abstract.html}
  {\bibfield  {journal} {\bibinfo  {journal} {Advances in Neural Information
  Processing Systems}\ }\textbf {\bibinfo {volume} {34}},\ \bibinfo {pages}
  {12661} (\bibinfo {year} {2021})}\BibitemShut {NoStop}%
\bibitem [{\citenamefont {Larocca}\ \emph
  {et~al.}(2022{\natexlab{a}})\citenamefont {Larocca}, \citenamefont {Czarnik},
  \citenamefont {Sharma}, \citenamefont {Muraleedharan}, \citenamefont
  {Coles},\ and\ \citenamefont {Cerezo}}]{larocca2021diagnosing}%
  \BibitemOpen
  \bibfield  {author} {\bibinfo {author} {\bibfnamefont {M.}~\bibnamefont
  {Larocca}}, \bibinfo {author} {\bibfnamefont {P.}~\bibnamefont {Czarnik}},
  \bibinfo {author} {\bibfnamefont {K.}~\bibnamefont {Sharma}}, \bibinfo
  {author} {\bibfnamefont {G.}~\bibnamefont {Muraleedharan}}, \bibinfo {author}
  {\bibfnamefont {P.~J.}\ \bibnamefont {Coles}},\ and\ \bibinfo {author}
  {\bibfnamefont {M.}~\bibnamefont {Cerezo}},\ }\bibfield  {title} {\bibinfo
  {title} {Diagnosing {B}arren {P}lateaus with {T}ools from {Q}uantum {O}ptimal
  {C}ontrol},\ }\href {https://doi.org/10.22331/q-2022-09-29-824} {\bibfield
  {journal} {\bibinfo  {journal} {{Quantum}}\ }\textbf {\bibinfo {volume}
  {6}},\ \bibinfo {pages} {824} (\bibinfo {year}
  {2022}{\natexlab{a}})}\BibitemShut {NoStop}%
\bibitem [{\citenamefont {Larocca}\ \emph
  {et~al.}(2022{\natexlab{b}})\citenamefont {Larocca}, \citenamefont {Sauvage},
  \citenamefont {Sbahi}, \citenamefont {Verdon}, \citenamefont {Coles},\ and\
  \citenamefont {Cerezo}}]{larocca2022group}%
  \BibitemOpen
  \bibfield  {author} {\bibinfo {author} {\bibfnamefont {M.}~\bibnamefont
  {Larocca}}, \bibinfo {author} {\bibfnamefont {F.}~\bibnamefont {Sauvage}},
  \bibinfo {author} {\bibfnamefont {F.~M.}\ \bibnamefont {Sbahi}}, \bibinfo
  {author} {\bibfnamefont {G.}~\bibnamefont {Verdon}}, \bibinfo {author}
  {\bibfnamefont {P.~J.}\ \bibnamefont {Coles}},\ and\ \bibinfo {author}
  {\bibfnamefont {M.}~\bibnamefont {Cerezo}},\ }\bibfield  {title} {\bibinfo
  {title} {Group-invariant quantum machine learning},\ }\href
  {https://doi.org/10.1103/PRXQuantum.3.030341} {\bibfield  {journal} {\bibinfo
   {journal} {PRX Quantum}\ }\textbf {\bibinfo {volume} {3}},\ \bibinfo {pages}
  {030341} (\bibinfo {year} {2022}{\natexlab{b}})}\BibitemShut {NoStop}%
\bibitem [{\citenamefont {K{\"o}kc{\"u}}\ \emph
  {et~al.}(2022{\natexlab{b}})\citenamefont {K{\"o}kc{\"u}}, \citenamefont
  {Steckmann}, \citenamefont {Wang}, \citenamefont {Freericks}, \citenamefont
  {Dumitrescu},\ and\ \citenamefont {Kemper}}]{kokcu2022fixed}%
  \BibitemOpen
  \bibfield  {author} {\bibinfo {author} {\bibfnamefont {E.}~\bibnamefont
  {K{\"o}kc{\"u}}}, \bibinfo {author} {\bibfnamefont {T.}~\bibnamefont
  {Steckmann}}, \bibinfo {author} {\bibfnamefont {Y.}~\bibnamefont {Wang}},
  \bibinfo {author} {\bibfnamefont {J.}~\bibnamefont {Freericks}}, \bibinfo
  {author} {\bibfnamefont {E.~F.}\ \bibnamefont {Dumitrescu}},\ and\ \bibinfo
  {author} {\bibfnamefont {A.~F.}\ \bibnamefont {Kemper}},\ }\bibfield  {title}
  {\bibinfo {title} {Fixed depth hamiltonian simulation via cartan
  decomposition},\ }\href {https://doi.org/10.1103/PhysRevLett.129.070501}
  {\bibfield  {journal} {\bibinfo  {journal} {Physical Review Letters}\
  }\textbf {\bibinfo {volume} {129}},\ \bibinfo {pages} {070501} (\bibinfo
  {year} {2022}{\natexlab{b}})}\BibitemShut {NoStop}%
\bibitem [{\citenamefont {Camps}\ \emph {et~al.}(2022)\citenamefont {Camps},
  \citenamefont {K{\"o}kc{\"u}}, \citenamefont {Bassman~Oftelie}, \citenamefont
  {De~Jong}, \citenamefont {Kemper},\ and\ \citenamefont
  {Van~Beeumen}}]{camps2022algebraic}%
  \BibitemOpen
  \bibfield  {author} {\bibinfo {author} {\bibfnamefont {D.}~\bibnamefont
  {Camps}}, \bibinfo {author} {\bibfnamefont {E.}~\bibnamefont
  {K{\"o}kc{\"u}}}, \bibinfo {author} {\bibfnamefont {L.}~\bibnamefont
  {Bassman~Oftelie}}, \bibinfo {author} {\bibfnamefont {W.~A.}\ \bibnamefont
  {De~Jong}}, \bibinfo {author} {\bibfnamefont {A.~F.}\ \bibnamefont
  {Kemper}},\ and\ \bibinfo {author} {\bibfnamefont {R.}~\bibnamefont
  {Van~Beeumen}},\ }\bibfield  {title} {\bibinfo {title} {An algebraic quantum
  circuit compression algorithm for hamiltonian simulation},\ }\href
  {https://doi.org/10.1137/21M1439298} {\bibfield  {journal} {\bibinfo
  {journal} {SIAM Journal on Matrix Analysis and Applications}\ }\textbf
  {\bibinfo {volume} {43}},\ \bibinfo {pages} {1084} (\bibinfo {year}
  {2022})}\BibitemShut {NoStop}%
\bibitem [{\citenamefont {K{\"o}kc{\"u}}\ \emph
  {et~al.}(2022{\natexlab{c}})\citenamefont {K{\"o}kc{\"u}}, \citenamefont
  {Camps}, \citenamefont {Oftelie}, \citenamefont {Freericks}, \citenamefont
  {de~Jong}, \citenamefont {Van~Beeumen},\ and\ \citenamefont
  {Kemper}}]{kokcu2022algebraic}%
  \BibitemOpen
  \bibfield  {author} {\bibinfo {author} {\bibfnamefont {E.}~\bibnamefont
  {K{\"o}kc{\"u}}}, \bibinfo {author} {\bibfnamefont {D.}~\bibnamefont
  {Camps}}, \bibinfo {author} {\bibfnamefont {L.~B.}\ \bibnamefont {Oftelie}},
  \bibinfo {author} {\bibfnamefont {J.~K.}\ \bibnamefont {Freericks}}, \bibinfo
  {author} {\bibfnamefont {W.~A.}\ \bibnamefont {de~Jong}}, \bibinfo {author}
  {\bibfnamefont {R.}~\bibnamefont {Van~Beeumen}},\ and\ \bibinfo {author}
  {\bibfnamefont {A.~F.}\ \bibnamefont {Kemper}},\ }\bibfield  {title}
  {\bibinfo {title} {Algebraic compression of quantum circuits for hamiltonian
  evolution},\ }\href {https://doi.org/10.1103/PhysRevA.105.032420} {\bibfield
  {journal} {\bibinfo  {journal} {Physical Review A}\ }\textbf {\bibinfo
  {volume} {105}},\ \bibinfo {pages} {032420} (\bibinfo {year}
  {2022}{\natexlab{c}})}\BibitemShut {NoStop}%
\bibitem [{\citenamefont {Zeier}\ and\ \citenamefont
  {Schulte-Herbr{\"u}ggen}(2011)}]{zeier2011symmetry}%
  \BibitemOpen
  \bibfield  {author} {\bibinfo {author} {\bibfnamefont {R.}~\bibnamefont
  {Zeier}}\ and\ \bibinfo {author} {\bibfnamefont {T.}~\bibnamefont
  {Schulte-Herbr{\"u}ggen}},\ }\bibfield  {title} {\bibinfo {title} {Symmetry
  principles in quantum systems theory},\ }\href
  {https://doi.org/https://doi.org/10.1063/1.3657939} {\bibfield  {journal}
  {\bibinfo  {journal} {Journal of mathematical physics}\ }\textbf {\bibinfo
  {volume} {52}},\ \bibinfo {pages} {113510} (\bibinfo {year}
  {2011})}\BibitemShut {NoStop}%
\bibitem [{\citenamefont {Jozsa}\ and\ \citenamefont
  {Miyake}(2008)}]{jozsa2008matchgates}%
  \BibitemOpen
  \bibfield  {author} {\bibinfo {author} {\bibfnamefont {R.}~\bibnamefont
  {Jozsa}}\ and\ \bibinfo {author} {\bibfnamefont {A.}~\bibnamefont {Miyake}},\
  }\bibfield  {title} {\bibinfo {title} {Matchgates and classical simulation of
  quantum circuits},\ }\href {https://doi.org/10.1098/rspa.2008.0189}
  {\bibfield  {journal} {\bibinfo  {journal} {Proceedings of the Royal Society
  A: Mathematical, Physical and Engineering Sciences}\ }\textbf {\bibinfo
  {volume} {464}},\ \bibinfo {pages} {3089} (\bibinfo {year}
  {2008})}\BibitemShut {NoStop}%
\bibitem [{\citenamefont {Wan}\ \emph {et~al.}(2023)\citenamefont {Wan},
  \citenamefont {Huggins}, \citenamefont {Lee},\ and\ \citenamefont
  {Babbush}}]{wan2022matchgate}%
  \BibitemOpen
  \bibfield  {author} {\bibinfo {author} {\bibfnamefont {K.}~\bibnamefont
  {Wan}}, \bibinfo {author} {\bibfnamefont {W.~J.}\ \bibnamefont {Huggins}},
  \bibinfo {author} {\bibfnamefont {J.}~\bibnamefont {Lee}},\ and\ \bibinfo
  {author} {\bibfnamefont {R.}~\bibnamefont {Babbush}},\ }\bibfield  {title}
  {\bibinfo {title} {Matchgate shadows for fermionic quantum simulation},\
  }\href {https://doi.org/10.1007/s00220-023-04844-0} {\bibfield  {journal}
  {\bibinfo  {journal} {Communications in Mathematical Physics}\ }\textbf
  {\bibinfo {volume} {404}},\ \bibinfo {pages} {629} (\bibinfo {year}
  {2023})}\BibitemShut {NoStop}%
\bibitem [{\citenamefont {Diaz}\ \emph {et~al.}(2023)\citenamefont {Diaz},
  \citenamefont {Garc{\'\i}a-Mart{\'\i}n}, \citenamefont {Kazi}, \citenamefont
  {Larocca},\ and\ \citenamefont {Cerezo}}]{diaz2023showcasing}%
  \BibitemOpen
  \bibfield  {author} {\bibinfo {author} {\bibfnamefont {N.~L.}\ \bibnamefont
  {Diaz}}, \bibinfo {author} {\bibfnamefont {D.}~\bibnamefont
  {Garc{\'\i}a-Mart{\'\i}n}}, \bibinfo {author} {\bibfnamefont
  {S.}~\bibnamefont {Kazi}}, \bibinfo {author} {\bibfnamefont {M.}~\bibnamefont
  {Larocca}},\ and\ \bibinfo {author} {\bibfnamefont {M.}~\bibnamefont
  {Cerezo}},\ }\bibfield  {title} {\bibinfo {title} {Showcasing a barren
  plateau theory beyond the dynamical lie algebra},\ }\href
  {https://arxiv.org/abs/2310.11505} {\bibfield  {journal} {\bibinfo  {journal}
  {arXiv preprint arXiv:2310.11505}\ } (\bibinfo {year} {2023})}\BibitemShut
  {NoStop}%
\bibitem [{\citenamefont {You}\ \emph {et~al.}(2022)\citenamefont {You},
  \citenamefont {Chakrabarti},\ and\ \citenamefont {Wu}}]{you2022convergence}%
  \BibitemOpen
  \bibfield  {author} {\bibinfo {author} {\bibfnamefont {X.}~\bibnamefont
  {You}}, \bibinfo {author} {\bibfnamefont {S.}~\bibnamefont {Chakrabarti}},\
  and\ \bibinfo {author} {\bibfnamefont {X.}~\bibnamefont {Wu}},\ }\bibfield
  {title} {\bibinfo {title} {A convergence theory for over-parameterized
  variational quantum eigensolvers},\ }\href {https://arxiv.org/abs/2205.12481}
  {\bibfield  {journal} {\bibinfo  {journal} {arXiv preprint arXiv:2205.12481}\
  } (\bibinfo {year} {2022})}\BibitemShut {NoStop}%
\bibitem [{\citenamefont {Cerezo}\ \emph {et~al.}(2025)\citenamefont {Cerezo},
  \citenamefont {Larocca}, \citenamefont {Garc{\'\i}a-Mart{\'\i}n},
  \citenamefont {Diaz}, \citenamefont {Braccia}, \citenamefont {Fontana},
  \citenamefont {Rudolph}, \citenamefont {Bermejo}, \citenamefont {Ijaz},
  \citenamefont {Thanasilp} \emph {et~al.}}]{cerezo2023does}%
  \BibitemOpen
  \bibfield  {author} {\bibinfo {author} {\bibfnamefont {M.}~\bibnamefont
  {Cerezo}}, \bibinfo {author} {\bibfnamefont {M.}~\bibnamefont {Larocca}},
  \bibinfo {author} {\bibfnamefont {D.}~\bibnamefont
  {Garc{\'\i}a-Mart{\'\i}n}}, \bibinfo {author} {\bibfnamefont {N.~L.}\
  \bibnamefont {Diaz}}, \bibinfo {author} {\bibfnamefont {P.}~\bibnamefont
  {Braccia}}, \bibinfo {author} {\bibfnamefont {E.}~\bibnamefont {Fontana}},
  \bibinfo {author} {\bibfnamefont {M.~S.}\ \bibnamefont {Rudolph}}, \bibinfo
  {author} {\bibfnamefont {P.}~\bibnamefont {Bermejo}}, \bibinfo {author}
  {\bibfnamefont {A.}~\bibnamefont {Ijaz}}, \bibinfo {author} {\bibfnamefont
  {S.}~\bibnamefont {Thanasilp}}, \emph {et~al.},\ }\bibfield  {title}
  {\bibinfo {title} {Does provable absence of barren plateaus imply classical
  simulability?},\ }\href {https://doi.org/10.1038/s41467-025-63099-6}
  {\bibfield  {journal} {\bibinfo  {journal} {Nature Communications}\ }\textbf
  {\bibinfo {volume} {16}},\ \bibinfo {pages} {7907} (\bibinfo {year}
  {2025})}\BibitemShut {NoStop}%
\bibitem [{aub(1980)}]{aubry1980analyticity}%
  \BibitemOpen
  \bibfield  {title} {\bibinfo {title} {Analyticity breaking and anderson
  localization in incommensurate lattices},\ }\href@noop {} {\bibfield
  {journal} {\bibinfo  {journal} {Ann. Israel Phys. Soc.}\ }\textbf {\bibinfo
  {volume} {3}},\ \bibinfo {pages} {133} (\bibinfo {year} {1980})}\BibitemShut
  {NoStop}%
\bibitem [{\citenamefont {Iyer}\ \emph {et~al.}(2013)\citenamefont {Iyer},
  \citenamefont {Oganesyan}, \citenamefont {Refael},\ and\ \citenamefont
  {Huse}}]{iyer2013many}%
  \BibitemOpen
  \bibfield  {author} {\bibinfo {author} {\bibfnamefont {S.}~\bibnamefont
  {Iyer}}, \bibinfo {author} {\bibfnamefont {V.}~\bibnamefont {Oganesyan}},
  \bibinfo {author} {\bibfnamefont {G.}~\bibnamefont {Refael}},\ and\ \bibinfo
  {author} {\bibfnamefont {D.~A.}\ \bibnamefont {Huse}},\ }\bibfield  {title}
  {\bibinfo {title} {Many-body localization in a quasiperiodic system},\ }\href
  {https://doi.org/10.1103/PhysRevB.87.134202} {\bibfield  {journal} {\bibinfo
  {journal} {Physical Review B}\ }\textbf {\bibinfo {volume} {87}},\ \bibinfo
  {pages} {134202} (\bibinfo {year} {2013})}\BibitemShut {NoStop}%
\bibitem [{\citenamefont {Larocca}\ \emph {et~al.}(2023)\citenamefont
  {Larocca}, \citenamefont {Ju}, \citenamefont {García-Martín}, \citenamefont
  {Coles},\ and\ \citenamefont {Cerezo}}]{larocca2021theory}%
  \BibitemOpen
  \bibfield  {author} {\bibinfo {author} {\bibfnamefont {M.}~\bibnamefont
  {Larocca}}, \bibinfo {author} {\bibfnamefont {N.}~\bibnamefont {Ju}},
  \bibinfo {author} {\bibfnamefont {D.}~\bibnamefont {García-Martín}},
  \bibinfo {author} {\bibfnamefont {P.~J.}\ \bibnamefont {Coles}},\ and\
  \bibinfo {author} {\bibfnamefont {M.}~\bibnamefont {Cerezo}},\ }\bibfield
  {title} {\bibinfo {title} {Theory of overparametrization in quantum neural
  networks},\ }\href
  {https://doi.org/https://doi.org/10.1038/s43588-023-00467-6} {\bibfield
  {journal} {\bibinfo  {journal} {Nature Computational Science}\ }\textbf
  {\bibinfo {volume} {3}},\ \bibinfo {pages} {542} (\bibinfo {year}
  {2023})}\BibitemShut {NoStop}%
\bibitem [{\citenamefont {Or{\'u}s}(2014)}]{orus2014practical}%
  \BibitemOpen
  \bibfield  {author} {\bibinfo {author} {\bibfnamefont {R.}~\bibnamefont
  {Or{\'u}s}},\ }\bibfield  {title} {\bibinfo {title} {A practical introduction
  to tensor networks: Matrix product states and projected entangled pair
  states},\ }\href {https://doi.org/10.1016/j.aop.2014.06.013} {\bibfield
  {journal} {\bibinfo  {journal} {Annals of Physics}\ }\textbf {\bibinfo
  {volume} {349}},\ \bibinfo {pages} {117} (\bibinfo {year}
  {2014})}\BibitemShut {NoStop}%
\bibitem [{\citenamefont {Feynman}(1982)}]{feynman1982simulating}%
  \BibitemOpen
  \bibfield  {author} {\bibinfo {author} {\bibfnamefont {R.~P.}\ \bibnamefont
  {Feynman}},\ }\bibfield  {title} {\bibinfo {title} {Simulating physics with
  computers},\ }\href {https://doi.org/10.1007/BF02650179} {\bibfield
  {journal} {\bibinfo  {journal} {International Journal of Theoretical
  Physics}\ }\textbf {\bibinfo {volume} {21}},\ \bibinfo {pages} {467}
  (\bibinfo {year} {1982})}\BibitemShut {NoStop}%
\bibitem [{\citenamefont {Gibbs}\ and\ \citenamefont
  {Cincio}(2025{\natexlab{a}})}]{gibbs2024deep}%
  \BibitemOpen
  \bibfield  {author} {\bibinfo {author} {\bibfnamefont {J.}~\bibnamefont
  {Gibbs}}\ and\ \bibinfo {author} {\bibfnamefont {L.}~\bibnamefont {Cincio}},\
  }\bibfield  {title} {\bibinfo {title} {Deep circuit compression for quantum
  dynamics via tensor networks},\ }\href
  {https://doi.org/10.22331/q-2025-07-09-1789} {\bibfield  {journal} {\bibinfo
  {journal} {Quantum}\ }\textbf {\bibinfo {volume} {9}},\ \bibinfo {pages}
  {1789} (\bibinfo {year} {2025}{\natexlab{a}})}\BibitemShut {NoStop}%
\bibitem [{\citenamefont {Zhang}\ \emph {et~al.}(2024)\citenamefont {Zhang},
  \citenamefont {Wiersema}, \citenamefont {Carrasquilla}, \citenamefont
  {Cincio},\ and\ \citenamefont {Kim}}]{zhang2024scalable}%
  \BibitemOpen
  \bibfield  {author} {\bibinfo {author} {\bibfnamefont {Y.}~\bibnamefont
  {Zhang}}, \bibinfo {author} {\bibfnamefont {R.}~\bibnamefont {Wiersema}},
  \bibinfo {author} {\bibfnamefont {J.}~\bibnamefont {Carrasquilla}}, \bibinfo
  {author} {\bibfnamefont {L.}~\bibnamefont {Cincio}},\ and\ \bibinfo {author}
  {\bibfnamefont {Y.~B.}\ \bibnamefont {Kim}},\ }\bibfield  {title} {\bibinfo
  {title} {Scalable quantum dynamics compilation via quantum machine
  learning},\ }\href {https://arxiv.org/abs/2409.16346} {\bibfield  {journal}
  {\bibinfo  {journal} {arXiv preprint arXiv:2409.16346}\ } (\bibinfo {year}
  {2024})}\BibitemShut {NoStop}%
\bibitem [{\citenamefont {Gibbs}\ and\ \citenamefont
  {Cincio}(2025{\natexlab{b}})}]{gibbs2025learning}%
  \BibitemOpen
  \bibfield  {author} {\bibinfo {author} {\bibfnamefont {J.}~\bibnamefont
  {Gibbs}}\ and\ \bibinfo {author} {\bibfnamefont {L.}~\bibnamefont {Cincio}},\
  }\bibfield  {title} {\bibinfo {title} {Learning circuits with infinite tensor
  networks},\ }\bibfield  {journal} {\bibinfo  {journal} {arXiv preprint
  arXiv:2506.02105}\ }\href {https://doi.org/10.48550/arXiv.2506.02105}
  {10.48550/arXiv.2506.02105} (\bibinfo {year}
  {2025}{\natexlab{b}})\BibitemShut {NoStop}%
\bibitem [{\citenamefont {Caro}\ \emph {et~al.}(2022)\citenamefont {Caro},
  \citenamefont {Huang}, \citenamefont {Cerezo}, \citenamefont {Sharma},
  \citenamefont {Sornborger}, \citenamefont {Cincio},\ and\ \citenamefont
  {Coles}}]{caro2021generalization}%
  \BibitemOpen
  \bibfield  {author} {\bibinfo {author} {\bibfnamefont {M.~C.}\ \bibnamefont
  {Caro}}, \bibinfo {author} {\bibfnamefont {H.-Y.}\ \bibnamefont {Huang}},
  \bibinfo {author} {\bibfnamefont {M.}~\bibnamefont {Cerezo}}, \bibinfo
  {author} {\bibfnamefont {K.}~\bibnamefont {Sharma}}, \bibinfo {author}
  {\bibfnamefont {A.}~\bibnamefont {Sornborger}}, \bibinfo {author}
  {\bibfnamefont {L.}~\bibnamefont {Cincio}},\ and\ \bibinfo {author}
  {\bibfnamefont {P.~J.}\ \bibnamefont {Coles}},\ }\bibfield  {title} {\bibinfo
  {title} {Generalization in quantum machine learning from few training data},\
  }\href {https://doi.org/10.1038/s41467-022-32550-3} {\bibfield  {journal}
  {\bibinfo  {journal} {Nature {C}ommunications}\ }\textbf {\bibinfo {volume}
  {13}},\ \bibinfo {eid} {4919} (\bibinfo {year} {2022})}\BibitemShut {NoStop}%
\bibitem [{\citenamefont {Caro}\ \emph {et~al.}(2023)\citenamefont {Caro},
  \citenamefont {Huang}, \citenamefont {Ezzell}, \citenamefont {Gibbs},
  \citenamefont {Sornborger}, \citenamefont {Cincio}, \citenamefont {Coles},\
  and\ \citenamefont {Holmes}}]{caro2022outofdistribution}%
  \BibitemOpen
  \bibfield  {author} {\bibinfo {author} {\bibfnamefont {M.~C.}\ \bibnamefont
  {Caro}}, \bibinfo {author} {\bibfnamefont {H.-Y.}\ \bibnamefont {Huang}},
  \bibinfo {author} {\bibfnamefont {N.}~\bibnamefont {Ezzell}}, \bibinfo
  {author} {\bibfnamefont {J.}~\bibnamefont {Gibbs}}, \bibinfo {author}
  {\bibfnamefont {A.~T.}\ \bibnamefont {Sornborger}}, \bibinfo {author}
  {\bibfnamefont {L.}~\bibnamefont {Cincio}}, \bibinfo {author} {\bibfnamefont
  {P.~J.}\ \bibnamefont {Coles}},\ and\ \bibinfo {author} {\bibfnamefont
  {Z.}~\bibnamefont {Holmes}},\ }\bibfield  {title} {\bibinfo {title}
  {Out-of-distribution generalization for learning quantum dynamics},\ }\href
  {https://doi.org/10.1038/s41467-023-39381-w} {\bibfield  {journal} {\bibinfo
  {journal} {Nature Communications}\ }\textbf {\bibinfo {volume} {14}},\
  \bibinfo {pages} {3751} (\bibinfo {year} {2023})}\BibitemShut {NoStop}%
\bibitem [{\citenamefont {Yoshida}\ \emph {et~al.}(2023)\citenamefont
  {Yoshida}, \citenamefont {Soeda},\ and\ \citenamefont
  {Murao}}]{yoshida2023reversing}%
  \BibitemOpen
  \bibfield  {author} {\bibinfo {author} {\bibfnamefont {S.}~\bibnamefont
  {Yoshida}}, \bibinfo {author} {\bibfnamefont {A.}~\bibnamefont {Soeda}},\
  and\ \bibinfo {author} {\bibfnamefont {M.}~\bibnamefont {Murao}},\ }\bibfield
   {title} {\bibinfo {title} {Reversing unknown qubit-unitary operation,
  deterministically and exactly},\ }\href
  {https://doi.org/10.1103/PhysRevLett.131.120602} {\bibfield  {journal}
  {\bibinfo  {journal} {Physical Review Letters}\ }\textbf {\bibinfo {volume}
  {131}},\ \bibinfo {pages} {120602} (\bibinfo {year} {2023})}\BibitemShut
  {NoStop}%
\bibitem [{\citenamefont {Chen}\ \emph {et~al.}(2024)\citenamefont {Chen},
  \citenamefont {Mo}, \citenamefont {Liu}, \citenamefont {Zhang},\ and\
  \citenamefont {Wang}}]{chen2024quantum}%
  \BibitemOpen
  \bibfield  {author} {\bibinfo {author} {\bibfnamefont {Y.-A.}\ \bibnamefont
  {Chen}}, \bibinfo {author} {\bibfnamefont {Y.}~\bibnamefont {Mo}}, \bibinfo
  {author} {\bibfnamefont {Y.}~\bibnamefont {Liu}}, \bibinfo {author}
  {\bibfnamefont {L.}~\bibnamefont {Zhang}},\ and\ \bibinfo {author}
  {\bibfnamefont {X.}~\bibnamefont {Wang}},\ }\bibfield  {title} {\bibinfo
  {title} {Quantum algorithm for reversing unknown unitary evolutions},\ }\href
  {https://arxiv.org/abs/2403.04704} {\bibfield  {journal} {\bibinfo  {journal}
  {arXiv preprint arXiv:2403.04704}\ } (\bibinfo {year} {2024})}\BibitemShut
  {NoStop}%
\bibitem [{\citenamefont {Diaz}\ \emph {et~al.}(2024)\citenamefont {Diaz},
  \citenamefont {Matera},\ and\ \citenamefont
  {Rossignoli}}]{diaz2023spacetime}%
  \BibitemOpen
  \bibfield  {author} {\bibinfo {author} {\bibfnamefont {N.~L.}\ \bibnamefont
  {Diaz}}, \bibinfo {author} {\bibfnamefont {J.~M.}\ \bibnamefont {Matera}},\
  and\ \bibinfo {author} {\bibfnamefont {R.}~\bibnamefont {Rossignoli}},\
  }\bibfield  {title} {\bibinfo {title} {Spacetime quantum and classical
  mechanics with dynamical foliation},\ }\href
  {https://doi.org/10.1103/PhysRevD.109.105008} {\bibfield  {journal} {\bibinfo
   {journal} {Phys. Rev. D}\ }\textbf {\bibinfo {volume} {109}},\ \bibinfo
  {pages} {105008} (\bibinfo {year} {2024})}\BibitemShut {NoStop}%
\bibitem [{\citenamefont {Diaz}\ and\ \citenamefont
  {Rossignoli}(2025)}]{diaz2025spacetime}%
  \BibitemOpen
  \bibfield  {author} {\bibinfo {author} {\bibfnamefont {N.~L.}\ \bibnamefont
  {Diaz}}\ and\ \bibinfo {author} {\bibfnamefont {R.}~\bibnamefont
  {Rossignoli}},\ }\bibfield  {title} {\bibinfo {title} {Spacetime quantum
  mechanics for bosonic and fermionic systems},\ }\href
  {https://arxiv.org/abs/2506.10250} {\bibfield  {journal} {\bibinfo  {journal}
  {arXiv preprint arXiv:2506.10250}\ } (\bibinfo {year} {2025})}\BibitemShut
  {NoStop}%
\bibitem [{\citenamefont {Kiefer}(2012)}]{kiefer2012quantum}%
  \BibitemOpen
  \bibfield  {author} {\bibinfo {author} {\bibfnamefont {C.}~\bibnamefont
  {Kiefer}},\ }\href {https://books.google.com/books?id=lnDQWIgmP7QC} {\emph
  {\bibinfo {title} {Quantum Gravity: Third Edition}}},\ International Series
  of Monographs on Physics\ (\bibinfo  {publisher} {OUP Oxford},\ \bibinfo
  {address} {Oxford},\ \bibinfo {year} {2012})\BibitemShut {NoStop}%
\bibitem [{\citenamefont {Green}\ \emph {et~al.}(2012)\citenamefont {Green},
  \citenamefont {Schwarz},\ and\ \citenamefont
  {Witten}}]{green2012superstring}%
  \BibitemOpen
  \bibfield  {author} {\bibinfo {author} {\bibfnamefont {M.~B.}\ \bibnamefont
  {Green}}, \bibinfo {author} {\bibfnamefont {J.~H.}\ \bibnamefont {Schwarz}},\
  and\ \bibinfo {author} {\bibfnamefont {E.}~\bibnamefont {Witten}},\
  }\href@noop {} {\emph {\bibinfo {title} {Superstring theory: volume 2, loop
  amplitudes, anomalies and phenomenology}}}\ (\bibinfo  {publisher} {Cambridge
  university press},\ \bibinfo {address} {Cambridge},\ \bibinfo {year}
  {2012})\BibitemShut {NoStop}%
\bibitem [{\citenamefont {DeWitt}(1967)}]{dewitt1967quantum}%
  \BibitemOpen
  \bibfield  {author} {\bibinfo {author} {\bibfnamefont {B.~S.}\ \bibnamefont
  {DeWitt}},\ }\bibfield  {title} {\bibinfo {title} {Quantum theory of gravity.
  i. the canonical theory},\ }\href {https://doi.org/10.1103/PhysRev.160.1113}
  {\bibfield  {journal} {\bibinfo  {journal} {Physical Review}\ }\textbf
  {\bibinfo {volume} {160}},\ \bibinfo {pages} {1113} (\bibinfo {year}
  {1967})}\BibitemShut {NoStop}%
\bibitem [{\citenamefont {Dirac}(1950)}]{dirac1950generalized}%
  \BibitemOpen
  \bibfield  {author} {\bibinfo {author} {\bibfnamefont {P.~A.~M.}\
  \bibnamefont {Dirac}},\ }\bibfield  {title} {\bibinfo {title} {Generalized
  hamiltonian dynamics},\ }\href {https://doi.org/10.4153/CJM-1950-012-1}
  {\bibfield  {journal} {\bibinfo  {journal} {Canadian journal of mathematics}\
  }\textbf {\bibinfo {volume} {2}},\ \bibinfo {pages} {129} (\bibinfo {year}
  {1950})}\BibitemShut {NoStop}%
\bibitem [{\citenamefont {Dirac}(1958)}]{dirac1958theory}%
  \BibitemOpen
  \bibfield  {author} {\bibinfo {author} {\bibfnamefont {P.~A.~M.}\
  \bibnamefont {Dirac}},\ }\bibfield  {title} {\bibinfo {title} {The theory of
  gravitation in hamiltonian form},\ }\href
  {https://doi.org/10.1098/rspa.1958.0142} {\bibfield  {journal} {\bibinfo
  {journal} {Proceedings of the Royal Society of London. Series A. Mathematical
  and Physical Sciences}\ }\textbf {\bibinfo {volume} {246}},\ \bibinfo {pages}
  {333} (\bibinfo {year} {1958})}\BibitemShut {NoStop}%
\bibitem [{Note3()}]{Note3}%
  \BibitemOpen
  \bibinfo {note} {This is a common example in both canonical quantum gravity
  \cite {kiefer2012quantum} and string theory \cite {green2012superstring}
  textbooks}\BibitemShut {NoStop}%
\bibitem [{\citenamefont {Marolf}(1995)}]{marolf1995quantum}%
  \BibitemOpen
  \bibfield  {author} {\bibinfo {author} {\bibfnamefont {D.}~\bibnamefont
  {Marolf}},\ }\bibfield  {title} {\bibinfo {title} {Quantum observables and
  recollapsing dynamics},\ }\href {https://doi.org/10.1088/0264-9381/12/5/011}
  {\bibfield  {journal} {\bibinfo  {journal} {Classical and Quantum Gravity}\
  }\textbf {\bibinfo {volume} {12}},\ \bibinfo {pages} {1199} (\bibinfo {year}
  {1995})}\BibitemShut {NoStop}%
\bibitem [{\citenamefont {Isham}\ and\ \citenamefont
  {Linden}(1995)}]{isham1995continuous}%
  \BibitemOpen
  \bibfield  {author} {\bibinfo {author} {\bibfnamefont {C.}~\bibnamefont
  {Isham}}\ and\ \bibinfo {author} {\bibfnamefont {N.}~\bibnamefont {Linden}},\
  }\bibfield  {title} {\bibinfo {title} {Continuous histories and the history
  group in generalized quantum theory},\ }\href
  {https://doi.org/10.1063/1.531267} {\bibfield  {journal} {\bibinfo  {journal}
  {Journal of Mathematical Physics}\ }\textbf {\bibinfo {volume} {36}},\
  \bibinfo {pages} {5392} (\bibinfo {year} {1995})}\BibitemShut {NoStop}%
\bibitem [{\citenamefont {Harper}\ \emph {et~al.}(2023)\citenamefont {Harper},
  \citenamefont {Mollabashi}, \citenamefont {Takayanagi}, \citenamefont {Taki}
  \emph {et~al.}}]{harper2023timelike}%
  \BibitemOpen
  \bibfield  {author} {\bibinfo {author} {\bibfnamefont {J.}~\bibnamefont
  {Harper}}, \bibinfo {author} {\bibfnamefont {A.}~\bibnamefont {Mollabashi}},
  \bibinfo {author} {\bibfnamefont {T.}~\bibnamefont {Takayanagi}}, \bibinfo
  {author} {\bibfnamefont {Y.}~\bibnamefont {Taki}}, \emph {et~al.},\
  }\bibfield  {title} {\bibinfo {title} {Timelike entanglement entropy},\
  }\href {https://doi.org/10.1007/JHEP05(2023)052} {\bibfield  {journal}
  {\bibinfo  {journal} {Journal of High Energy Physics}\ }\textbf {\bibinfo
  {volume} {2023}},\ \bibinfo {pages} {1} (\bibinfo {year} {2023})}\BibitemShut
  {NoStop}%
\bibitem [{\citenamefont {Narayan}\ and\ \citenamefont
  {Saini}(2023)}]{narayan2023notes}%
  \BibitemOpen
  \bibfield  {author} {\bibinfo {author} {\bibfnamefont {K.}~\bibnamefont
  {Narayan}}\ and\ \bibinfo {author} {\bibfnamefont {H.~K.}\ \bibnamefont
  {Saini}},\ }\bibfield  {title} {\bibinfo {title} {Notes on time entanglement
  and pseudo-entropy},\ }\href {https://arxiv.org/abs/2303.01307} {\bibfield
  {journal} {\bibinfo  {journal} {arXiv preprint arXiv:2303.01307}\ } (\bibinfo
  {year} {2023})}\BibitemShut {NoStop}%
\bibitem [{\citenamefont {Chu}\ and\ \citenamefont
  {Parihar}(2023)}]{chu2023time}%
  \BibitemOpen
  \bibfield  {author} {\bibinfo {author} {\bibfnamefont {C.-S.}\ \bibnamefont
  {Chu}}\ and\ \bibinfo {author} {\bibfnamefont {H.}~\bibnamefont {Parihar}},\
  }\bibfield  {title} {\bibinfo {title} {Time-like entanglement entropy in
  ads/bcft},\ }\href {https://arxiv.org/abs/2304.10907} {\bibfield  {journal}
  {\bibinfo  {journal} {arXiv preprint arXiv:2304.10907}\ } (\bibinfo {year}
  {2023})}\BibitemShut {NoStop}%
\bibitem [{\citenamefont {Zhao}\ \emph {et~al.}(2018)\citenamefont {Zhao},
  \citenamefont {Pisarczyk}, \citenamefont {Thompson}, \citenamefont {Gu},
  \citenamefont {Vedral},\ and\ \citenamefont {Fitzsimons}}]{zhao2018geometry}%
  \BibitemOpen
  \bibfield  {author} {\bibinfo {author} {\bibfnamefont {Z.}~\bibnamefont
  {Zhao}}, \bibinfo {author} {\bibfnamefont {R.}~\bibnamefont {Pisarczyk}},
  \bibinfo {author} {\bibfnamefont {J.}~\bibnamefont {Thompson}}, \bibinfo
  {author} {\bibfnamefont {M.}~\bibnamefont {Gu}}, \bibinfo {author}
  {\bibfnamefont {V.}~\bibnamefont {Vedral}},\ and\ \bibinfo {author}
  {\bibfnamefont {J.~F.}\ \bibnamefont {Fitzsimons}},\ }\bibfield  {title}
  {\bibinfo {title} {Geometry of quantum correlations in space-time},\ }\href
  {https://doi.org/10.1103/PhysRevA.98.052312} {\bibfield  {journal} {\bibinfo
  {journal} {Physical Review A}\ }\textbf {\bibinfo {volume} {98}},\ \bibinfo
  {pages} {052312} (\bibinfo {year} {2018})}\BibitemShut {NoStop}%
\bibitem [{\citenamefont {Hoehn}\ \emph {et~al.}(2023)\citenamefont {Hoehn},
  \citenamefont {Russo},\ and\ \citenamefont {Smith}}]{hoehn2023matter}%
  \BibitemOpen
  \bibfield  {author} {\bibinfo {author} {\bibfnamefont {P.~A.}\ \bibnamefont
  {Hoehn}}, \bibinfo {author} {\bibfnamefont {A.}~\bibnamefont {Russo}},\ and\
  \bibinfo {author} {\bibfnamefont {A.~R.}\ \bibnamefont {Smith}},\ }\bibfield
  {title} {\bibinfo {title} {Matter relative to quantum hypersurfaces},\ }\href
  {https://arxiv.org/abs/2308.12912} {\bibfield  {journal} {\bibinfo  {journal}
  {arXiv preprint arXiv:2308.12912}\ } (\bibinfo {year} {2023})}\BibitemShut
  {NoStop}%
\bibitem [{\citenamefont {Anza}\ \emph {et~al.}(2018)\citenamefont {Anza},
  \citenamefont {Gogolin},\ and\ \citenamefont {Huber}}]{anza2018eigenstate}%
  \BibitemOpen
  \bibfield  {author} {\bibinfo {author} {\bibfnamefont {F.}~\bibnamefont
  {Anza}}, \bibinfo {author} {\bibfnamefont {C.}~\bibnamefont {Gogolin}},\ and\
  \bibinfo {author} {\bibfnamefont {M.}~\bibnamefont {Huber}},\ }\bibfield
  {title} {\bibinfo {title} {Eigenstate thermalization for degenerate
  observables},\ }\href {https://doi.org/10.1103/PhysRevLett.120.150603}
  {\bibfield  {journal} {\bibinfo  {journal} {Physical Review Letters}\
  }\textbf {\bibinfo {volume} {120}},\ \bibinfo {pages} {150603} (\bibinfo
  {year} {2018})}\BibitemShut {NoStop}%
\bibitem [{\citenamefont {Wilde}(2011)}]{wilde2011classical}%
  \BibitemOpen
  \bibfield  {author} {\bibinfo {author} {\bibfnamefont {M.~M.}\ \bibnamefont
  {Wilde}},\ }\bibfield  {title} {\bibinfo {title} {From classical to quantum
  shannon theory},\ }\href@noop {} {\bibfield  {journal} {\bibinfo  {journal}
  {arXiv preprint arXiv:1106.1445}\ } (\bibinfo {year} {2011})}\BibitemShut
  {NoStop}%
\bibitem [{\citenamefont {Lieb}\ \emph {et~al.}(1961)\citenamefont {Lieb},
  \citenamefont {Schultz},\ and\ \citenamefont {Mattis}}]{lieb1961two}%
  \BibitemOpen
  \bibfield  {author} {\bibinfo {author} {\bibfnamefont {E.}~\bibnamefont
  {Lieb}}, \bibinfo {author} {\bibfnamefont {T.}~\bibnamefont {Schultz}},\ and\
  \bibinfo {author} {\bibfnamefont {D.}~\bibnamefont {Mattis}},\ }\bibfield
  {title} {\bibinfo {title} {Two soluble models of an antiferromagnetic
  chain},\ }\href {https://doi.org/10.1016/0003-4916(61)90115-4} {\bibfield
  {journal} {\bibinfo  {journal} {Annals of Physics}\ }\textbf {\bibinfo
  {volume} {16}},\ \bibinfo {pages} {407} (\bibinfo {year} {1961})}\BibitemShut
  {NoStop}%
\bibitem [{\citenamefont {Canosa}\ and\ \citenamefont
  {Rossignoli}(2007)}]{canosa2007entanglement}%
  \BibitemOpen
  \bibfield  {author} {\bibinfo {author} {\bibfnamefont {N.}~\bibnamefont
  {Canosa}}\ and\ \bibinfo {author} {\bibfnamefont {R.}~\bibnamefont
  {Rossignoli}},\ }\bibfield  {title} {\bibinfo {title} {Entanglement between
  distant qubits in cyclic x x chains},\ }\href
  {https://doi.org/10.1103/PhysRevA.75.032350} {\bibfield  {journal} {\bibinfo
  {journal} {Physical Review A}\ }\textbf {\bibinfo {volume} {75}},\ \bibinfo
  {pages} {032350} (\bibinfo {year} {2007})}\BibitemShut {NoStop}%
\bibitem [{\citenamefont {Wilkinson}(1984)}]{wilkinson1984critical}%
  \BibitemOpen
  \bibfield  {author} {\bibinfo {author} {\bibfnamefont {M.}~\bibnamefont
  {Wilkinson}},\ }\bibfield  {title} {\bibinfo {title} {Critical properties of
  electron eigenstates in incommensurate systems},\ }\href
  {https://doi.org/10.1098/rspa.1984.0016} {\bibfield  {journal} {\bibinfo
  {journal} {Proceedings of the Royal Society of London. A. Mathematical and
  Physical Sciences}\ }\textbf {\bibinfo {volume} {391}},\ \bibinfo {pages}
  {305} (\bibinfo {year} {1984})}\BibitemShut {NoStop}%
\bibitem [{\citenamefont {Sinai}(1987)}]{sinai1987anderson}%
  \BibitemOpen
  \bibfield  {author} {\bibinfo {author} {\bibfnamefont {Y.~G.}\ \bibnamefont
  {Sinai}},\ }\bibfield  {title} {\bibinfo {title} {Anderson localization for
  one-dimensional difference schr{\"o}dinger operator with quasiperiodic
  potential},\ }\href {https://doi.org/10.1007/BF01011146} {\bibfield
  {journal} {\bibinfo  {journal} {Journal of statistical physics}\ }\textbf
  {\bibinfo {volume} {46}},\ \bibinfo {pages} {861} (\bibinfo {year}
  {1987})}\BibitemShut {NoStop}%
\bibitem [{\citenamefont {Bezanson}\ \emph {et~al.}(2017)\citenamefont
  {Bezanson}, \citenamefont {Edelman}, \citenamefont {Karpinski},\ and\
  \citenamefont {Shah}}]{bezanson2017julia}%
  \BibitemOpen
  \bibfield  {author} {\bibinfo {author} {\bibfnamefont {J.}~\bibnamefont
  {Bezanson}}, \bibinfo {author} {\bibfnamefont {A.}~\bibnamefont {Edelman}},
  \bibinfo {author} {\bibfnamefont {S.}~\bibnamefont {Karpinski}},\ and\
  \bibinfo {author} {\bibfnamefont {V.~B.}\ \bibnamefont {Shah}},\ }\bibfield
  {title} {\bibinfo {title} {Julia: A fresh approach to numerical computing},\
  }\href {https://doi.org/10.1137/141000671} {\bibfield  {journal} {\bibinfo
  {journal} {SIAM {R}eview}\ }\textbf {\bibinfo {volume} {59}},\ \bibinfo
  {pages} {65} (\bibinfo {year} {2017})}\BibitemShut {NoStop}%
\bibitem [{\citenamefont {Kingma}\ and\ \citenamefont
  {Ba}(2015)}]{kingma2015adam}%
  \BibitemOpen
  \bibfield  {author} {\bibinfo {author} {\bibfnamefont {D.~P.}\ \bibnamefont
  {Kingma}}\ and\ \bibinfo {author} {\bibfnamefont {J.}~\bibnamefont {Ba}},\
  }\bibfield  {title} {\bibinfo {title} {Adam: {A} method for stochastic
  optimization},\ }in\ \href {http://arxiv.org/abs/1412.6980} {\emph {\bibinfo
  {booktitle} {Proceedings of the 3rd International Conference on Learning
  Representations (ICLR)}}}\ (\bibinfo {year} {2015})\BibitemShut {NoStop}%
\bibitem [{\citenamefont {Hubbard}(1963)}]{hubbard1963electron}%
  \BibitemOpen
  \bibfield  {author} {\bibinfo {author} {\bibfnamefont {J.}~\bibnamefont
  {Hubbard}},\ }\bibfield  {title} {\bibinfo {title} {Electron correlations in
  narrow energy bands},\ }\href {https://doi.org/10.1098/rspa.1963.0204}
  {\bibfield  {journal} {\bibinfo  {journal} {Proceedings of the Royal Society
  of London. Series A. Mathematical and Physical Sciences}\ }\textbf {\bibinfo
  {volume} {276}},\ \bibinfo {pages} {238} (\bibinfo {year}
  {1963})}\BibitemShut {NoStop}%
\bibitem [{\citenamefont {Schollw{\"o}ck}(2011)}]{schollwock2011density}%
  \BibitemOpen
  \bibfield  {author} {\bibinfo {author} {\bibfnamefont {U.}~\bibnamefont
  {Schollw{\"o}ck}},\ }\bibfield  {title} {\bibinfo {title} {The density-matrix
  renormalization group in the age of matrix product states},\ }\href
  {https://www.sciencedirect.com/science/article/pii/S0003491610001752?via%3Dihub}
  {\bibfield  {journal} {\bibinfo  {journal} {Annals of physics}\ }\textbf
  {\bibinfo {volume} {326}},\ \bibinfo {pages} {96} (\bibinfo {year}
  {2011})}\BibitemShut {NoStop}%
\bibitem [{\citenamefont {Vidal}(2003)}]{Vidal2003Efficient}%
  \BibitemOpen
  \bibfield  {author} {\bibinfo {author} {\bibfnamefont {G.}~\bibnamefont
  {Vidal}},\ }\bibfield  {title} {\bibinfo {title} {Efficient classical
  simulation of slightly entangled quantum computations},\ }\href
  {https://doi.org/10.1103/PhysRevLett.91.147902} {\bibfield  {journal}
  {\bibinfo  {journal} {Phys. Rev. Lett.}\ }\textbf {\bibinfo {volume} {91}},\
  \bibinfo {pages} {147902} (\bibinfo {year} {2003})}\BibitemShut {NoStop}%
\bibitem [{\citenamefont {White}(1992)}]{white1992density}%
  \BibitemOpen
  \bibfield  {author} {\bibinfo {author} {\bibfnamefont {S.~R.}\ \bibnamefont
  {White}},\ }\bibfield  {title} {\bibinfo {title} {Density matrix formulation
  for quantum renormalization groups},\ }\href
  {https://doi.org/10.1103/PhysRevLett.69.2863} {\bibfield  {journal} {\bibinfo
   {journal} {Physical Review Letters}\ }\textbf {\bibinfo {volume} {69}},\
  \bibinfo {pages} {2863} (\bibinfo {year} {1992})}\BibitemShut {NoStop}%
\end{thebibliography}

\makeatletter
\close@column@grid
\makeatother
\cleardoublepage
\newpage
\onecolumngrid
\appendix

\renewcommand\figurename{Supplementary Figure}
\setcounter{figure}{0}


\section*{APPENDIXES}

Here we present derivations, additional details for the main results, and further general discussions on quantum time notions. The appendices are organized as follows: In Appendix \ref{sec:literat} we give an overview of quantum time approaches and describe connections between our work and the literature. In Appendix \ref{sec:equilibr} we explain in more detail the relation between quantum time and the equilibration problem and prove the related Theorems in the main body. 
In Appendix \ref{app:circ} we give detailed proofs of all  the different circuits we proposed in the main body. In Appendix \ref{app:proof-theo-5} we prove Theorem \ref{theo:scaling} about circuit depths in the sequential and parallel-in-time schemes.
In Appendix \ref{app:algebra} we prove Proposition \ref{prop:generating-set} about the algebra of the $XY$ model. In Appendix \ref{app:proofdepths} we prove Theorem \ref{theo:depths} regarding the depth of the total protocols involving both Hamiltonian diagonalization and history states. Finally, in Appendix \ref{App:methods} we provide details about the methods employed in the numerics (sec.\ \ref{sec:numerics}).

\section{Connection to literature and Overview of Quantum Time approaches}
\label{sec:literat}
Here we briefly discuss the broader conceptual picture of ``quantum time-related proposals''. This should also clarify the range of applicability of our current ideas. 
Regarding the Page and Wootters approach \cite{page1983evolution}, it is worth noting that their original idea was to replace dynamics with quantum correlations. The physical picture was a static universe from which evolution \emph{emerges} from a convenient separation between the system, and the rest. In this sense the ``time'' Hilbert space corresponds to a ``cosmological'' clock. One main motivation for these ideas was the discussion about time in quantum gravity \cite{kuchavr2011time}, where the Wheeler-deWitt equation \cite{dewitt1967quantum} suggests a static universe. 
Such equation appears as a constraint induced by the ``gauge'' freedom in the choice of coordinates in general relativity, and can be understood within the framework of Dirac's generalization of Hamiltonian dynamics to constrained systems \cite{dirac1950generalized, dirac1958theory}.

However, there are important conceptual and mathematical differences between the PaW's and Dirac's approaches which are at the core of our ability to leverage the first and not the second to develop useful computational schemes. To clarify this statement, let us briefly discuss a simple but archetypal \footnote{This is a common example in both canonical quantum gravity \cite{kiefer2012quantum} and string theory \cite{green2012superstring} textbooks} application of Dirac's approach:
in classical mechanics, one can accommodate time itself and its conjugate variable $p_t$ in an extended phase-space by introducing a new variable $\tau$ which parametrizes phase-space variables, including $t=t(\tau)$, $p_t=p_t(\tau)$. The Poisson brackets are also extended by imposing $\{t,p_t\}=1$ with $p_t\neq H$ which for a single particle and $x^0\equiv t$, $p_t\equiv p_0$ completes the algebra $\{x^\mu,p_\nu\}=\delta^\mu_{\;\nu}$, which now is explicitly covariant. Part of the quantization scheme now consists of the replacement
\begin{equation}\label{eq:extalg}
    \{x^\mu,p_\nu\}=\delta^\mu_{\;\nu}\to [x^\mu,p_\nu]=i\hbar\delta^\mu_{\;\nu}
\end{equation}
which for $\mu=\nu=0$ is just $[T,P_T]=i\hbar$  (when the system is a particle), i.e. $x^0\equiv T$ with $[T,H]=0$ for $H$ some Hamiltonian of the particle. On the other hand, the independence of physical quantities on the way the $\tau$-parametrization is chosen leads to $\mathcal{J}|\Psi\rangle=0$ after quantization. This reparametrization invariance  is analogous to the general covariance of general relativity, and the constraint is analogous to the Wheeeler-deWitt equation.

However, contrary to our main body discussion, in Dirac's approach 
the constraint equation defines the so-called physical Hilbert space which is regarded as different from the previous ``kinematic'' Hilbert space. Moreover, the physical Hilbert space is not treated as a subspace of the latter, instead the kinematical one appears only as an auxiliary step in the whole quantization process but not in the final construction. As a consequence, the time operator is ruled out in the final formalism for not being a physical observable (see e.g. \cite{marolf1995quantum} for discussions about this procedure and the related ``Hilbert space problem'').  As we have shown in the main body, in the PaW approach the interpretation and treatment of the constraint are completely different: the time operator is not disregarded since it corresponds to an actual observable of the clock system. The complete Hilbert space e.g. as defined by \eqref{eq:extalg} is preserved, while the state of the system is recovered by conditioning. 
The theoretical ideas underlying our manuscript clearly exploit the ``extended'' Hilbert space associated with the PaW approach and are mostly inspired by the recent developments in the context of quantum information \cite{giovannetti2015quantum, boette2016system, diaz2019history, diaz2019historystate}, and in particular in the discrete-time formulation \cite{boette2016system}, which was further extended in \cite{boette2018history}. 

Let us also stress that for the purposes of our manuscript it is not relevant whether explicit covariance is achieved via the definition of a time operator. This means that we can consider arbitrary many-body systems as we have shown in the main body. This also includes relativistic systems (properly discretized quantum field theories) but their simulation  corresponds to the evolution as seen from a fixed reference frame: in this case there is no simple rule to relate a history state in a given reference frame to another. In this sense, we regard the PaW formalism as non-explicitly-relativistic. There is however a simple situation where such rule is straightforward: for a single relativistic particle Lorentz transformations can be introduced explicitly and geometrically (as space-time ``rotations'' independently of the theory) \cite{diaz2019history,diaz2019historystate}. For this to be actually achieved one needs to preserve the kinematical space and deal with an extended inner product, a result which is again highlighting the advantages of the extended scheme. This approach was recently developed in \cite{diaz2019history} for Dirac's particles and in \cite{diaz2019historystate} for scalar particles where the new inner products were successfully related to the ``physical'' ones while preserving the extended Hilbert spaces. As a consequence, the associated history states can still be realized by the means presented in this manuscript (adopting some discretization of space-time), and in principle, Lorentz transformations may be introduced as non-local gates acting on both the system and time qubits. 

Another interesting feature of these single-particle quantum time formalisms (which preserve the extended Hilbert space) is that their ``second quantization'', as introduced in \cite{diaz2019historystate, diaz2021spacetime, diaz2021path} naturally leads to a new approach to many-body scenarios and to quantum field theories in which space and time are \emph{explicitly} on equal footing (see also the closely related ``quantum mechanics of events'' proposal \cite{giovannetti2023geometric}). In particular, the approach in \cite{diaz2021spacetime} also leads to a redefinition of the Path Integral formulation \cite{diaz2021path}. In fact, the real challenge to treat space and time on equal footing at the Hilbert space level  in  relativistic settings has to do with another more subtle asymmetry in the treatment of time \cite{isham1994quantum,isham1995continuous, diaz2021spacetime,diaz2021path}: joint systems separated in space are described by the tensor product of the corresponding Hilbert spaces. No such rule is applied to time. This asymmetry is particularly evident in the case of quantum field theories, in which case space is treated as a site, which e.g. for bosons is equivalent to a tensor product structure in space. While time is a parameter, it is not a site index and there is no associated tensor product structure. This is manifest in the \emph{equal time} canonical  algebra imposed on the fields which requires a fixed foliation. We notice also that this is an obstacle in defining a notion of ``time-like entanglement'' (see however \cite{diaz2021spacetime,diaz2021path, harper2023timelike,narayan2023notes,chu2023time} for recent related discussions), 
 a consideration which applies to all QM: the previous asymmetry  is present in any quantum mechanical system, as also discussed in \cite{isham1994quantum,zhao2018geometry,horsman2017can,cotler2018superdensity,diaz2021spacetime,diaz2021path}. 

For these reasons, it is not sufficient to define a quantum time operator to solve ``the problem of time'' or more precisely to have an explicit space-time symmetric version of QM. 
This renders the aforementioned second quantization of ``PaW particles'' particularly interesting and relevant in the context of \emph{quantum field theories}, as shown recently in \cite{diaz2023spacetime, diaz2025spacetime} (for a different extension of the PaW scheme, developed along the lines of parameterized field theories see \cite{hoehn2023matter}). While the description of the results relating the PaW mechanism with field theories exceeds the purpose of this manuscript, we can speculate that these developments may provide additional quantum computational and informational tools, just as the PaW approach inspired the various algorithms of our manuscript.

\section{More on the Relation between Quantum Time and the Equilibration Problem: Proof of Theorem \ref{theo:deph}}\label{sec:equilibr}

In this section we highlight the relations between the current quantum time approach (the PaW-inspired formalism), its associated concept of system-time entanglement, and the problem of equilibration of an isolated quantum system. In particular, we provide a prove and analysis of Theorem \ref{theo:deph}. Let us mention that this connection has not been explored in the literature (see however the recent article \cite{favalli2022peaceful}).

Let us recall first that a system is said to equilibrate 
if it evolves towards some particular in general mixed state
 and remains in that state or
close to it for almost all times \cite{linden2009quantum}. More precisely, this holds for either a small subsystem of a large quantum system evolving unitarily \cite{linden2009quantum} (subsystem equilibration), or for expectation values associated with ``experimentally realistic conditions'' \cite{reimann2008foundation} (observable equilibration). The last interpretation refers to the fact that the probability  that  mean values computed with $\dya{\psi(t)}$ differ from  mean values computed with the equilibrium state becomes exponentially vanishing for typical experimentally accessible observables \cite{reimann2008foundation, anza2018eigenstate}. 
In both scenarios, the dephased state $\bar{\rho}$ defined in Eq.~\eqref{eq:averagedst} plays a central role since it is identified with the equilibrium state (equivalently, the subsystems equilibrium state is a partial trace of this global state). Moreover, the theoretical bound \cite{reimann2008foundation} used to prove this is related to $\bar{\LC}$ precisely because $
    \Tr [\bar{\rho}^2]=\bar{\LC}\,,
$
i.e. the purity of the time-averaged state is $\bar{\LC}$. The same quantity provides a bound for the (time average of) the distance between the subsystem state and (the partial trace of) $\bar{\rho}$ \cite{linden2009quantum}.

Now returning to the history state, we have seen that the quadratic entanglement involves the purity of $\rho_S=\Tr_T[ |\Psi\rangle \langle \Psi|]$. The explicit expansion of $\rho_S$ is
\begin{equation}\label{eq:rhos}
    \rho_S=\frac{1}{N}\sum_{t} |\psi(\varepsilon t)\rangle \langle \psi(\varepsilon t)|\,.
\end{equation}
In order to relate this state with $\bar{\rho}$  we can use the energy basis $|k\rangle$ so that $H|k\rangle=E_k |k\rangle$. Any initial state has an expansion $|\psi_0\rangle=\sum_k c_k |k\rangle$. Notice that even if there is degeneracy  we can always write a fixed pure state as before, with
$
    |c_k|^2:=\sum_j |\psi_{kj}|^2\,,\;\;\; |k\rangle:=c_k^{-1}\sum_j \psi_{kj}|kj\rangle\,,
$
and $H|kj\rangle=E_k |kj\rangle$, i.e. $j$ is a degeneracy index.
The quantity $c_k |k\rangle$ is the projection of $|\psi\rangle$ onto the subspace of states with energy $E_k$. Thus we obtain
\begin{equation}
    \rho_S=\frac{1}{N}\sum_t \sum_{k,k'} c_k c_{k'}^\ast e^{-i(E_k-E_{k'})t\varepsilon}|k\rangle \langle k'|\,,
\end{equation}
which is precisely the discrete time version of $\bar{\rho}$ as defined in Eq.~\ref{eq:averagedst}, i.e.
\begin{equation}
    \rho_S=\tilde{\rho}\,.
\end{equation}
Notice that while the purity of $\bar{\rho}$ coincides with $\bar{\LC}$, the purity of $\rho_S$ is not $\widetilde{\LC}$. Remarkably, the natural discrete-time generalization of $\bar{\LC}=\Tr [\bar{\rho}^2]$ is  $\Tr [\rho_S^2]$ and not $\widetilde{\mathcal{L}}$: the system-time entanglement provides the proper discrete-time version of time average related bounds (it is worth remarking that the \emph{discrete-time entanglement} provides strict bounds to the \emph{infinite and continuum} time averages). 
This is captured by Theorem \ref{theo:deph} and its corollaries.

We are now in a position to give the proof of Theorem \ref{theo:deph}:
\begin{proof}
Let us first rewrite $\rho_S$ as
\begin{equation}\rho_S=\sum_{k,k'} c_k c_{k'}^\ast \Delta_{kk'}|k\rangle \langle k'|\,,\end{equation} with $\Delta_{kk'}:=\frac{1}{N}\sum_t e^{-i(E_k-E_{k'})t\varepsilon}$. Notice that $\Delta_{kk}=1$. Being a quantum state, we can diagonalize $\rho_S$ in some basis as
\begin{equation}\label{eq:rhosdelta}
    \rho_S=\sum_l p_l |l\rangle \langle l|\,.
\end{equation}
In fact, it is clear that this is what is obtained by tracing  over the clocks with $|\Psi\rangle$ in its Schmidts decomposition \eqref{eq:schmidt}. 

We can combine these two expansions to write
\begin{equation}
    \langle k|\rho |k\rangle=|c_k|^2=\sum_l p_l |\langle l|k\rangle|^2\,.
\end{equation}
The quantity $|\langle l|k\rangle|^2$ defines a double stochastic matrix thus yielding the desired majorization relation
\begin{equation}
    \{|c_k|^2\}\prec \{ p_l\}
\end{equation}
thus implying the desired majorization relation between states (we recall that $\bar{\rho}=\sum_k |c_k|^2 |k\rangle \langle k|$).
Notice that similar ideas have been used in \cite{boette2016system}. 

It is also clear that by taking the limits $\varepsilon\to 0$ and then $T\to \infty$ the energy dephasing is recovered: the first limit can be considered by writing $$\Delta_{kk'}=\frac{1}{\varepsilon N}\sum_t \varepsilon\, e^{-i(E_k-E_{k'})t\varepsilon}\to \int_0^T \frac{dt}{T}e^{-i(E_k-E_{k'})t}\,,$$
with $dt\equiv \varepsilon$. The large $T$ limit now is the familiar limit used conventionally (which yields a delta), thus leading again to the asymptotic relation
\begin{equation}
  \rho_S\to \bar{\rho} \,. 
\end{equation}

Instead, in the periodic case we can use the fact that the periodicity condition requires $E_k\equiv E_{l_k}=2\pi l_k/T$ for $l_k$ an integer. This means that $$\Delta_{kk'}=\frac{1}{N}\frac{1-e^{i (E_k-E_{k'})T}}{1-e^{i (E_k-E_{k'})\varepsilon}}=\delta_{kk'}$$
with the last equality holding only in the periodic case. By replacing this in Eq.\ \eqref{eq:rhosdelta} we obtain precisely $\rho_S=\bar{\rho}$.

\end{proof}
Notice that the Corollaries \ref{theo:ent-loschmidt} and \ref{theo:ent-periodic} follow immediately from the Schur-concavity of functions defining any entropy. These Corollaries are the particular case of the linear entropy. Interestingly, new bounds can be obtained by just considering other entropies.

Let us also make more precise the main body statement that the quantum time formalism gives a new interpretation to the loss of coherences induced by a time average: as we said, since the system is ``entangled with time'',  by ignoring the ``clock qubits'' we lose information. For discrete time, the information loss induces $\rho\to \tilde{\rho}$ which is a \emph{quantum channel} with Krauss operators $K_t:=e^{-iHt\varepsilon}/\sqrt{N}$ such that
\begin{equation}\label{eq:channel}
    \tilde{\rho}=\sum_t K_t^\dag \rho K_t\,.
\end{equation}
As usual this quantum channel can be purified by using the isometric extension \cite{wilde2011classical} $U[K_t]:=\sum_t |t\rangle \otimes K_t$ so that the channel is recovered by tracing over the ``environment'' (here the clock) in a global state $U[K_t] |\psi\rangle$.  This global state is precisely the history state of Eq.~\eqref{eq:phystate}, i.e.,
\begin{equation}
   |\Psi\rangle=U[K_t] |\psi_0\rangle\,.
\end{equation}
We see that one can ``rediscover'' the quantum time formalism from the natural purification of the channel~\eqref{eq:channel}. Just as in the general case, where a nontrivial channel is induced by correlations between the system and an environment, the system is correlated (entangled) with time. 
Moreover, the quantum channel's theory  \cite{wilde2011classical} implies that there is a unitary $V$ such that
\begin{equation}
    |\Psi\rangle=V|0\rangle\otimes |\psi\rangle\,.
\end{equation}
One possible unitary is provided by the circuit of figure \ref{fig:hist-state}.
We should remark however that the operation of tracing over the clock degrees of freedom is very different from measuring on the clock register and conditioning the state of the system. When conditioning one has access to the clock, as it is required for implementing a projection at a given time state. In summary, the history state contains both information about evolution at specific times, recovered from conditioning, and about the ``equilibration channel'' \eqref{eq:channel}, recovered by ignoring the clock.

\section{Proofs involving circuits}\label{app:circ}
\subsection{Proof of Proposition ~\ref{prop:sequential}}
\begin{proof}
Consider the circuit in Fig.~\ref{fig:circuit-F-sequential}. For ease of calculation, we will assume that $\rho=\dya{\psi_0}$, for some initial state $\ket{\psi_0}$. Moreover, we recall that both $O_1$ and $O_2$ are Pauli operators. The input state to the circuit is $\ket{\psi_0}\ket{0}$, where the single qubit state initialized in the zero state is an ancilla used to perform a Hadamard test. The action of the Hadamard gates is to map
\begin{equation}
    \ket{\psi_0}\ket{0}\rightarrow\ket{\psi_0}\ket{+}\,.
\end{equation}
Assuming the colored dashed gate in Fig.~\eqref{fig:circuit-F-sequential} is an identity, we next have a controlled $O_2$ operation. This produces the state
\begin{equation}
   \ket{\psi_0}\ket{+}\rightarrow\frac{1}{\sqrt{2}}(\ket{\psi_0}\ket{0}+O_2\ket{\psi_0}\ket{1})\,.
\end{equation}
Next, in the $t$-th experiment, we evolve the state with a unitary $U(\varepsilon t)$, leading to 
\begin{equation}
    \frac{1}{\sqrt{2}}(\ket{\psi_0}\ket{0}+O_2\ket{\psi_0}\ket{1})\rightarrow\frac{1}{\sqrt{2}}(U(\varepsilon t)\ket{\psi_0}\ket{0}+U(\varepsilon t)O_2\ket{\psi_0}\ket{1})\,.
\end{equation}
The next controlled $O_1$ gate leads to 
\begin{equation}
    \frac{1}{\sqrt{2}}(U(\varepsilon t)\ket{\psi_0}\ket{0}+U(\varepsilon t)O_2\ket{\psi_0}\ket{1})\rightarrow\frac{1}{\sqrt{2}}(U(\varepsilon t)\ket{\psi_0}\ket{0}+O_1 U(\varepsilon t)O_2\ket{\psi_0}\ket{1})\,.
\end{equation}
The final Hadamard gate produces the state
\small
\begin{align}
    \frac{1}{\sqrt{2}}(U(\varepsilon t)\ket{\psi_0}\ket{0}+O_1 U(\varepsilon t)O_2\ket{\psi_0}\ket{1})\rightarrow&\frac{1}{2}(U(\varepsilon t)\ket{\psi_0}(\ket{0}+\ket{1})+O_1 U(\varepsilon t)O_2\ket{\psi_0}(\ket{0}-\ket{1}))\nonumber\\
    =&\frac{1}{2}\left((U(\varepsilon t)\ket{\psi_0}+O_1 U(\varepsilon t)O_2\ket{\psi_0})\ket{0}+(U(\varepsilon t)\ket{\psi_0}-O_1 U(\varepsilon t)O_2\ket{\psi_0})\ket{1}\right)\,.
\end{align}
\normalsize
Then, the probability of measuring the ancilla qubit in the zero state is 
\begin{align}
    p(0)&=\frac{1}{4}\left((\bra{\psi_0}U\ad(\varepsilon t)+\bra{\psi_0}O_2 U\ad(\varepsilon t)O_1)\right)\left((U(\varepsilon t)\ket{\psi_0}+O_1 U(\varepsilon t)O_2\ket{\psi_0})\right)\nonumber\\
    &=\frac{1}{4}(2+ \bra{\psi_0}U\ad(\varepsilon t)O_1U(\varepsilon t)O_2\ket{\psi_0}+\bra{\psi_0}O_2 \ad U\ad(\varepsilon t)O_1\ad U(\varepsilon t)\ket{\psi_0})\nonumber\\
    &=\frac{1}{2}(1+
    \Re[\langle O_1(t)O_2\rangle_{\psi_0}])\,.\label{eq:prob-anc-zero-1}
\end{align}
Similarly, the probability of measuring the ancilla qubit in the one state is 
\begin{align}
    p(1)&=\frac{1}{2}(1-
     \Re[\langle O_1(t)O_2\rangle_{\psi_0}])\,.\label{eq:prob-anc-one-1}
\end{align}
Combining Eqs.~\eqref{eq:prob-anc-zero-1} and~\eqref{eq:prob-anc-one-1} shows that the expectation value of the $Z$ operator on the ancilla qubit is
\begin{align}
    \langle Z\rangle=p(1)-p(0)&=
    \Re[\langle O_1(t)O_2\rangle_{\psi_0}]\,.
\end{align}
By adding, an $S\ad$ gate in place of the colored dashed gate in Fig.~\ref{fig:circuit-F-sequential} one finds 
\begin{align}
    \langle Z\rangle=
    \Im[ \langle O_1(t)O_2\rangle_{\psi_0}]\,.
\end{align}

Since this procedure needs to be repeated $N$ times, and since we want to estimate the expectation values up to precision $\delta$, then one needs to perform $\OC(N/\delta^2)$ experiments. 
\end{proof}

\subsection{Proof of Theorem~\ref{theo:parallel1}}
\begin{proof}
Consider the circuit in Fig.~\ref{fig:circuit-F-parallel}. For ease of calculation, we will assume that $\rho=\dya{\psi_0}$ for some initial state $\ket{\psi_0}$. We also recall that  both $O_1$ and $O_2$ are Pauli operators. The input state to the circuit is $\ket{0}^{\log{N}}\ket{\psi_0}\ket{0}$, where we recall that $\ket{0}^{\log{N}}$ is the initial state of the clock-qubits. The action of the Hadamard gates is to map
\begin{equation}
    \ket{0}^{\log{N}}\ket{\psi_0}\ket{0}\rightarrow\frac{1}{\sqrt{N}}\sum_{t=0}^{N-1}\ket{t}\ket{\psi_0}\ket{+}\,.
\end{equation}
Here we have used the identity $\ket{+}^{ \log(N)}=\frac{1}{\sqrt{N}}\otimes_{j=1}^{\log{N}}(\ket{0_j}+\ket{1_j})=\frac{1}{\sqrt{N}}\sum_{t=0}^{N-1}\ket{t}$ which follows by expressing $t$ in its binary form $t=\sum_{j=1}^{\log{N}}t_j2^{j-1}$. Assuming the colored dashed gate in Fig.~\ref{fig:circuit-F-parallel} is an identity, we next have a controlled $O_2$ operation. This produces the state
\begin{equation}
    \frac{1}{\sqrt{N}}\sum_{t=0}^{N-1}\ket{t}\ket{\psi_0}\ket{+}\rightarrow\frac{1}{\sqrt{2N}}\sum_{t=0}^{N-1}\ket{t}(\ket{\psi_0}\ket{0}+O_2\ket{\psi_0}\ket{1})\,.
\end{equation}
Next, the sequence of  $\log{N}$ controlled gates $U(2^{j-1}\frac{T}{N})=U(\frac{T}{N})^{2^{j-1}}$ for $j=1,\ldots,\log{N}$ perform the operations
\begin{equation}
    \frac{1}{\sqrt{2N}}\sum_{t=0}^{N-1}\ket{t}(\ket{\psi_0}\ket{0}+O_2\ket{\psi_0}\ket{1})\rightarrow\frac{1}{\sqrt{2N}}\sum_{t=0}^{N-1}\ket{t}(U(\varepsilon t)\ket{\psi_0}\ket{0}+U(\varepsilon t)O_2\ket{\psi_0}\ket{1})\,.
\end{equation}
The next controlled $O_1$ gate leads to 
\begin{equation}
    \frac{1}{\sqrt{2N}}\sum_{t=0}^{N-1}\ket{t}(U(\varepsilon t)\ket{\psi_0}\ket{0}+U(\varepsilon t)O_2\ket{\psi_0}\ket{1})\rightarrow\frac{1}{\sqrt{2N}}\sum_{t=0}^{N-1}\ket{t}(U(\varepsilon t)\ket{\psi_0}\ket{0}+O_1U(\varepsilon t)O_2\ket{\psi_0}\ket{1})\,,
\end{equation}
while the controlled phase gates perform the map 
\begin{equation}
    \frac{1}{\sqrt{2N}}\sum_{t=0}^{N-1}\ket{t}(U(\varepsilon t)\ket{\psi_0}\ket{0}+O_1U(\varepsilon t)O_2\ket{\psi_0}\ket{1})\rightarrow\frac{1}{\sqrt{2N}}\sum_{t=0}^{N-1}( \ket{t}U(\varepsilon t)\ket{\psi_0}\ket{0}+e^{-i \omega \epsilon t}\ket{t} O_1U(\varepsilon t)O_2\ket{\psi_0}\ket{1})\,.
\end{equation}
The final Hadamard gate gives rise to 
\small
\begin{align}
    \frac{1}{\sqrt{2N}}\sum_{t=0}^{N-1}( \ket{t}U(\varepsilon t)\ket{\psi_0}\ket{0}+e^{-i \omega \varepsilon t} \ket{t}O_1U(\varepsilon t)O_2\ket{\psi_0}\ket{1})\rightarrow&\frac{1}{2\sqrt{N}}\sum_{t=0}^{N-1}( \ket{t}U(\varepsilon t)\ket{\psi_0}(\ket{0}+\ket{1})+e^{-i \omega \varepsilon t}\ket{t} O_1U(\varepsilon t)O_2\ket{\psi_0}(\ket{0}-\ket{1}))\nonumber\\
    =&\frac{1}{2\sqrt{N}}\sum_{t=0}^{N-1}( \ket{t}U(\varepsilon t)\ket{\psi_0}+e^{-i \omega \varepsilon t}\ket{t} O_1U(\varepsilon t)O_2\ket{\psi_0})\ket{0} \nonumber\\
    &+ (\ket{t} U(\varepsilon t)\ket{\psi_0}-e^{-i \omega \varepsilon t}\ket{t} O_1U(\varepsilon t)O_2\ket{\psi_0})\ket{1}\,.
\end{align}
\normalsize
Thus, the probability of measuring the ancilla qubit in the zero state is 
\begin{align}
    p(0)&=\frac{1}{4N}\sum_{t=0}^{N-1}\sum_{t'=0}^{N-1}( \bra{t'}\bra{\psi_0}U\ad(t')+e^{i \omega \varepsilon t'} \bra{t'}\bra{\psi_0}O_2\ad U\ad(t')O_1\ad)( \ket{t}U(\varepsilon t)\ket{\psi_0}+e^{-i \omega \varepsilon t}\ket{t} O_1U(\varepsilon t)O_2\ket{\psi_0})\nonumber\\
    &=\frac{1}{4N}\sum_{t=0}^{N-1}(2+ e^{-i \omega \varepsilon t}\bra{\psi_0}U\ad(\varepsilon t)O_1U(\varepsilon t)O_2\ket{\psi_0}+e^{i \omega \varepsilon t} \bra{\psi_0}O_2 \ad U\ad(\varepsilon t)O_1\ad U(\varepsilon t)\ket{\psi_0})\nonumber\\
    &=\frac{1}{2}(1+
    \Re[\widetilde{F}(O_1,O_2,\omega)])\,.\label{eq:prob-anc-zero}
\end{align}
Similarly, the probability of measuring the ancilla qubit in the one state is 
\begin{align}
    p(1)&=\frac{1}{2}(1-
    \Re[\widetilde{F}(O_1,O_2,\omega)])\,.\label{eq:prob-anc-one}
\end{align}
Combining Eqs.~\eqref{eq:prob-anc-zero} and~\eqref{eq:prob-anc-one} shows that the expectation value of the $Z$ operator on the ancilla qubit is
\begin{align}
    \langle Z\rangle=p(1)-p(0)&=
    \Re[\widetilde{F}(O_1,O_2,\omega)]\,.
\end{align}
By adding, an $S\ad$ gate in place of the colored dashed gate in Fig.~\ref{fig:circuit-F-parallel} one finds 
\begin{align}
    \langle Z\rangle=
    \Im[\widetilde{F}(O_1,O_2,\omega)]\,.
\end{align}

If, one wishes to estimate $\langle Z\rangle$ up to precision $\delta$, then one needs to perform $\OC(1/\delta^2)$ measurements. This results in $\OC(1/\delta^2)$ experiments. 
\end{proof}

\subsection{Proof of Proposition~\ref{prop:sequential-losch}}

\begin{proof}
Consider the circuit in Fig.~\ref{fig:circuit-loschmidt-sequential}. We can readily see that the circuit therein is nothing but the circuit for computing the overlap between two quantum state $\rho$ and $\sigma$ derived in Ref.~\cite{cincio2018learning} and also shown in Sup. Fig.~\ref{fig:state-overlap}. 

\begin{figure}[t!]
\centering
\includegraphics[width=.6\linewidth]{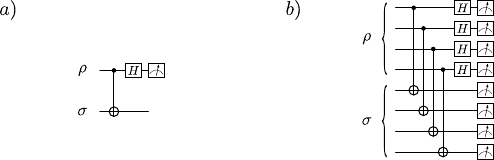}
\caption{\textbf{Algorithm for computing the overlap between two quantum states.}  We show an algorithm which takes as input two arbitrary $n$-qubit quantum states $\rho$ and $\sigma$ and which estimates the overlap $\Tr[\rho\sigma]$. In a) we show the algorithm for the case when $\rho$ and $\sigma$ and single qubit states, and in b) the generalization for larger qubit sizes. We can see that in all cases the circuit depth is equal to two, and hence independent of $n$.   }
\label{fig:state-overlap}
\end{figure}

To understand the algorithm of Ref.~\cite{cincio2018learning}, let us consider first the case of $\rho$ and $\sigma$ being single qubit states. We know that the following identity holds
\begin{equation}
    \Tr[\rho\sigma]=\Tr[(\rho\otimes \sigma) {\rm SWAP}]\,,
\end{equation}
where ${\rm SWAP}$ denotes the SWAP operator, whose action can be defined in the computational basis as  ${\rm SWAP}\ket{ij}=\ket{ji}$. Then, let us define the Bell basis:
\begin{equation}
    \ket{\Phi_1}=\frac{1}{\sqrt{2}}(\ket{01}+\ket{10})\,,\quad \ket{\Phi_2}=\frac{1}{\sqrt{2}}(\ket{00}+\ket{11})\,,\quad
    \ket{\Phi_3}=\frac{1}{\sqrt{2}}(\ket{00}-\ket{11})\,,\quad
    \ket{\Phi_4}=\frac{1}{\sqrt{2}}(\ket{01}-\ket{10})\,.
\end{equation}
It is not hard to see that the SWAP operator is diagonal in the Bell basis, and that it can be expressed as 
\begin{equation}
    {\rm SWAP}=\dya{\Phi_1}+\dya{\Phi_2}+\dya{\Phi_3}-\dya{\Phi_4}\,.
\end{equation}
Thus, we can estimate the expectation value of the SWAP operator over the state $\rho\otimes \sigma$ 
\begin{equation}
    \langle {\rm SWAP}\rangle_{\rho\otimes \sigma}=\Tr[(\rho\otimes \sigma){\rm SWAP}]=P(\Phi_1)+P(\Phi_2)+P(\Phi_3)-P(\Phi_4)\,,
\end{equation}
where here we defined the probabilities $P(\Phi_i)=\langle \Phi_i|\rho\otimes \sigma|\Phi_i\rangle$. That is, $P(\Phi_i)$ denotes the probability of measuring  the state $\rho\otimes \sigma$ in the Bell basis, and obtaining the measurement outcome $\ket{\Phi_i}$. This result shows that we cane estimate $\Tr[\rho\sigma]$ by preparing $\rho\otimes \sigma$  and measuring in the Bell basis. Crucially, one can readily prove the such a measurement can be performed with the circuit in Sup. Fig.~\ref{fig:state-overlap}(a), which is composed of a $CNOT$ gate and a Hadamard gate in the first qubit (i.e., with the inverse of the circuit used to prepare a Bell state).

A similar result will follow when $\rho\otimes \sigma$ are $n$-qubit states, but now one has to measure the expectation value of the operator $ \prod_{j=1}^n{\rm SWAP}_{j_A,j_B}$, where here ${\rm SWAP}_{j_A,j_B}$ denotes the operator that swaps the $j$-th qubit of $\rho$ with the $j$-th qubit of $\sigma$. Since each operator ${\rm SWAP}_{j_A,j_B}$ can be expanded in a (local) Bell basis, the circuit used to measure in the eigenbasis of $ \prod_{j=1}^n{\rm SWAP}_{j_A,j_B}$, and hence to estimate $\Tr[\rho\sigma]$ is precisely that which is shown in Sup. Fig.~\ref{fig:state-overlap}(b). Finally, since we are computing an expectation value, if we want to reach a precision $\delta$ for each $N$, we need to run $\OC(N/\delta^2)$ experiments. 

\end{proof}

\subsection{Proof of Theorem~\ref{theo:parallel-losch}}

\begin{proof}

The proof of this theorem follows very closely that of Proposition~\ref{prop:sequential-losch}. In particular, we can see from Fig.~\ref{fig:circuit-loschmidt-parallel} that we are performing a SWAP test, i.e., measuring the overlap via Bell basis measurements, between the reduced state $\rho_S$ of the history state on the system qubits, and the state $\dya{\psi_0}$. On the other hand, one can readily see by expanding $\rho_S$ as in Eq.\ \eqref{eq:rhos} that the overlap between $\rho_S$ and $\dya{\psi_0}$ is precisely $\widetilde{\LC}(\psi_0)$. Hence, we can use the circuit in Fig.~\ref{fig:circuit-loschmidt-parallel} to estimate $\widetilde{\LC}(\psi_0)$ up to $\delta$ precision with $\OC(1/\delta^2)$ experiments.

\end{proof}

\section{Proof of Theorem \ref{theo:scaling}}\label{app:proof-theo-5}
\begin{proof}
Let us remark first that the gates we consider here are only those directly related to implementing the unitary evolution. The other parts of the circuit are common to both the sequential and parallel in time approaches, meaning that it is advantageous to have less runs (thus the parallel approach is always more convenient). The only exception is when we add the phases in Figure \ref{fig:circuit-F-parallel} but this only adds $\log(N)$ gates which is negligible in this analysis. A fully explicit depth analysis taking into account all gates for each circuit is provided in Appendix \ref{app:proofdepths} in the context of the variational diagonalization scheme.

Let us discuss first the conventional sequential approach. 
The maximum number of gates one requires in a single run is the ``last one'' corresponding to $U(T)$. In this case, one needs $\#_g(T)\in \gamma l T^\alpha$ gates. A more relevant quantity is the total number of gates $\#^{\text{seq}}$ one needs to implement since this will determine the actual computational time. By summing over the amount of gates of each run one obtains 
\begin{equation}
 \#^{\text{seq}}=\gamma l \sum_{t=1}^{N-1}(\varepsilon t)^\alpha\in \gamma l\mathcal{O}(N^{\alpha+1})  \,,
\end{equation}
where we retain the information on $l$ for a proper comparison.

Now let us consider the parallel-in-time approach. In this scheme each $U(t)$ is first ``Trotterized'' and each individual gate is controlled. This gives an additional overall multiplicative factor proportional to $l$, say $\beta l$. For local Hamiltonians, this is logarithmic in the system size. On the hand, the parallelization reduces the number of evolution gates exponentially. The total number of gates $\#^{\text{par}}$ is in fact now
\begin{equation}
    \#^{\text{par}}=\gamma \beta l^2 \sum_{j=0}^{\log(N)-1}(\varepsilon 2^j)^\alpha\in  \beta \gamma l^2\mathcal{O}( N^\alpha)\,.
\end{equation}

Notice that contrary to the sequential case, in the parallel in time scheme we sum over a logarithmic amount of times, with $U(T/2)$ the gate with the longest evolution in the whole protocol ($T/2=\varepsilon 2^{\log(N)-1}$). 
By comparing the previous equations the last part of the Theorem immediately follows.

\end{proof}

\section{Proof of proposition~\ref{prop:generating-set}} \label{app:algebra}

\begin{proof}
Note that $\mf{k}$ admits a \textit{bodyness-graded} basis
\small
\begin{align}
B = \bigcup_{k=2}^n B_k \,,\quad \text{where $B_k=\{ \widehat{X_i Y}_{i+k-1} , \widehat{Y_i X}_{i+k-1} \}_{i=1}^{n-k} $}
\end{align}
\normalsize
Clearly, $|B_k|=2(n-k)$ and $2 \sum_{k=2}^n (n-k) = 2 {n \choose 2} =  n(n-1) = \dim(\mf{k})$. The claim is that the set $B_2\subset B$ generates $\mf{k}$.

We prove this by induction. First we show $B_2$ generates $B_3$. For all $i\in 1,\ldots, n-2$ we obtain each element in $B_3$ from
\begin{align}
&[X_iY_{i+1},X_{i+1}Y_{i+2}] \propto X_i Z_{i+1} Y_{i+2} \\
&[Y_iX_{i+1},Y_{i+1}X_{i+2}] \propto Y_i Z_{i+1} Z_{i+2}
\end{align}
Now lets see that, in general, $B_2$ and $B_k$ suffice to generate $B_{k+1}$. Specifically, each element in $B_{k+1}$ follows from
\begin{align}
&[X_iY_{i+1}, \widehat{X_{i+1}Y}_{i+k-1}] \propto \widehat{X_i Y}_{i+k} \\
&[Y_iX_{i+1},\widehat{Y_{i+1}X}_{i+k-1}] \propto\widehat{Y_i X}_{i+k}
\end{align}
for $i\in 1\ldots n-(k-1)$.
\end{proof}

\section{Proof of Theorem~\ref{theo:depths}}\label{app:proofdepths}

\begin{proof}
In what follows we will assume that the history state preparation subroutine is replaced as in Fig.~\ref{fig:diag-history} where $W$ and $D$ are respectively given by the optimized ansatzes $W(\vec{\alpha})$ and $W(\vec{\beta})$. Moreover, we recall from Ref.~\cite{larocca2021diagnosing} that for $W(\vec{\theta})$ to be overparametrized it needs a number of parameters in $\OC(\dim(\mf{k}))=\OC(n^2)$. Since the ansatz in Eq.~\eqref{eq:local-gen} has $(2n-2)$ parameters per layer (see also Fig.~\ref{fig:dansatz-W}) then, we need $L$ to be in $\OC(\dim(\mf{k}))/(2n-2))=\OC(n)$.

We begin with the history state preparation circuit in Fig.~\ref{fig:circuit-F-parallel}. Let us list the gates which we need to implement and their respective depth. Gates that are listed on the same bullet can be parallelized.
\begin{itemize}
    \item Hadamard gates on the clock qubits and $W(\vec{\alpha})$ on the system qubits. The depth is $\OC(n)$ as it is dominated by that of $W(\vec{\alpha})$.
    \item $\log{N}$  controlled gates $e^{-i D 2^{j-1}\varepsilon}$ for $j=1,\ldots,\log{N}$. Each ccontrolled gate has a depth in $\OC(n)$ as we need to control single qubit gates (see Fig.~\ref{fig:diag-history}(b)) . The depth is then $\OC(\log(M)n)$.
    \item A gate $W\ad(\vec{\alpha})$ on the system qubits. The depth is $\OC(n)$.
\end{itemize}
Hence, the total depth to implement the circuit in Fig.~\ref{fig:circuit-F-parallel} is $\OC(\log(M)n+2n)=\OC(\log(M)n)$.

Next, let us consider the circuit in Fig.~\ref{fig:circuit-loschmidt-parallel}. Assuming as in Theorem~\ref{theo:parallel1} that $O_1$ and $O_2$ are Pauli operators we have. 
\begin{itemize}
    \item Hadamard gates on the clock and ancilla qubits and $W(\vec{\alpha})$ on the system qubits. The depth is $\OC(n)$ as it is dominated by that of $W(\vec{\alpha})$.
    \item Apply a controlled $O_2$ gate. If $O_2$ acts non-trivially on at most  $k_2$-qubits (with $k_2\leq n$), then the controlled $O_2$ gate is implemented with $\OC(k_2)$ gates, or more generally with depth in  $\OC(n)$.
    \item $\log{N}$  controlled gates $e^{-i D 2^{j-1}\varepsilon}$ for $j=1,\ldots,\log{N}$. Each ccontrolled gate has a depth in $\OC(n)$ as we need to control single qubit gates (see Fig.~\ref{fig:diag-history}(b)) . The depth is then $\OC(\log(M)n)$.
    \item A gate $W\ad(\vec{\alpha})$ on the system qubits. The depth is $\OC(n)$.
    \item Apply a controlled $O_1$ gate. If $O_1$ acts non-trivially on at most  $k_1$-qubits (with $k_1\leq n$), then the controlled $O_1$ gate is implemented with $\OC(k_1)$ gates, or more generally with depth in  $\OC(n)$. 
    \item Apply $\log(M)$ controlled phase gates. Their depth is $\OC(\log(M))$.
\end{itemize}
The total depth of the circuit is then $\OC(\log(M)n+4n+\log(M))=\OC(\log(M)n)$. Note that if $O_1$ and $O_2$ are instead expressed as a sum of Pauli operators, then the experiment complexity changes, but not the circuit depth.

We now consider the circuit in Fig.~\ref{fig:circuit-Ent_Overlap}. Here, we know that we need to prepare two copies of the history states, which can be done with depth in $\OC(\log(M)n)$ plus implement a depth-two state overlap circuit~\cite{cincio2018learning}. Hence, the depth is still in $\OC(\log(M)n)$.

Finally, the circuit in Fig.~\ref{fig:circuit-Ent_Rand} is composed of the history state preparation circuit plus a random unitary on each qubit. The depth is then in $\OC(\log(M)n)$.
\end{proof}

\section{Numerical methods}  
Here we discuss the analytical techniques used to reduce the computational cost in the simulations of Section \ref{sec:numerics}. \label{App:methods}

\subsection{Simulations via Jordan Wigner}

Consider first the non-uniform $XX$ Hamiltonian of Eq.~\eqref{eq:hxx}, for $n$ sites and periodic boundary conditions (PBC). We have employed the well-known Jordan Wigner (JW) map  
 $c_j^\dag=s_j^+\exp(-i\pi \sum_{k=1}^{j-1}s_k^+s_k^-)$ \cite{lieb1961two} (with $s_j^{\pm}=(X_j\pm iY_j)/2$)
in order to write
\begin{align}
    H_\sigma&=\frac{J}{2}\sum_{j=1}^n (c_j^\dag c_{j+1}+h.c.)+\lambda \sum_{j=1}^n \cos(2\pi \alpha j)c^\dag_jc_j\nonumber\\
   &-J\delta_{\sigma 1}(c^\dag_n c_{n+1}+c^\dag_{n+1}c_n)\,,
\end{align}
which up to the border term is the Aubry-Andre Hamiltonian \cite{aubry1980analyticity}, with $\{c_j,c^\dag_{j'}\}=\delta_{jj'}$ and the other anticommutators vanishing. Here $\sigma$ indicates the parity dependence of $H_\sigma$ on the sector of states in which acts, with $\sigma=\pm 1\equiv e^{i\pi N}$ \cite{canosa2007entanglement} and with the convention $c_{n+1}=c_1$. 
As usual the vacuum state $|0\rangle$ is mapped to $|\downarrow\dots \downarrow\rangle$ with $Z_j=2 c_j^\dag c_j -1$.

Notice that for each parity we can write $H_\sigma=\textbf{c}^\dag M_\sigma \textbf{c}$ with $\textbf{c}=(c_1c_2\dots)^t$ and $M_\sigma$ an $n\times n$ matrix. 
For general ``single particle'' (sp) states  $|\psi\rangle=\sum_j \psi_j c_j^\dag|0\rangle\equiv \psi_1|\uparrow \downarrow\dots \rangle+\psi_2|\downarrow \uparrow\dots \rangle+ \dots$ this allows us to write
\begin{equation}\label{eq:jwexp}
   \mathcal{L}(t)= |\langle \psi|e^{-iHt}|\psi\rangle|^2=|\bm{\psi}^\dag e^{-iM_{-1} t}\bm{\psi}|^2\,,
\end{equation}
with $\bm{\psi}=(\psi_1\psi_2\dots)^t$. Thus we only need to exponentiate a matrix of $n\times n$ dimensions, rather than the original Hamiltonian of size $2^n\times 2^n$. The states we employed in Section \ref{sec:numerics} are mapped to sp states, allowing us to reach large values of $n$. In addition, one has to perform the sums and/or integrals in time. Let us also mention that the spectrum of the model exhibits remarkable (fractal) properties \cite{aubry1980analyticity,wilkinson1984critical,sinai1987anderson} rendering numerical treatments almost mandatory for our purposes.

To obtain $\widetilde{\LC}(\psi_0)$ we basically summed the expression~\eqref{eq:jwexp} over discrete times. Instead, in order to obtain the exact Loschmidt's echo average we used 
\begin{equation}
    \overline{\LC}(\psi_0)=\sum_k |\langle k|\psi\rangle|^4
\end{equation}
 for  $|k\rangle$ of the eigenbasis $H$. Notice that we don't need the complete basis $|k\rangle$ but only those eigenstates corresponding to sp excitations so that $|k\rangle \equiv  \bm{\phi}_k \bm{c}^\dag|0\rangle$ implying $\langle k|\psi\rangle=\bm{\phi}_k^\dag \bm{\psi}$, for $|\psi\rangle$ a sp state. 

For computing the infinite temporal variance of observables we used a similar strategy. One can prove that for a general observable and state \cite{reimann2008foundation} $\sigma_O^2=\sum_{k\neq k'}|\rho_{kk'}|^2|O_{kk'}|^2$ with $O_{kk'}=\langle k|O|k'\rangle$, $\rho_{kk'}=\langle k|\rho|k'\rangle$. In the main body example we used a single particle operator which in fermionic notation can be written as $O=c^\dag_{L/2}c_{L/2+1}+h.c.= \sum_{i,j}M_{ij}c_i^\dag c_j$. Thus one can replace $|O_{kk'}|^2$ with $|M_{kk'}|^2:=|\bf{\phi}_k^\dag M \bf{\phi}_{k'}|^2$ and $|\rho_{kk'}|^2$ with $|\bm{\phi}_k^\dag \bm{\psi}|^2 |\bm{\phi}_{k'}^\dag \bm{\psi}|^2$.

Regarding now the numerical computation of $E_2$, notice first that
\begin{equation}\label{eq:rhot}
   {\rm Tr}\rho^2_S= {\rm Tr}({\rm Tr}_T |\Psi\rangle\langle \Psi|)^2=\frac{1}{N^2}\sum_{t,t'}|\langle \psi(\varepsilon t')|\psi(\varepsilon t)\rangle|^2\,,
\end{equation}
i.e., the quantity ${\rm Tr}\rho^2_T={\rm Tr}\rho^2_S$ also depends on the overlap of the state at different times. 
For sp states, these overlaps have the expression
\begin{equation}
    \langle \psi(t')|\psi(t)\rangle=\bm{\psi}^\dag e^{-iM(t-t')}\bm{\psi}\,,
\end{equation}
meaning again that we only need to exponentiate a matrix of $m\times m$ size. For time-independent Hamiltonians only the time differences matter. This allows one to write the $\rho_S$ purity as the single sum
\begin{equation}\label{eq:onesumpur}
     {\rm Tr}\rho^2_S=\frac{2}{N^2}\sum_{t}(N-t)\mathcal{L}(\varepsilon t)-\frac{1}{N}\,,
\end{equation}
which holds for general time-independent Hamiltonians. For our numerics, we have combined Eqs.~\eqref{eq:onesumpur} and~\eqref{eq:jwexp}.

\subsection{Variational Hamiltonian diagonalization}

Here we present the details of the variational Hamiltonian diagonalization algorithm. All the simulations were  conducted  with the \textit{Julia} programming language \cite{bezanson2017julia}  with the \textit{LinearAlgebra} and \textit{SparseArrays} packages. 

We trained the parameters in the ansatz by using  gradient-based minimization, resorting to the \textit{ADAM} optimizer \cite{kingma2015adam} for both the diagonalization unitary parameters $\vec{\alpha}$ and the diagonal, eigenvalues, matrix ones $\vec{\beta}$. We used two different instances of the ADAM optimizer, both initialized with the same learning rate $\eta_{\vec{\alpha}, \vec{\beta}}=0.1$. Although unusually large (the standard value for ADAM's initial learning rate is $\eta=0.001$) this learning rate proved to be very effective. A training instance of our model comprises $n_r=10$ randomly initialized runs, each lasting at most $n_I = 10^{5}$ training iterations. In practice, runs that achieve loss values less than $10^{-14}$ are stopped and deemed successful. 
We studied the $N=6$ sites non-uniform XX model defined by the Hamiltonian in Eq.~\eqref{eq:hxx} for the value $\alpha=\frac{\sqrt{5}-1}{2}$ that makes it possible, in the thermodynamic limit, for a delocalization-localization transition at the critical point $\lambda=J$ to exist

\subsection{Simulation of the interacting case}
Consider now the fully interacting model of Eq.\ \eqref{eq:hxxnum}. Under the JW map, and assuming open boundary conditions, the term $H_{\text{int}}=\sum_{j=1}^{n-1}Z_j Z_{j+1}$ is mapped to $H_{\text{int}}=4\sum_{j=1}^{n-1}(c^\dag_{j}c_j-1/2)(c^{\dag}_{j+1}c_{j+1}-1/2)$ which from the fermionic perspective corresponds to interactions, essentially of the Hubbard model type \cite{hubbard1963electron}. The cost of exactly simulating the evolution of arbitrary states classically becomes exponential for this model. However, considering that the Hamiltonian preserves the number of fermions, if we restrict our study to a low number of particles we can diagonalize the Hamiltonian in polynomial subspaces.  

We restrict our study to a total number of particles $N'=2$. As noticed in the main body, this simple scenario already captures the phenomenon of single particle localization surviving interactions. In this subspace we can write 
\begin{equation}
    H_{\text{int}}\equiv 4\sum_{j=1}^{n-1}c_j^\dag c_j c^\dag_{j+1}c_{j+1}-c_1^\dag c_1-c_n^\dag c_n\,.
\end{equation}
Now we introduce the notation $|k,l\rangle=c_k^\dag c_l^\dag|0\rangle$ and define the $n^2\times n^2$ matrix $\bar{H}_{ij; kl}:=\langle ij|H|kl\rangle $ with $(i,j)$ and $(k,l)$ joined indices leading to $n^2$ distinct values, i.e. we can write $(i,j)\to a=n(i-1)+j$ for $a=1,\dots, n^2$. 
The entries of $\bar{H}$, although cumbersome, are straightforwardly obtained via fermionic contractions, yielding 
\begin{align}
  \langle i,j|\sum_{i',j'}h_{i'j'}c_{i'}^\dag c_{j'} |k,l\rangle&=\delta_{jl} h_{ik}-\delta_{il} h_{jk}-\delta_{jk} h_{il}+\delta_{ik} h_{jl}\\
   \langle i,j|H_{\text{int}}|k,l\rangle&=2 \{\delta _{j1} \delta _{k1} \delta _{il}-\delta _{i1} \delta _{k1} \delta _{jl}-\delta _{j1} \delta _{l1} \delta _{ik}+\delta _{i1} \delta _{l1} \delta _{jk}-(-\delta _{in} \delta _{jl} \delta _{kn}+\delta _{il} \delta _{jn} \delta _{kn}-\delta _{ik} \delta _{jn} \delta _{ln}+\delta _{in} \delta _{jk} \delta _{ln})\}\nonumber \\ &+4 \{\delta _{ji+1} (\delta _{ik} \delta _{i+1l}-\delta _{i+1k} \delta _{il})-\delta _{ij+1} (\delta _{jk} \delta _{j+1l}-\delta _{j+1k} \delta _{jl})\}\,,
\end{align}
from which $\bar{H}$ is immediately obtained by choosing $h_{ij}$ as in \eqref{eq:hxxnum}, adding the interacting part multiplied by $\Delta$ and properly reshaping the indices.

Given the polynomial size, one can diagonalize $\bar{H}$ exactly. One obtains numerically $\sum_{b}\bar{H}_{a;b}\Psi^{(\mu)}_{b}=\lambda^{\mu}\Psi^{(\mu)}_{a}$ and then aims to define fermionic states which are eigenstates of $H$. In order to do so, an additional projection is generally needed so that $\Psi^{\mu}_{kl}=-\Psi^{\mu}_{lk}$. Having projected the $n^2\times 1$ vector $\Psi^{(\mu)}$ accordingly, it is reshaped to an $n\times n$ matrix allowing to define
\begin{equation}
|\mu\rangle:=\sum_{k,l}\Psi^{(\mu)}_{kl}|k,l\rangle\,,
\end{equation}
satisfying $H|\mu\rangle=\lambda_\mu |\mu\rangle$. 
It is now straightforward to write $\LC(t)=\sum_\mu \langle \psi|\mu\rangle \langle \mu| \psi\rangle e^{iE \lambda_\mu}=\sum_\mu |\langle \psi|\mu\rangle |^2 e^{iE \lambda_\mu}$ since the states $|\mu\rangle$ provide a complete basis of the $N'=2$ subspace. In addition, for $|\psi\rangle=|i,j\rangle$ one obtains $|\langle \psi|\mu\rangle |^2=|\Psi_{ij}^{(\mu)}|^2$. 
With these results at hand, one can write the temporal sums involved in computing $\bar{\LC}(i,j)$, namely the average of the Loschmidt's echo for the initial state $|\psi\rangle=|i,j\rangle$, in closed form:
\begin{equation}\label{eq:Linteracting}
    \widetilde{\LC}(i,j)=\frac{4}{N}\sum_{\mu\leq \mu'=1}^{n^2}|\Psi_{ij(\mu)}^{(\mu)}|^2|\Psi_{ij(\mu')}^{(\mu')}|^2\left[1-\cos (T (\lambda_\mu-\lambda_{\mu'}))+\cot \left(\frac{\varepsilon}{2}  (\lambda_\mu-\lambda_{\mu'})\right) \sin (T (\lambda_\mu-\lambda_{\mu'}))\right]\,.
\end{equation}
Similarly, and using \eqref{eq:onesumpur} we obtain
\begin{equation}
    \Tr[\rho^{2}_S]= \frac{4}{N}\sum_{\mu\leq\mu'=1}^{n^2}|\Psi_{ij(\mu)}^{(\mu)}|^2|\Psi_{ij(\mu')}^{(\mu')}|^2\left[\frac{\csc ^2\left(\frac{\varepsilon}{4} (\lambda_\mu-\lambda_{\mu'})\right) \left(N \left(-\cos \left(\frac{\varepsilon}{2}  (\lambda_\mu-\lambda_{\mu'})\right)\right)-\cos \left(\frac{T}{2}  (\lambda_\mu-\lambda_{\mu'})\right)+N+1\right)}{N^2}-\frac{1}{N}\right]\,.
\end{equation}
In these expressions $ij(\mu)$ represents those $ij$ for which $m=n (i-1)+j$. 
These are the final expressions we employed in section \ref{sec:numerics} with the eigenvalues $\lambda_\mu$, and eigenfunctions $\Psi^{(\mu)}_{ij}$ obtained via numerical diagonalization as described above. 
Finally, notice that by similar means we can compute $\bar{\LC}(\psi_0)$ using
\begin{equation}
    \bar{\LC}(\psi_0)=\sum_\mu |\langle \psi|\mu\rangle|^4=4\sum_\mu |\Psi_{ij(\mu)}^{(\mu)}|^4\,,
\end{equation}
which may be recovered from \eqref{eq:Linteracting} and taking the limit $T\to \infty$.

Lastly, we provide yet another method of simulation of our chain based on Tensor Network (TN) methods \cite{orus2014practical,schollwock2011density,Vidal2003Efficient,white1992density}.
In the case of one-dimensional systems, as the one studied throughout our manuscript, Matrix Product States (MPS) methods are the best tool TNs have to offer.
MPSs are a powerful tool for simulating the dynamics of quantum many-body systems as they provide an efficient classical representation of quantum states with limited entanglement, which is often the case for ground states and low-energy excitations of local Hamiltonians. 
In a quantum system with $N$ sites, the state $\ket{\psi}$ is typically described by a rank-$N$ tensor $\Psi_{i_1 i_2 \ldots i_N}$, where each index $i_k$ corresponds to a local basis state at site $k$. For a system of $d$-dimensional local spaces, this tensor has $d^N$ elements, which becomes infeasible to store and manipulate directly for large $N$.
The key idea of MPS is to express this large tensor as a contraction of a series of smaller tensors, each associated with a site
\begin{equation}
\Psi_{i_1 i_2 \ldots i_N} = \sum_{\alpha_1, \alpha_2, \ldots, \alpha_{N-1}} A^{[1]}_{i_1 \alpha_1} A^{[2]}_{\alpha_1 i_2 \alpha_2} \cdots A^{[N]}_{\alpha_{N-1} i_N}
\end{equation}
Here, each $A^{[k]}$ is a tensor with three indices: $i_k$ is the physical index corresponding to the local basis at site $k$, and $\alpha_{k-1}$, $\alpha_k$ are the bond indices that connect neighboring sites. The bond dimension $\chi$ is the maximum dimension of the $\alpha$ indices, determining the amount of entanglement the MPS can represent. Larger $\chi$ values allow for the representation of more entangled states but increase computational costs.

To simulate the time evolution of an initial quantum state $|\psi_0\rangle$ under a given Hamiltonian $H$, one can use Trotterization. The goal is to approximate the time evolution operator $e^{-iHt}$, which describes how the state evolves over time $t$, as a sequence of local gates. This is achieved by breaking the evolution into smaller time steps $\delta t$ and decomposing the Hamiltonian into local terms:

\begin{equation}
e^{-iHt} \approx \left(e^{-iH_1 \delta t} e^{-iH_2 \delta t} \cdots e^{-iH_N \delta t}\right)^n
\end{equation}

where $n = t/\delta t$ and $H = H_1 + H_2 + \cdots + H_N$. The second-order Trotter-Suzuki decomposition is commonly used due to its balance between accuracy and computational efficiency. The error for a single time step in second-order Trotterization scales as $\mathcal{O}((\delta t)^3)$, meaning that the total error over the entire simulation time is $\mathcal{O}(t(\delta t)^2)$. Notice that studying Trotterization and its performace is crucial, as it could play a key role in implementing quantum evolution algorithms on actual quantum computers.

Hence, one can encode the initial state $|\psi_0\rangle$ in an MPS, and repeatedly updating it via the gates resulting from trotterizing the evolution operator. To apply a gate to an MPS one can resort to the Time-Evolving Block Decimation (TEBD) algorithm. TEBD takes advantage of the local nature of interactions in the Hamiltonian to update the MPS tensors sequentially by first joining two adjacent tensors $A^{[k]}, A^{[k+1]}$ into one, contracting along the shared bond index, contracting the resulting tensor with the gate, and ultimately performing singular value decomposition to split the updated tensor back into two local ones. The computational cost of TEBD is $\mathcal{O}(N\chi^3)$, where $N$ is the number of sites and $\chi$ is the bond dimension.
It is important to notice that the bond dimension $\chi$ is a critical parameter in MPS simulations. Besides determining the maximum amount of entanglement that the MPS can capture it also controls the feasibility of the chosen algorithm. While in principle one could not constraint $\chi$, letting it increase during the evolution to account for the possibly increasing correlations, most often one decides for a maximum value $\chi_{\rm max}$ after which the MPS method begins to be approximate.
In summary, MPS combined with Trotterization and the TEBD algorithm provides an efficient framework for simulating time evolution in one-dimensional quantum systems. The choice of the bond dimension $\chi_{\rm max}$ and the time step $\delta t$ are key factors in managing computational cost and error.

\begin{figure}[t!]
    \centering
    \includegraphics[width=0.7\columnwidth]{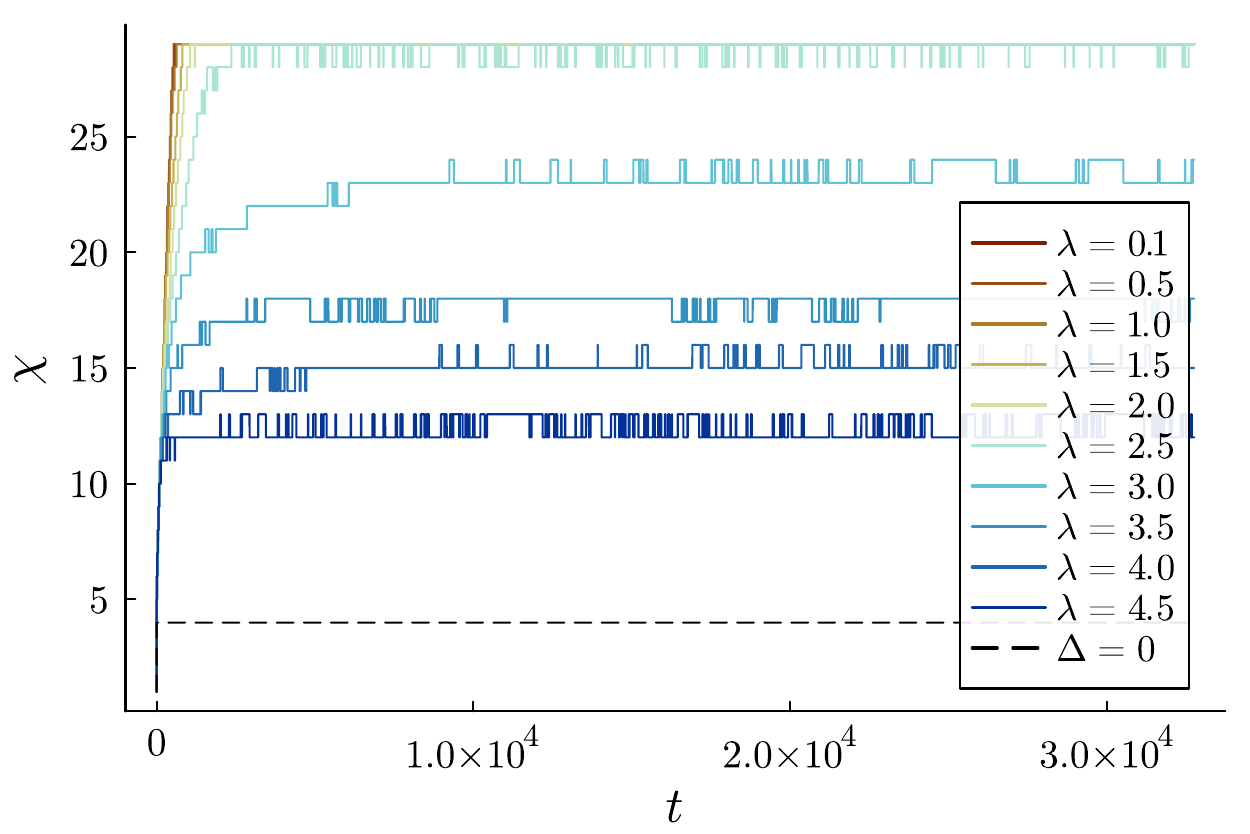}
    \caption{\textbf{Scaling of Bond Dimension with time.} We report the scaling with time of the bond dimension $\chi$ of the MPS used to represent the interacting model state, with $\Delta=0.05$. The black, dashed line shows the non-interacting model ($\Delta=0$) bond dimension, which does not depend on $\lambda$ and is equal to $\chi=4$ throughout the whole system's evolution. Instead, in the interacting case, the maximum value of $\chi$ becomes smaller as the localization strength $\lambda$ increases, confirming that the stronger the localization the easier the description of the system.}
    \label{fig:chi_scaling}
\end{figure}

Here we showcase how such an MPS based method fares when facing the interacting model studied in the main text. Particularly, we are interested in the scaling of the unconstrained $\chi$ with the number of time steps, and in how much the trotterization error makes the results deviate from the ones obtained via exact diagonalization.
Particularly, Fig.~\ref{fig:chi_scaling} shows how $\chi$ grows with time in the interacting model with $\Delta=0.05$ for the different values of $\lambda$ considered in the main text. Compared to the non interacting case, for which we find $\chi=4$ through the whole evolution, now the less the system is localized (small $\lambda$) the more the correlations grow. However, we can see that the bond dimension eventually caps off at values smaller than $\chi=30$, ensuring that the method remains very efficient.

\begin{figure*}[htb]
  \centering
  \includegraphics[width=\linewidth]{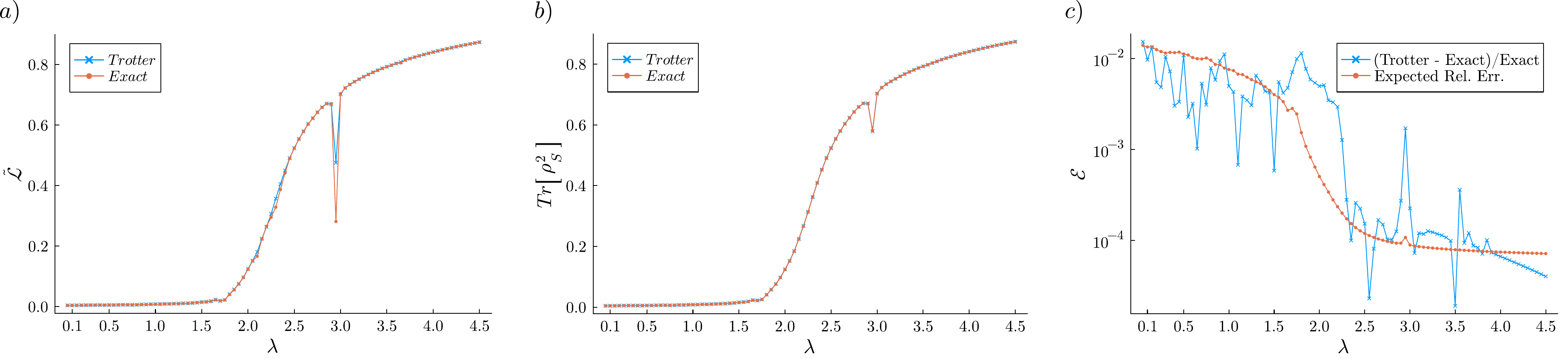}
  \caption{\textbf{Trotterization Error for Loschmidt echo and Entanglement.} Comparison between the \textit{Exact} diagonalization technique and the Trotterized (\textit{Trotter}) MPS method for the interacting model with $\Delta=0.05$. Panels $a)$ and $b)$ show the Loschmidt echo $\tilde{\LC}$ and system-time entranglement $\Tr[\rho_S^2]$, respectively. We can numerically observe how the finer information encoded in the former cannot be exactly caputerd via Trotterization, whereas we get perfet accordance in the entanglement behavior. Lastly, panel $c)$ shows that the numerical relative error between the two techniques for computing the system-time entanglement matches the expected error introduced by second-order Trotterization.
  }
  \label{fig:trotter_errors}
\end{figure*}

Finally, in Fig.~\ref{fig:trotter_errors} we show a comparison between the values of the Loschmidt echo and of the sytem-time entanglement obtained through exact diagonalization and through MPS simulation in the interacting case, where we keep $\Delta=0.05$. As expected, the Loschmidt echo is more susceptible to errors, and while the trend is reproduced, trotterization fails at capturing the finer details. This does not happen for the system time entaglement, which get exactly reproduced. The last panel of Fig.~\ref{fig:trotter_errors} confirms the predicted error coming from trotterization, showing that no further sources of uncertainty are affecting the simulation.

\end{document}